\pdfoutput=1 
\documentclass[cernpreprint,english,USenglish]{na61doc}

\usepackage{lineno, blindtext}
\usepackage{amsmath}
\usepackage{enumerate}
\usepackage{placeins}
\usepackage{appendix}
\usepackage[T1]{fontenc}
\usepackage[utf8]{inputenc}
\usepackage{lmodern}
\usepackage{chngcntr}
\usepackage{microtype}
\usepackage{lineno}
\usepackage{color}
\usepackage{epstopdf}
\usepackage{colortbl}
\usepackage{tabularx}

\usepackage{subcaption}
\usepackage{booktabs}
\usepackage{url}
\usepackage{bbding}

\definecolor{kOrange+8}{RGB}{255,102,51}
\definecolor{kBlue}{RGB}{0,0,204}
\definecolor{kRed+2}{RGB}{153,0,0}
\definecolor{kGreen}{RGB}{0,153,0}
\definecolor{kAuAu}{RGB}{255,153,153}

\newcommand{\eV}{\ensuremath{\mbox{e\kern-0.1em V}}\xspace}
\newcommand{\GeV}{\ensuremath{\mbox{Ge\kern-0.1em V}}\xspace}
\newcommand{\MeV}{\ensuremath{\mbox{Me\kern-0.1em V}}\xspace}
\newcommand{\GeVc}{\ensuremath{\mbox{Ge\kern-0.1em V}\!/\!c}\xspace}
\newcommand{\GeVcc}{\ensuremath{\mbox{Ge\kern-0.1em V}\!/\!c^2}\xspace}
\newcommand{\AGeV}{\ensuremath{A\,\mbox{Ge\kern-0.1em V}}\xspace}
\newcommand{\AGeVc}{\ensuremath{A\,\mbox{Ge\kern-0.1em V}\!/\!c}\xspace}
\newcommand{\MeVc}{\ensuremath{\mbox{Me\kern-0.1em V}/c}\xspace}

\newcommand{\y}{\ensuremath{y}\xspace}

\newcommand{\pt}{\ensuremath{p_{T}}\xspace}

\newcommand{\GeantThree}{{\scshape Geant3}\xspace}

\newcommand{\Epos}{{\scshape Epos}\xspace}

\newcommand{\Crmc}{{\scshape Crmc}\xspace}

\def\avg#1{\langle{#1}\rangle}

\newcommand{\NASixtyOne}{NA61\slash SHINE\xspace}
\newcommand{\CernVM}{\textsc{Cern\-\kern-0.05emVM}\xspace}

\newcommand{\pim}{\ensuremath{\pi^-}\xspace}
\newcommand{\pip}{\ensuremath{\pi^+}\xspace}

\newcommand{\pp}{\mbox{\textit{p+p}}\xspace}
\newcommand{\NN}{\mbox{\textit{N+N}}\xspace}

\newcommand{\dEdx}{$\text{d}E/\text{d}x$ }
\newcommand{\coordinate}[1]{{\fontfamily{lmss}\selectfont#1}}

\definecolor{darkred}{rgb}{0.5,0,0}
\definecolor{darkblue}{rgb}{0,0,0.5}
\definecolor{firebrick}{rgb}{0.75,0.125,0.125}
\definecolor{darkgreen}{rgb}{0,0.5,0}
\definecolor{kPink+2}{RGB}{204,102,153}
\definecolor{kOrange+8}{RGB}{255,102,51}
\definecolor{kGreen+2}{RGB}{0,153,0}
\definecolor{kCyan+2}{RGB}{0,153,153}
\definecolor{kBlue+2}{RGB}{0,0,153}
\definecolor{kRed+1}{RGB}{204,0,0}
\definecolor{kBlue}{RGB}{0,0,204}
\definecolor{kBlue-9}{RGB}{153,153,255}
\definecolor{kGreen}{RGB}{0,153,0}
\definecolor{kRed}{RGB}{204,0,0}
\definecolor{kCyan}{RGB}{51,204,204}
\definecolor{kMagenta}{RGB}{153,0,153}
\definecolor{kPink}{RGB}{204,0,102}
\definecolor{kGray}{RGB}{204,204,204}
\definecolor{kBlack}{RGB}{0,0,0}

\definecolor{kRed+3}{RGB}{102,0,0}
\definecolor{kRed+2}{RGB}{153,0,0}
\definecolor{kRed-4}{RGB}{255,51,51}
\definecolor{kRed-7}{RGB}{255,102,102}
\definecolor{kRed-9}{RGB}{255,153,153}

\definecolor{color150Comparison}{RGB}{204,0,0}
\definecolor{color75Comparison}{RGB}{204,204,0}
\definecolor{color40Comparison}{RGB}{0,153,0}
\definecolor{color30Comparison}{RGB}{51,153,255}
\definecolor{color19Comparison}{RGB}{0,0,204}

\newcommand{\markerNineteenComparison}{\CircleSolid}
\newcommand{\markerThirtyComparison}{\TriangleDown}
\newcommand{\markerFourtyComparison}{\Plus}
\newcommand{\markerSeventyFifeComparison}{\TriangleUp}
\newcommand{\markerOneFiftyComparison}{\SquareSolid}

\ShineTitle{
Measurements of $\pi^-$ production in $^7$Be+$^9$Be collisions at beam momenta from 19$A$ to 150\AGeVc 
in the \NASixtyOne experiment at the CERN SPS 
}

\PreprintIdNumber{CERN-EP-2020-150}

\ShineJournal{Eur. Phys. J. C}
\ShineAbstract{ 
The \NASixtyOne collaboration studies at the CERN Super Proton Synchrotron (SPS) the onset
of deconfinement in hadronic matter by the measurement of particle production in collisions of nuclei
with various sizes at a set of energies covering the SPS energy range. This paper presents results
on inclusive double-differential spectra and mean multiplicities of $\pi^{-}$ mesons produced in the 5\% most \textit{central}
$^7$Be+$^9$Be collisions at beam momenta of 19$A$, 30$A$,
40$A$, 75$A$ and 150\AGeVc obtained by the so-called $h^-$ method
which does not require any particle identification.

The shape of the transverse mass spectra differs from the shapes measured in  
central Pb+Pb collisions and inelastic \pp interactions.
The normalized width of the rapidity distribution decreases with increasing collision energy and
is in between the results for inelastic nucleon-nucleon and central Pb+Pb collisions. 
The mean multiplicity of pions per wounded nucleon in \textit{central} $^7$Be+$^9$Be collisions 
is close to that in central Pb+Pb collisions up to 75\AGeVc. However, at the top SPS energy the result 
lies between those for nucleon-nucleon and Pb+Pb interactions.

The results are discussed in the context of predictions for the onset of deconfinement at the CERN SPS
collision energies.
}
\begin{document}

\maketitle


\newpage 
{\Large The \NASixtyOne Collaboration}
\bigskip

\noindent
A.~Acharya$^{\,9}$,
H.~Adhikary$^{\,9}$,
A.~Aduszkiewicz$^{\,15}$,
K.K.~Allison$^{\,25}$,
E.V.~Andronov$^{\,21}$,
T.~Anti\'ci\'c$^{\,3}$,
V.~Babkin$^{\,19}$,
M.~Baszczyk$^{\,13}$,
S.~Bhosale$^{\,10}$,
A.~Blondel$^{\,4}$
% ...


\noindent
% ...
$^{3}$~Ru{\dj}er Bo\v{s}kovi\'c Institute, Zagreb, Croatia\\
$^{4}$~LPNHE, University of Paris VI and VII, Paris, France\\
% ...
$^{9}$~Jan Kochanowski University in Kielce, Poland\\
$^{10}$~Institute of Nuclear Physics, Polish Academy of Sciences, Cracow, Poland\\
% ...
$^{13}$~AGH - University of Science and Technology, Cracow, Poland\\
% ...
$^{15}$~University of Warsaw, Warsaw, Poland\\
% ...
$^{19}$~Joint Institute for Nuclear Research, Dubna, Russia\\
% ...
$^{21}$~St. Petersburg State University, St. Petersburg, Russia\\
% ...
$^{25}$~University of Colorado, Boulder, USA\\
% ...


\section{Introduction}
This paper presents measurements of inclusive spectra
and mean multiplicities of $\pi^{-}$ mesons produced in \textit{central}  $^7$Be+$^9$Be collisions at beam
momenta of 19$A$, 30$A$, 40$A$, 75$A$ and 150\AGeVc ($\sqrt{s_{\textit{NN}}}$ = 6.1, 7.6, 8.8, 11.9 and 16.8 GeV) 
performed by the \NASixtyOne collaboration.
These results are part of the strong interactions studies proposed by the \NASixtyOne collaboration~\cite{Antoniou:2006mh} to
investigate the properties of the onset of deconfinement and to search for the possible existence
of a critical point in the phase diagram of strongly interacting matter. 
The first goal of the programme is motivated by
the observation of rapid changes of hadron production properties in central Pb+Pb
collisions at about 30\AGeVc by the NA49 experiment~\cite{Afanasiev:2002mx,Alt:2007aa} -
a sharp peak in the kaon to pion ratio ("horn"), the start of a plateau in the inverse slope parameter for kaons ("step"), and a steepening of the increase of pion production per
wounded nucleon with increasing collision energy ("kink"). These findings
were predicted and interpreted as the onset of deconfinement~\cite{Gazdzicki:1998vd,Gazdzicki:2010iv}. 
They were recently confirmed by the RHIC beam
energy scan programme~\cite{Adamczyk:2017iwn}, and the interpretation is supported by the LHC
results (see Ref.~\cite{Rustamov:2012np} and references therein). 
Experimentally the goals of the \NASixtyOne strong interaction programme are persued by
a two dimensional scan in collision energy and nuclear mass number of colliding nuclei. 
The scan allows 
to explore systematically the phase diagram of strongly interacting matter~\cite{Antoniou:2006mh}. 
In particular, the analysis of the existing data within the framework of statistical models suggests that by increasing collision
energy one increases temperature and decreases baryon chemical potential  of strongly interacting matter at freeze-out~\cite{Becattini:2005xt}, whereas by increasing nuclear mass number of the colliding nuclei one decreases the temperature~\cite{Alt:2007uj,Becattini:2005xt,Vovchenko:2015idt}.

Within this programme \NASixtyOne recorded data on \pp, Be+Be, Ar+Sc, Xe+La and Pb+Pb collisions.
Further high statistics measurements of Pb+Pb collisions are planned with an upgraded detector 
starting in 2021~\cite{PbAddendum}.
Results on particle spectra and multiplicities have already been published from
\pp interactions~\cite{Abgrall:2013pp_pim,Aduszkiewicz:2017sei,Aduszkiewicz:2019zsv} which represent the basic reference.
This paper reports \NASixtyOne results from the next step in size of the collision system namely
measurements of $\pi^-$ production for the 5\% most \textit{central} $^7$Be+$^9$Be collisions. The data were 
recorded in 2011, 2012 and 2013 using a secondary $^7$Be beam produced by fragmentation of the primary Pb beam 
from the CERN SPS~\cite{Abgrall:7Bebeam}.
The $^7$Be+$^9$Be collisions play a special role in the \NASixtyOne scan programme. First, it was predicted within 
the statistical models~\cite{Poberezhnyuk:2015wea,Motornenko:2018gdc} that the yield ratio of strange hadrons to pions in these collisions should be close to those in central
Pb+Pb collisions and significantly higher than in \pp interactions.
Second, the collision system composed of a $^7$Be and a $^9$Be nucleus has eight protons and eight neutrons, and thus is isospin symmetric. 
Within the \NASixtyOne scan programme the $^7$Be+$^9$Be collisions serve as the lowest mass isospin symmetric reference needed to study collisions of medium and large mass nuclei. This is of particular importance when data on proton-proton, neutron-proton and neutron-neutron are not available to construct the nucleon-nucleon reference~\cite{Gazdzicki:1991ih}.

In this paper the so-called $h^-$ method is used for determining
$\pi^{-}$ production since it provides the largest phase space coverage. This procedure utilizes the fact
that negatively charged particles are predominantly $\pi^{-}$ mesons with a small admixture
(of order 10\%) of $K^-$ mesons and anti-protons which can be subtracted reliably.

The paper is organized as follows: after this introduction the experiment is briefly described
in Sec.~\ref{sec:experiment}. The analysis procedure is discussed in Sec.~\ref{sec:analysis}. 
Section~\ref{sec:results} describes the results of the analysis. In Sec.~\ref{sec:discussion} the new measurements are discussed 
and compared to model calculations. A summary closes the paper.

The following variables and definitions are used in this paper. The particle rapidity \y is calculated
in the collision center of mass system (cms), $\y=0.5 \cdot ln{[(E+p_{L})/(E-p_{L})]}$, where $E$
and $p_{L}$ are the particle energy and longitudinal momentum, respectively. The transverse component
of the momentum is denoted as $p_T$, and the transverse mass $m_T$ is defined as
$m_T = \sqrt{m^2 + (c p_T)^2}$
where $m$ is the particle mass. The momentum in the laboratory frame is denoted $p_{\text{lab}}$ and the
collision energy per nucleon pair in the center of mass by $\sqrt{s_{\textit{NN}}}$.

Be+Be collisions can be characterized by the energy detected in the region populated by projectile spectators. Low values of this forward energy are referred to
\textit{central} collision and a selection of collisions based on the 
forward energy is called a \textit{centrality} selection.
Although for Be+Be collisions the forward energy is not tightly correlated with 
the impact parameter of the collision, the terms \textit{central} and \textit{centrality} are adopted following the convention widely used in heavy-ion physics.

\section{Experimental setup}\label{sec:experiment}

\subsection{Detector}

\begin{figure} 
        \centering
        \includegraphics[width=1\linewidth]{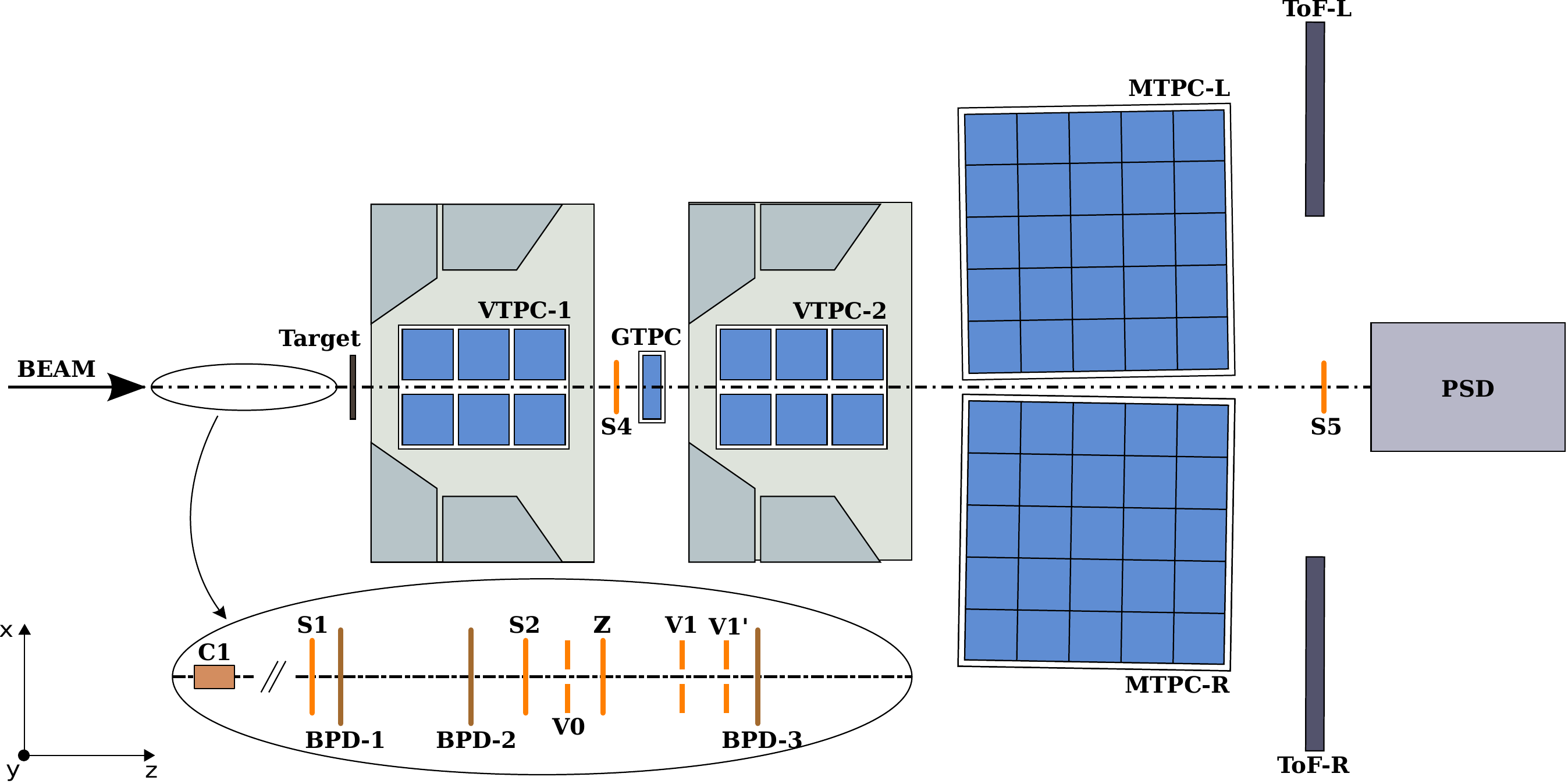}
        \vspace{0.3cm}
        \caption{The schematic layout of the \NASixtyOne experiment at the CERN SPS ~\cite{Abgrall:2014fa}
                 showing the components used for the Be+Be energy scan (horizontal cut, not to scale).
                 The beam instrumentation is sketched in the inset (see also Fig.~\ref{fig:beamAndTriggerDetectors} 
                 below). Alignment of the chosen coordinate system as shown in the figure;
                 its origin lies in the middle of VTPC-2, on the beam axis.
                 The \coordinate{z} axis is along the nominal beam direction. The magnetic field 
                 bends charged particle trajectories in the \coordinate{x}--\coordinate{z} (horizontal) plane.
                 The drift direction in the TPCs is along the \coordinate{y} (vertical) axis.}
        \label{fig:detectorSetup}
\end{figure}

The \NASixtyOne experiment is a multi-purpose facility designed to measure particle production in
nucleus-nucleus, hadron-nucleus and \pp interactions~\cite{Abgrall:2014fa}. The detector
is situated at the CERN Super Proton Synchrotron (SPS) in the H2 beamline of the North experimental area.
A schematic diagram of the setup during Be+Be data taking is shown in Fig.~\ref{fig:detectorSetup}.
The main components of the produced particle detection system are four large volume
Time Projection Chambers (TPC). Two of them, called Vertex TPCs (VTPC), are located
downstream of the target inside superconducting magnets with maximum combined bending power of 9~Tm.
The magnetic field was scaled down in proportion to the beam momentum in order to obtain
similar phase space acceptance at all energies. The main TPCs (MTPC) and two walls of
pixel Time-of-Flight (ToF-L/R) detectors are placed symmetrically to the beam line downstream of the magnets.
The fifth small TPC (GAP-TPC) is placed between VTPC1 and VTPC2 directly on the beam line.
The TPCs are filled with Ar:CO$_{2}$ gas mixtures in proportions 90:10 for the VTPCs and the GAP-TPC, 
and 95:5 for the MTPCs. 

The Projectile Spectator Detector (PSD), which covers the region into which the projectile spectators are emitted 
is positioned 20.5 m (16.7 m) downstream of the MTPCs at 75$A$ and 150\AGeVc
(19$A$, 30$A$, 40\AGeVc) centered in the transverse plane on
the deflected position of the beam. The PSD allows to select the \textit{centrality}  of the
collision by imposing an upper limit on the measured forward energy.

\begin{figure}[!ht]
\centering
\includegraphics[width=\textwidth]{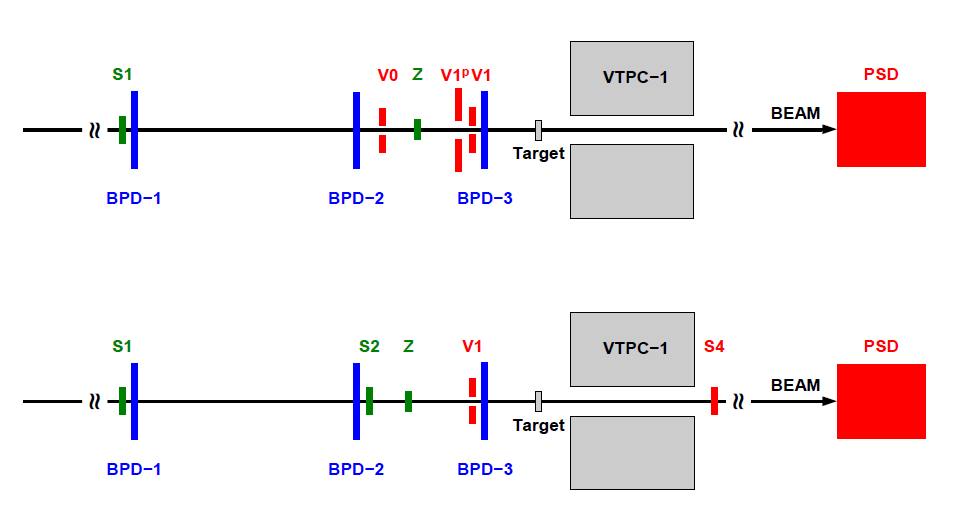}
\caption{The schematic of the placement of the beam and trigger detectors in high-momentum \emph{(top)} and 
         low-momentum \emph{(bottom)} data taking configurations showing scintillation counters S,
         veto counters V, charge Cherekov counter Z and beam position detectors BPD.}
\label{fig:beamAndTriggerDetectors}
\end{figure}

The beam line instrumentation is schematically depicted in Fig.~\ref{fig:beamAndTriggerDetectors}.
A set of scintillation counters as well as beam position detectors (BPDs)~\cite{Abgrall:2014fa} upstream of the target
provide timing reference, selection, identification and precise measurement of the position
and direction of individual beam particles.

The target was a 12~mm thick plate of $^9$Be placed $\approx$ 80 cm upstream of VTPC1. 
Mass concentrations of impurities in the target were measured at 0.3\%
resulting in an estimated increase of the produced pion multiplicity by less than 0.5\%
~\cite{Banas:2018sak}. No correction was applied for
this negligible contamination. 
Data were taken with target inserted (denoted I, 90\%) and target removed (denoted R, 10\%). 

\subsection{$^7$Be Beam}

The beam line of \NASixtyOne experiment is designed to provide good momentum resolution and particle identification even with secondary ion beams.
The beam instrumentation (see Fig.~\ref{fig:beamAndTriggerDetectors}) consists of scintillator counters (S) 
used for triggering and beam particle identification, 
veto scintillation counters (V) with a hole in the middle for rejection of upstream interactions 
and beam halo particles, and a Cherenkov charge detector Z built based on quartz glass radiator for the measurement of the secondary beam charge. Additionally the three 
Beam Position Detectors (BPDs) are used for determination of the charge of individual beam particles.

\begin{figure}[ht!]
        \centering
        \includegraphics[width=0.45\linewidth]{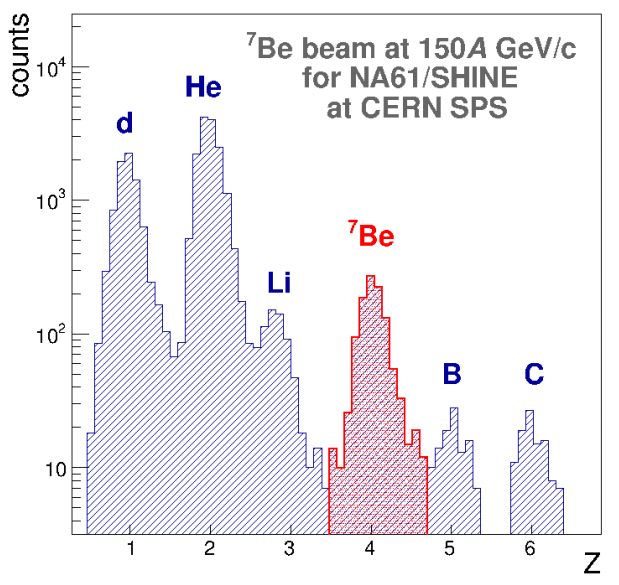}
        \caption{Charge of the beam particles measured by the Z detector.}
        \label{fig:beamComposition}
\end{figure}

\begin{figure}[ht!]
        \centering
        \includegraphics[width=1\linewidth]{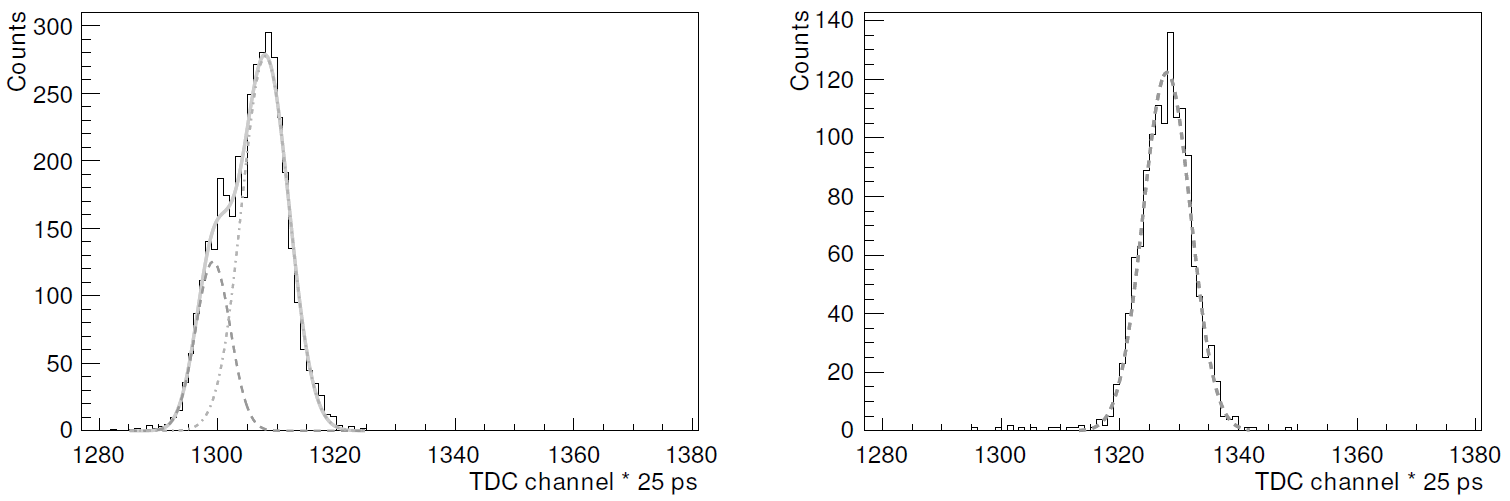}
        \caption{Mass of fragments of Z/A with momentum of 13.9\AGeVc. Left: carbon ions show double Gaussian 
         structure due to two isotopes of carbon in the beam. Right: beryllium ions show single Gaussian distribution, 
         indicating isotopic purity of the beryllium in the beam. Charge of the beam particle was selected by the measurement of scintillation counters.}
        \label{fig:beamAdet}
\end{figure}

This paragraph provides a brief description of the $^7$Be beam properties (see \cite{Abgrall:7Bebeam}). 
Primary Pb$^{82+}$ ions extracted from the SPS were steered toward a 180 mm long beryllium fragmentation target 
placed 535~m upstream of the \NASixtyOne experiment. An interaction of a Pb-ion with the fragmentation target produces a mixture of nuclear fragments with a large fraction of so-called spectator nucleons which originate from the Pb nucleus but did not participate in the collision. Their momenta per nucleon $p_{N}$ are equal to the beam momentum per nucleon smeared by Fermi motion.
The field strength in the bending magnets of the beam line define the rigidity of the transported charged particles: $B\rho = 3.33 \cdot p_\text{beam}/Z$, 
where $B\rho$ can be adjusted by setting the current in the dipole magnets 
and $p_\text{beam}=A \cdot p_{N}$ is the beam momentum and $Z$ the charge of the beam particle.
Thus the beam line selects particles with the wanted $A/Z$ ratio. 
Figure~\ref{fig:beamComposition} shows the charge spectrum of a fragment beam with a rigidity corresponding to fully stripped $^7$Be ions with a momentum of 150~\AGeVc measured by the Z detector. 
A well separated peak for charge $Z$ equal 4 is visible. In a special run taken at beam momentum of 13.9\AGeVc it was possible to also
measure the time-of-flight of the beam particles. As demonstrated in Fig.~\ref{fig:beamAdet} the selected Be
fragments are pure $^7$Be.

\subsection{Trigger}

The schematic of the placement of the beam and trigger detectors can be seen in Fig.~\ref{fig:beamAndTriggerDetectors}. 
The trigger detectors consist of a set of scintillation counters recording the presence of the beam particle (S1, S2), 
a set of veto scintillation counters with a hole used to reject beam particles passing far from the centre of 
the beamline (V0, V1), and a charge detector (Z). Beam particles were defined by the coincidence
T1 = $\text{S1} \cdot \text{S2} \cdot \overline{\text{V1}} \cdot \text{Z(Be)}$ and
T1 = $\text{S1} \cdot \overline{\text{V0}} \cdot \overline{\text{V1}} \cdot \overline{\text{V1'}} \cdot \text{Z(Be)}$ 
for low and high momentum data taking respectively. 
In addition, for the two lower energies an interaction trigger detector (S4) was used to check whether the beam particle changed charge after 
passing through the target. \textit{Central} collisions were selected by requiring an energy signal 
below a set threshold from the 16 central modules of the PSD. The event trigger condition thus was 
T2 = T1$\cdot \overline{\text{S4}} \cdot \overline{\text{PSD}}$ and T2 = T1$\cdot\overline{\text{PSD}}$ for the lower and higher energies, respectively. The PSD threshold was set to
retain from $\approx$~70\% to $\approx$~40\% of inelastic interactions 
at beam momenta from 19$A$ to 150\AGeVc, respectively. 
The statistics of recorded events are summarised in Tab.~\ref{tab:eventStat}.

\begin{table}
   \caption{
Basic beam properties, number of events recorded, and number of events selected for the analysis
for $^7$Be+$^9$Be interactions of 5~\% most \textit{central} collisions at incident momenta of
19$A$, 30$A$, 40$A$, 75$A$ and 150\AGeVc.
}
\vspace{0.3cm}
\centering
\begin{tabular}{| c | c | c | c |}
  \hline\hline
    $p_\text{beam}\ (\AGeVc)$ & $\sqrt{s_{NN}}$ $(\GeV)$ & \parbox[][1.5cm][c]{2.5cm}{\centering Recorded events (all triggers)} & \parbox[][1.5cm][c]{2cm}{\centering Number of selected events} \\
    \hline
    19  & 6.1  & $3.46\cdot10^{6}$ & $0.33\cdot10^{5}$  \\
    30  & 7.6  & $5.41\cdot10^{6}$ & $0.37\cdot10^{5}$  \\
    40  & 8.8  & $3.42\cdot10^{6}$ & $0.99\cdot10^{5}$  \\
    75  & 11.9 & $5.24\cdot10^{6}$ & $1.00\cdot10^{5}$ \\
    150 & 16.8 & $2.93\cdot10^{6}$ & $0.81\cdot10^{5}$ \\\hline\hline
  \end{tabular}
  \label{tab:eventStat}
\end{table}

\section{Analysis procedure}\label{sec:analysis}

\begin{figure}
        \centering 
        \includegraphics[width=0.45\textwidth]{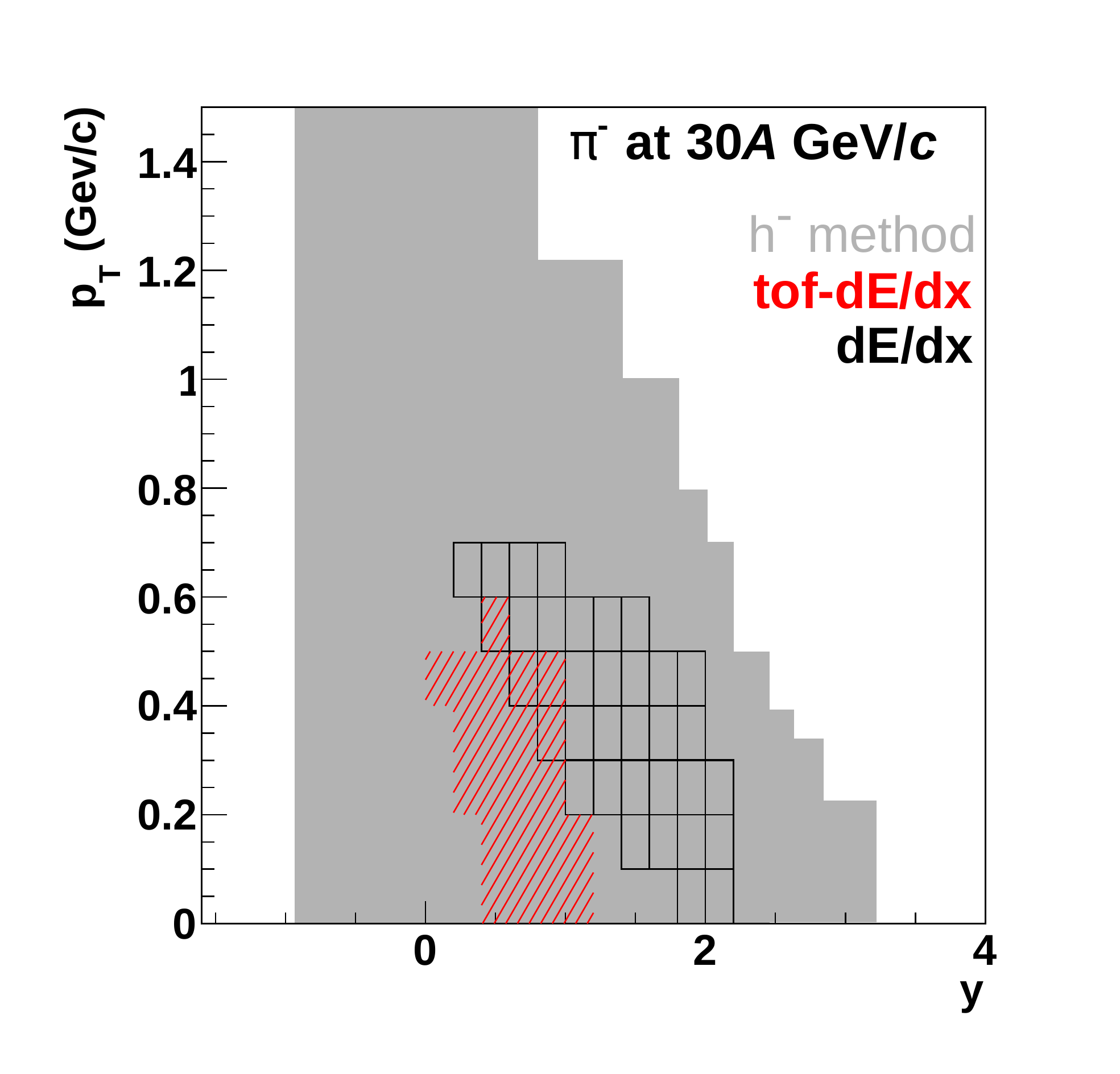}
        \includegraphics[width=0.45\textwidth]{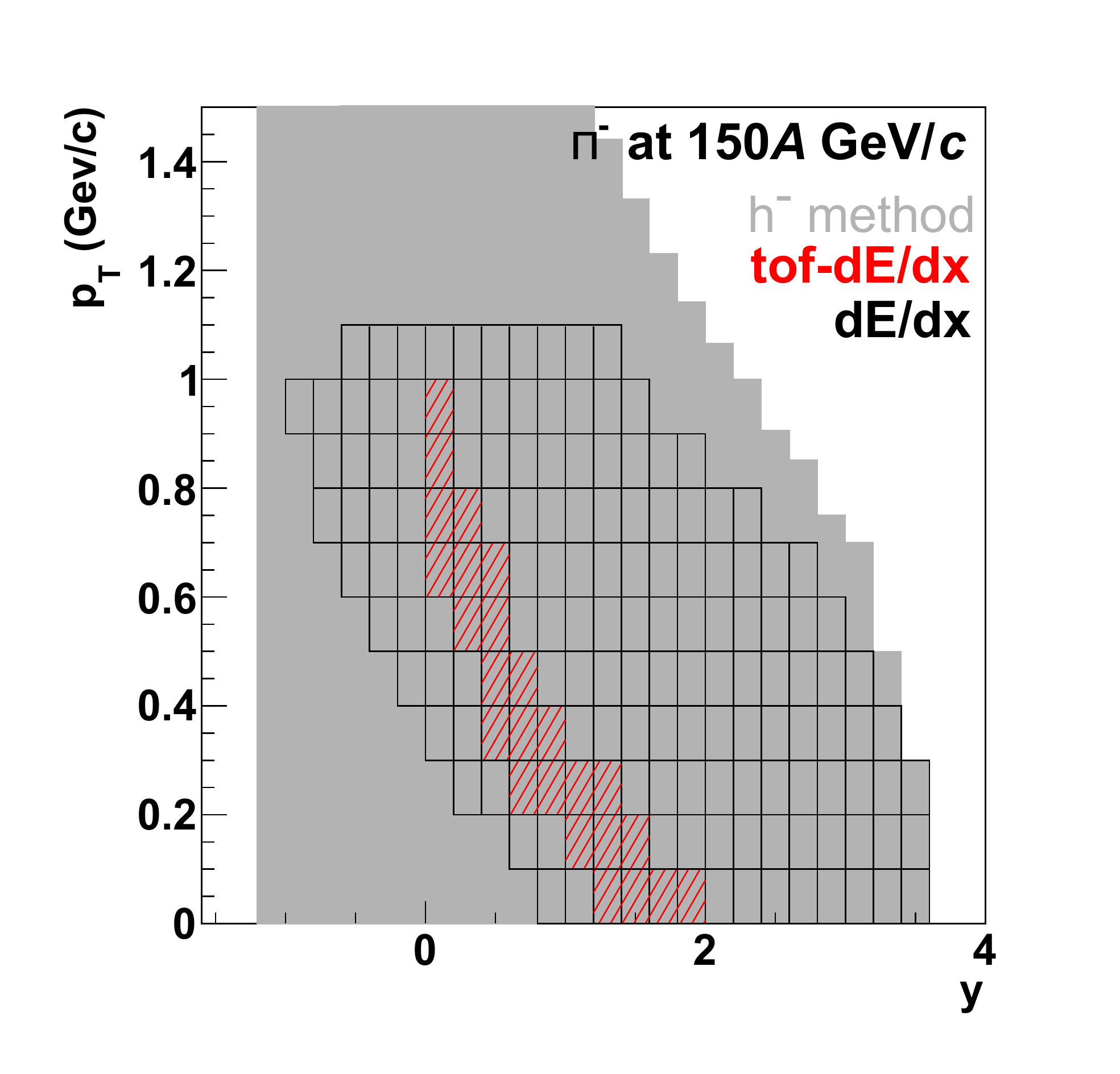}   
        \caption{ Acceptance in \y and \pt of analysis techniques used by \NASixtyOne to obtain
        multiplicities of $\pi^{-}$ produced in Be+Be collisions at 30$A$ (\textit{left}) 
        and 150\AGeVc (\textit{right}). Acceptance for the $h^{-}$ method is shown as gray area, for the \dEdx
        identification technique as black boxes and for the $tof$-\dEdx identification as red hatching.}
        \label{fig:acc}
\end{figure}

In this paper the so-called $h^-$ method is used for determining
$\pi^{-}$ production utilizing the fact that negatively charged particles are predominantly $\pi^{-}$ mesons 
with a small admixture (of order 10\%) of $K^-$ mesons and anti-protons which can be subtracted reliably.
Compared to the other analysis strategies used by \NASixtyOne, aiming at identifying particles based on measuring 
energy loss in the TPCs and time-of-flight, the $h^-$ method provides the largest phase space coverage. 
The acceptance of the $h^-$ analysis technique for $\pi^{-}$ produced in Be+Be collisions is shown in
Fig.~\ref{fig:acc} for 30$A$ and 150\AGeVc. It covers almost the full forward and part of the backward hemisphere of \y and \pt down to zero.

This section gives a brief overview of the data analysis procedure and the applied corrections.
It also defines to which class of particles the final results correspond.
A description of the calibration and the track and vertex reconstruction procedure can be found in
Ref.~\cite{Abgrall:2013pp_pim}.

The analysis procedure consists of the following steps:
\begin{enumerate}[(i)]

  \item application of event and track selection criteria,
  \item determination of spectra of negatively charged hadrons
        using the selected events and tracks,
  \item evaluation of corrections to the spectra based on
        experimental data and simulations,
  \item calculation of the corrected spectra and mean multiplicities,
  \item calculation of statistical and systematic uncertainties.

\end{enumerate}

Corrections for the following biases were evaluated:
\begin{enumerate}[(i)]
 \item contribution from off-target interactions,
 \item bias of selection procedure of \textit{central} collisions,
 \item geometrical acceptance,
 \item contribution of particles other than \emph{primary} (see below)
       negatively charged pions produced in Be+Be interactions,
 \item losses of produced negatively charged pions
       due to their decays and secondary interactions.
\end{enumerate}

Correction (i) was not applied due to insufficient statistics of the target removed data. The
contamination of the target inserted data was estimated from the \coordinate{z} distribution of fitted
vertices (see Fig.~\ref{fig:vertexZzoomout}) and found to be very small
($\approx$~0.35\%) for the selected \textit{central} interactions.
Correction (ii) was estimated to be small and was therefore included in the systematic uncertainty (see Sec.~\ref{sec:centrality}).
Corrections (iii)-(v) were estimated by simulations, see Sec.~\ref{sec:corrections} below.

The final results refer to $\pi^{-}$ produced in \textit{central} Be+Be collisions
by strong interaction processes and in electromagnetic
decays of produced hadrons. Such hadrons are referred to as \emph{primary} hadrons.
The definition and the selection procedure of \textit{central} collisions is given in Sec.~\ref{sec:centrality}.

The analysis was performed independently in (\y, \pt) bins.
The bin sizes were selected taking into account the statistical uncertainties
and the resolution of the momentum reconstruction~\cite{Abgrall:2013pp_pim}.
Corrections as well as statistical and systematic uncertainties
were calculated for each bin.

\subsection{\textit{Central} collisions}
\label{sec:centrality}

The term \textit{centrality} of the collision is related in the simplest models to the impact parameter $b$ or the number of wounded
nucleons $W$. Neither quantity is experimentally measurable and one uses instead the number $N$ of produced particles or the
energy emitted into the forward spectator region to characterise the \textit{centrality} of the collision. The first choice may
bias the measurements of particle production probabilities whereas such a bias is avoided by the second choice.
Therefore final results presented in this paper refer to the 5\% of Be+Be collisions with the lowest value of 
the forward energy $E_F$ (\textit{central} collisions). The quantity $E_F$ is defined as the total energy in the laboratory system of all particles produced in a Be+Be collision via strong and electromagnetic processes in the forward momentum region defined by the acceptance map in Ref.~\cite{PSD_acceptance}. Results on \textit{central} collisions defined as above allow a precise comparison with predictions of models without any additional information about the \NASixtyOne setup and used magnetic field.
Using this definition the mean number of wounded nucleons $\langle W \rangle$ and the mean collision impact parameter $\langle b \rangle$  
were calculated within the Wounded Nucleon Model~\cite{Bialas:1976ed} implemented in \Epos, see Sec.~\ref{sec:wounded}.

Negatively charged pion production was studied in event ensembles the centrality of which was selected by upper limits of  the energy $E_{PSD}$ measured by a subset of PSD modules. 
This subset was optimised for best sensitivity to projectile spectators (see Sec.~\ref{sec:PSDenergy} 
for details). For each collision energy the upper limit value was adjusted to select 5\% of all inelastic interactions. 
The forward momentum acceptance in the definition of $E_F$ corresponds to the acceptance of the optimised subset of PSD modules. 
Based on simulations the results for the $E_{PSD}$ selected collisions were corrected to correspond to the $E_F$ selected results.

The details of the described procedures are given below.

\subsubsection{Event selection based on the PSD energy}
\label{sec:PSDenergy}

\begin{figure}
        \centering 
        \includegraphics[width=0.35\textwidth]{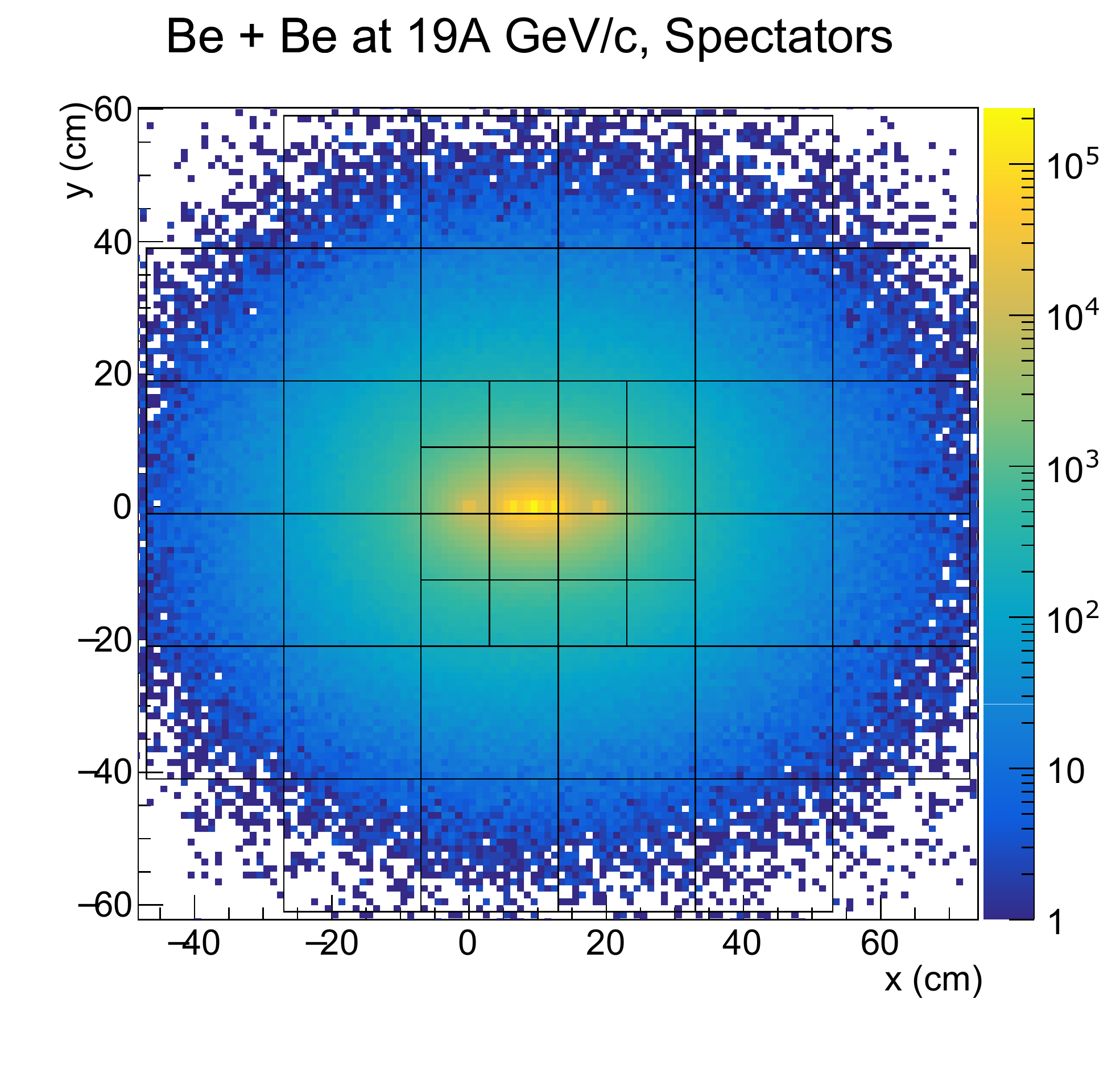}
        \includegraphics[width=0.35\textwidth]{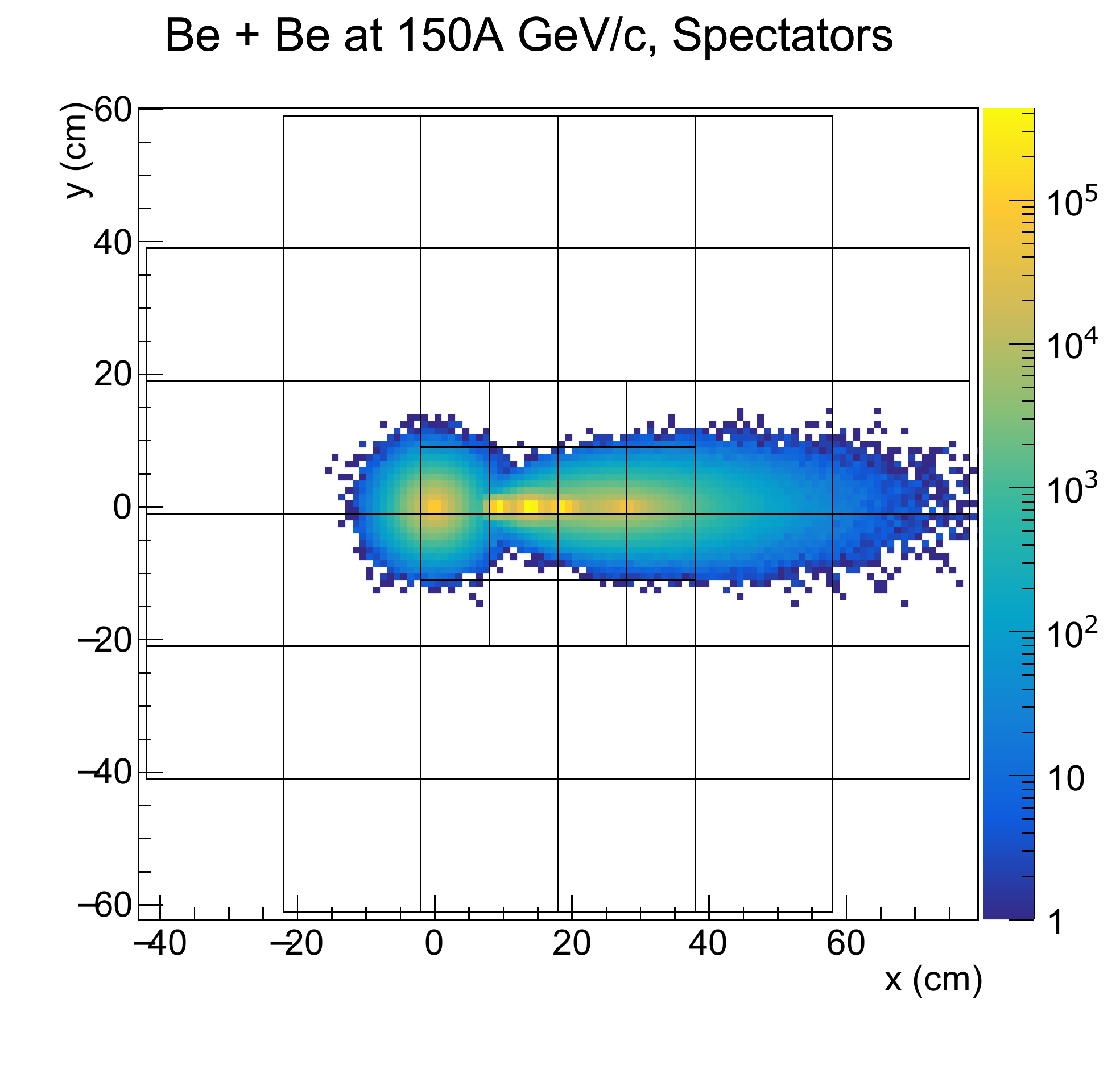}\\     
        \includegraphics[width=0.35\textwidth]{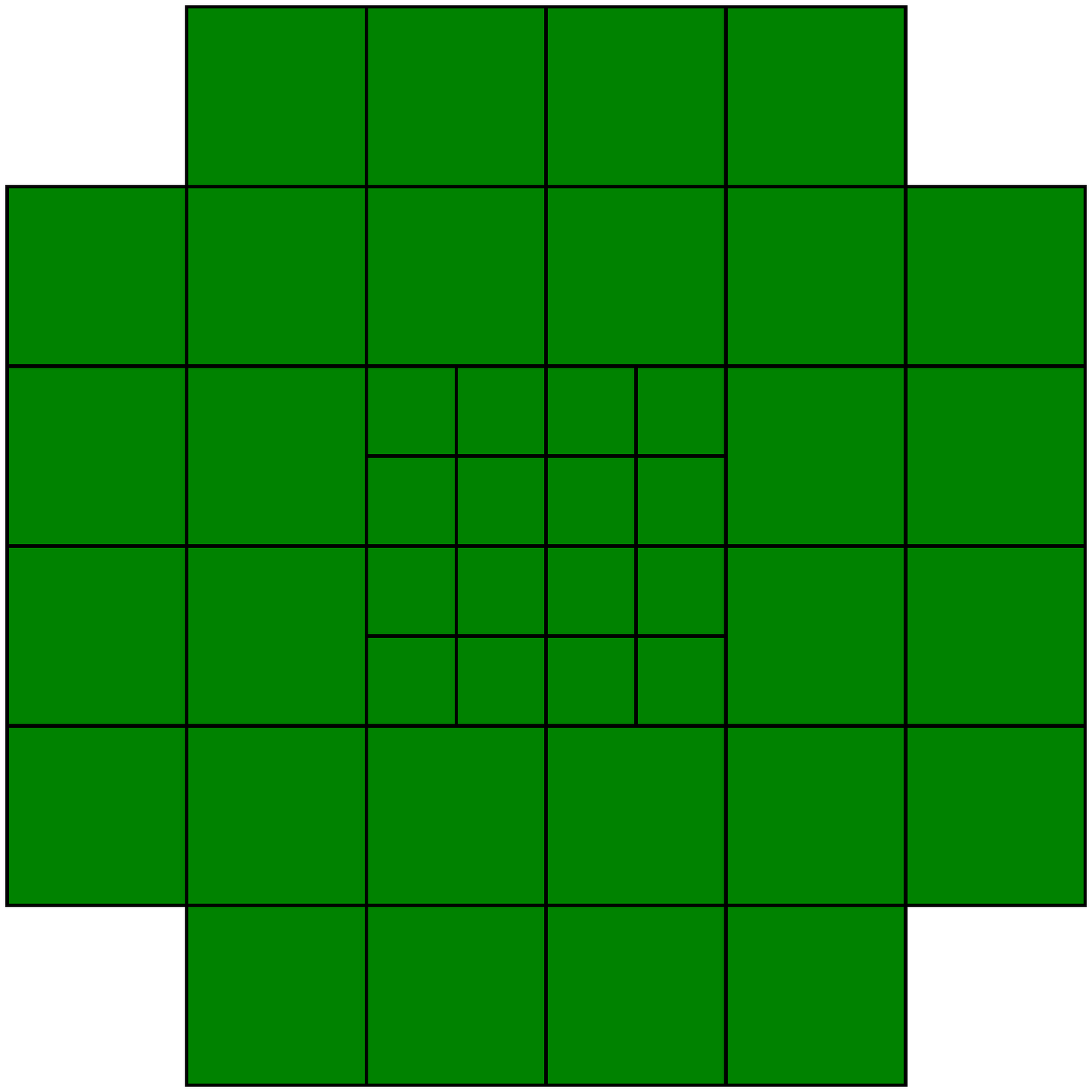}
        \includegraphics[width=0.35\textwidth]{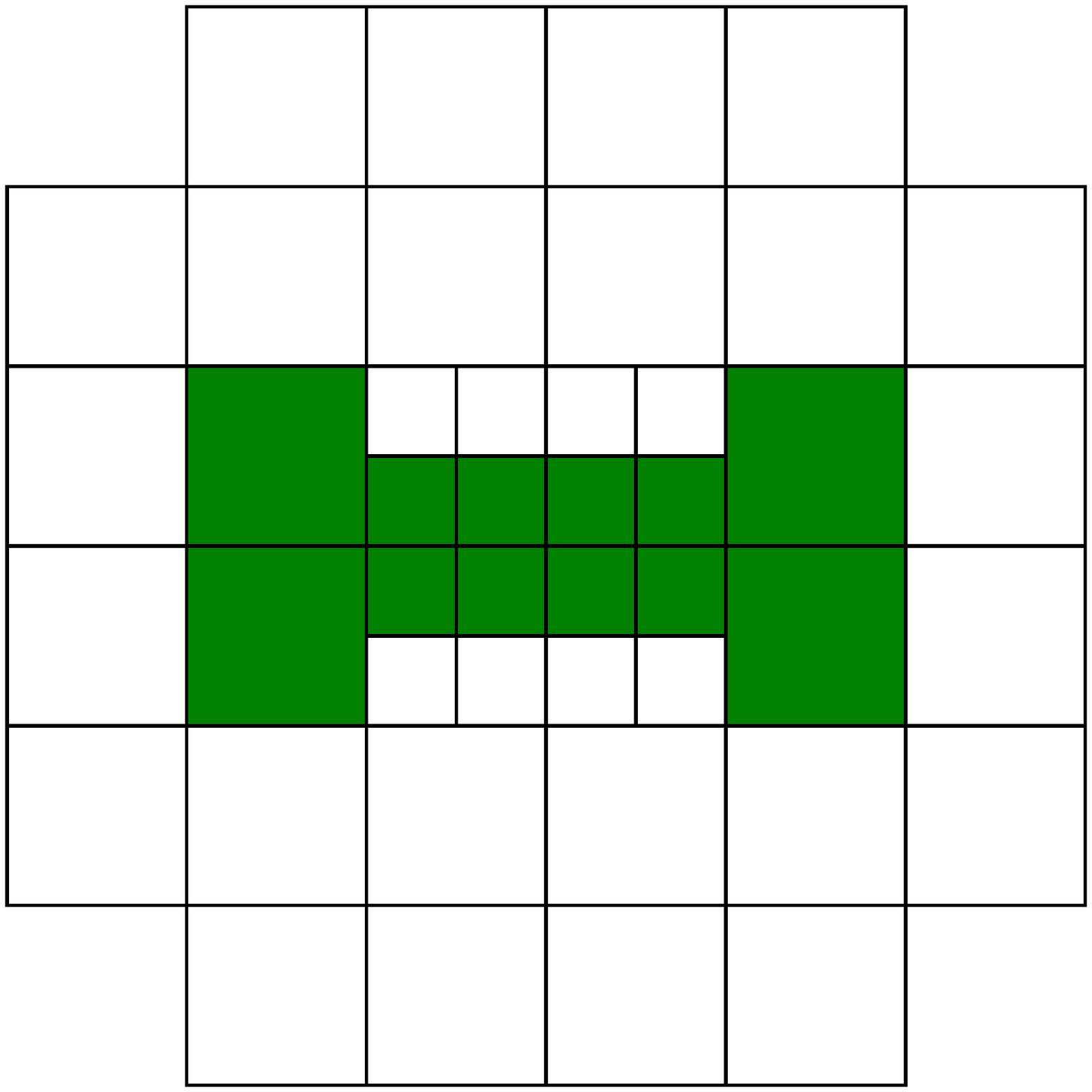}
        \vspace{-2.2cm}
        \caption{ 
        \textit{Upper row:} Simulated impact points of particles on the front face of the PSD for beam momentum of 19\AGeVc (\textit{left}) and 150\AGeVc (\textit{right}).
        \textit{Lower row:} PSD modules included in the calculation of the projectile spectator energy $E_{PSD}$
        used for event selection  for beam momenta of 19$A$, and 30\AGeVc (\textit{left}) and for 40$A$, 75$A$ and 150\AGeVc (\textit{right})}
        \label{fig:PSDAllModuleSelections}
\end{figure}

In order to optimize the
sensitivity to projectile spectators, only a subset of PSD modules was included in the calculation
of $E_{PSD}$~\cite{Kaptur:2007}. Figure~\ref{fig:PSDAllModuleSelections} (\textit{top}) shows the impact points 
of projectile spectator nucleons on the front face of the PSD obtained from the internal Glauber model of \Epos~\cite{Pierog:2018} including Fermi motion.  Figure~\ref{fig:PSDAllModuleSelections} (\textit{bottom}) depicts the
modules selected for the summation of $E_{PSD}$. Online event selection by the hardware trigger (T2) 
used a threshold on the sum of energies over the 16 central modules of the PSD.

\begin{figure}[!ht]
        \centering
        \includegraphics[width=0.45\textwidth]{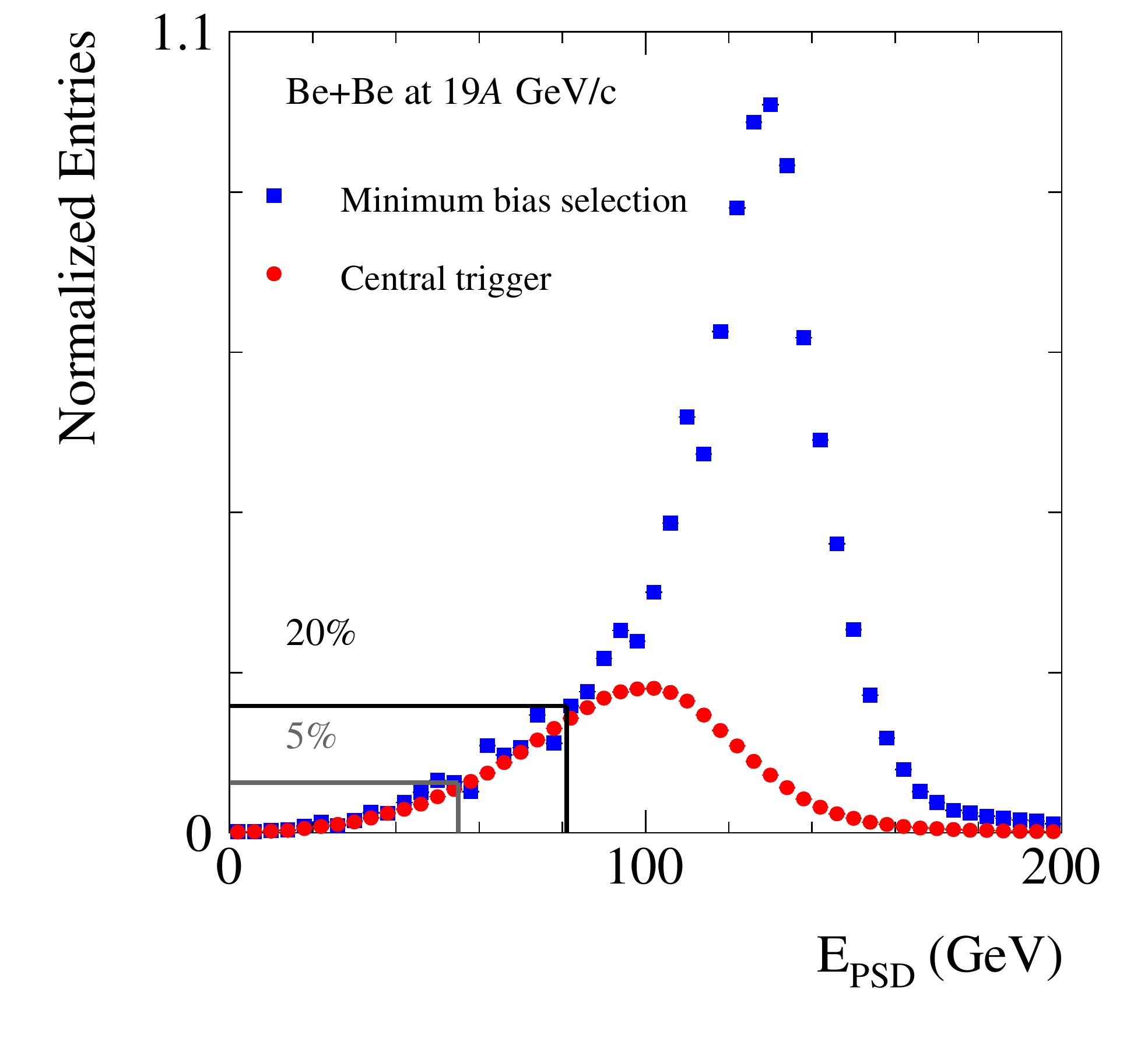}
        \includegraphics[width=0.45\textwidth]{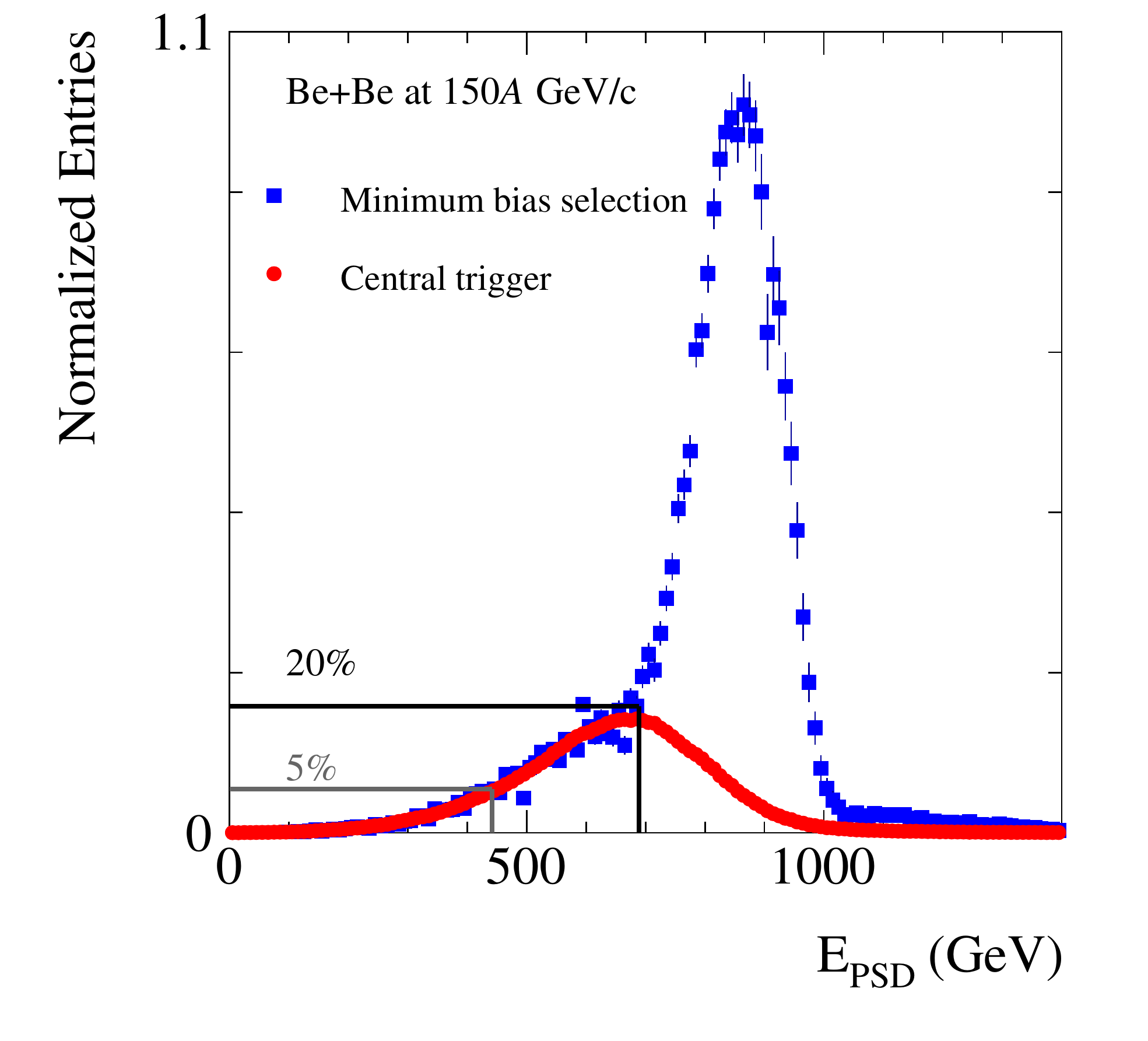}
        \caption{Two examples of the measured $E_{PSD}$ distribution for minimum-bias selected (blue data points) and T2 selected (red data points) events at
                 19\AGeVc (\textit{left}) and 150\AGeVc (\textit{right}) beam momentum.
                 Histograms are normalized to agree in the overlap region.
                 The limits used to select events are shown by black lines and they correspond to $\approx$~5\% and $\approx$~20\% of inelastic collisions.
                }
        \label{fig:PSDEnergy_cent}
\end{figure}

Measured distributions of $E_{PSD}$ for minimum-bias and T2 trigger selected events, calculated in the offline analysis, are shown 
in Fig.~\ref{fig:PSDEnergy_cent} at beam momenta of 19\AGeVc and 150\AGeVc, respectively. 
Also drawn are vertical lines which define the
$E_{PSD}$ corresponding to the 5\% and 20\% of events with the lowest $E_{PSD}$ values.
A minimum-bias distribution was obtained using the data from the beam trigger T1 with offline selection of events by 
requiring an event vertex in the target region and a cut on the ionisation energy detected in the GTPC to exclude Be beams. 
A properly normalized spectrum for target removed events was subtracted.

\subsubsection{Scaling to $E_F$ selected results }

Comparison of presented experimental results with other data require a realistic implementation of described centrality selection.  This is more easily realized using a quantity "forward energy" ($E_F$) instead of $E_{PSD}$, since the latter requires detailed knowledge of its response. The acceptance map provided in Ref.~\cite{PSD_acceptance} gives the recipe for the computation of ~$E_F$.
Both $E_F$ and $E_{PSD}$
were calculated in simulations using the \Epos model, which employed a dedicated software package which tracks particles through the magnetic fields and simulates the response of the PSD modules. 
A global factor $c_{cent}$ was then calculated as
the ratio of mean multiplicities of negatively charged pions obtained with the two selection procedures in the 5\% most \textit{central} events. A possible dependence of the scaling factor on rapidity and transverse momentum was neglected.
The resulting factors $c_\mathrm{cent}$ range between 1.00 and 1.03 (see Table~\ref{tab:w}) corresponding to only a small correction compared to the systematic uncertainties of the measured $\pi^-$ multiplicities. The correction was therefore not applied but instead included as a contribution 
to the systematic uncertainties.

\subsubsection{Mean number of wounded nucleons and collision impact parameter}
\label{sec:wounded}

\begin{figure}[h]
  \centering
     \includegraphics[width=0.49\textwidth]{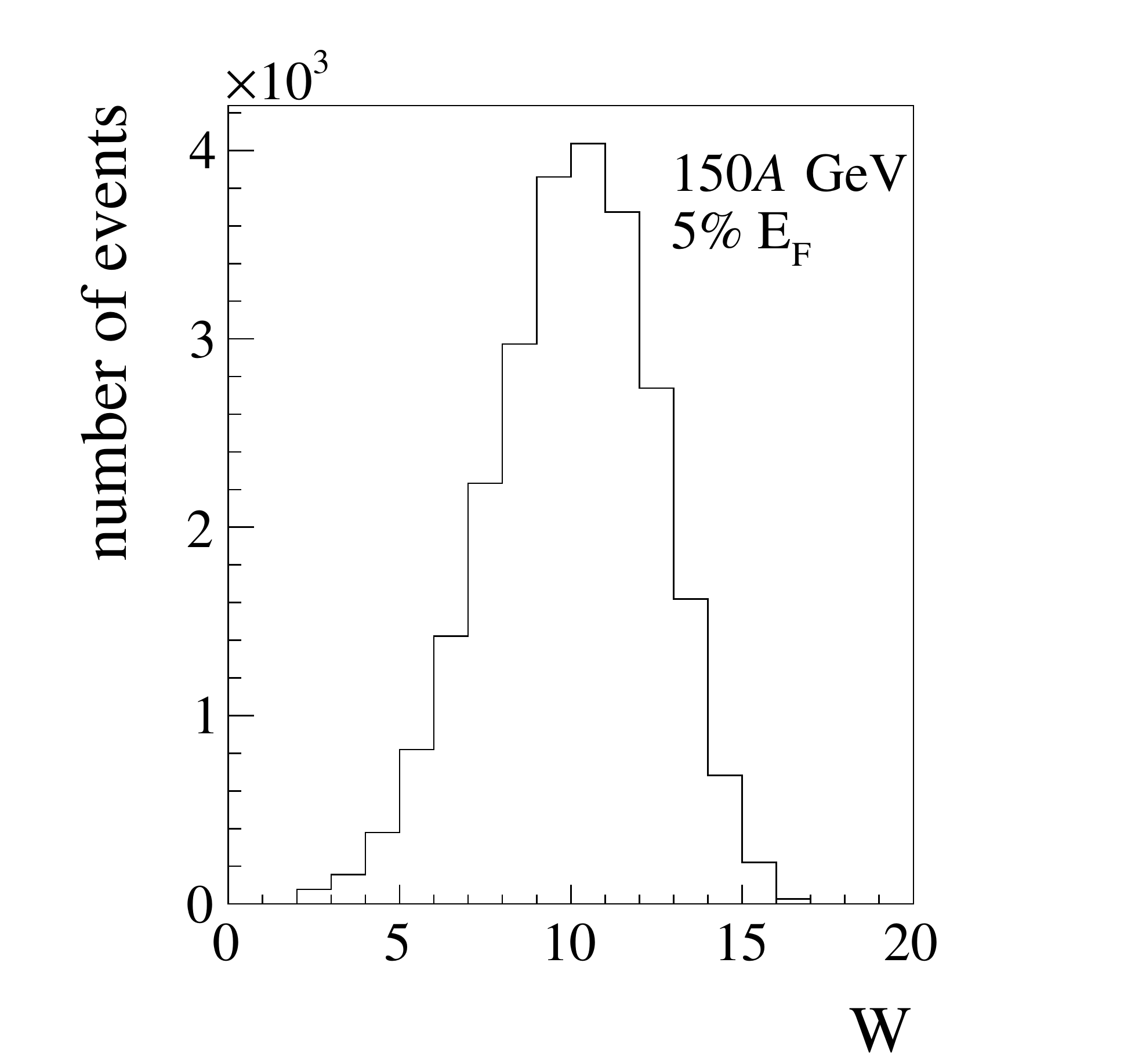}
     \includegraphics[width=0.49\textwidth]{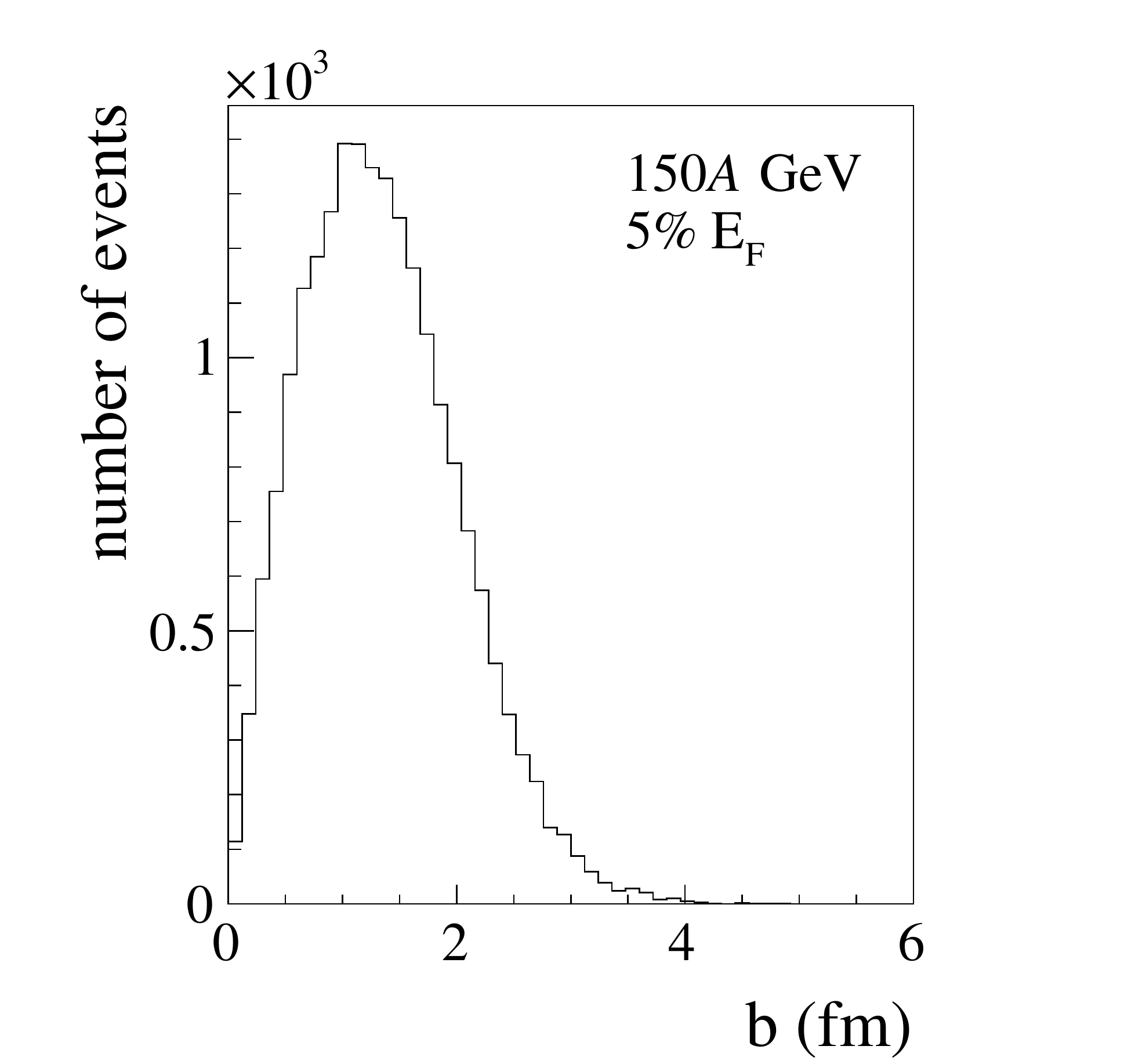}\\
     \includegraphics[width=0.49\textwidth]{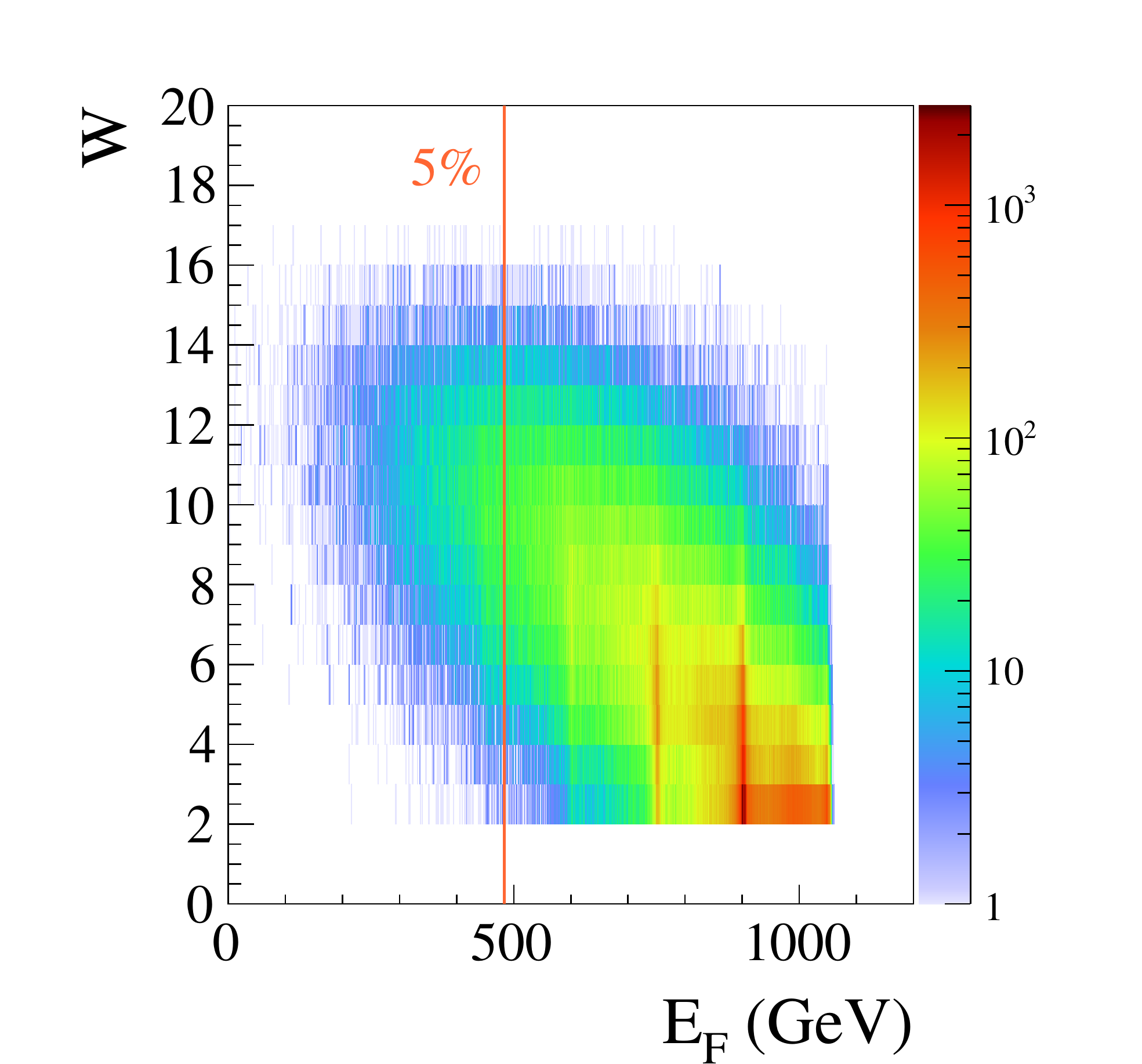}
     \includegraphics[width=0.49\textwidth]{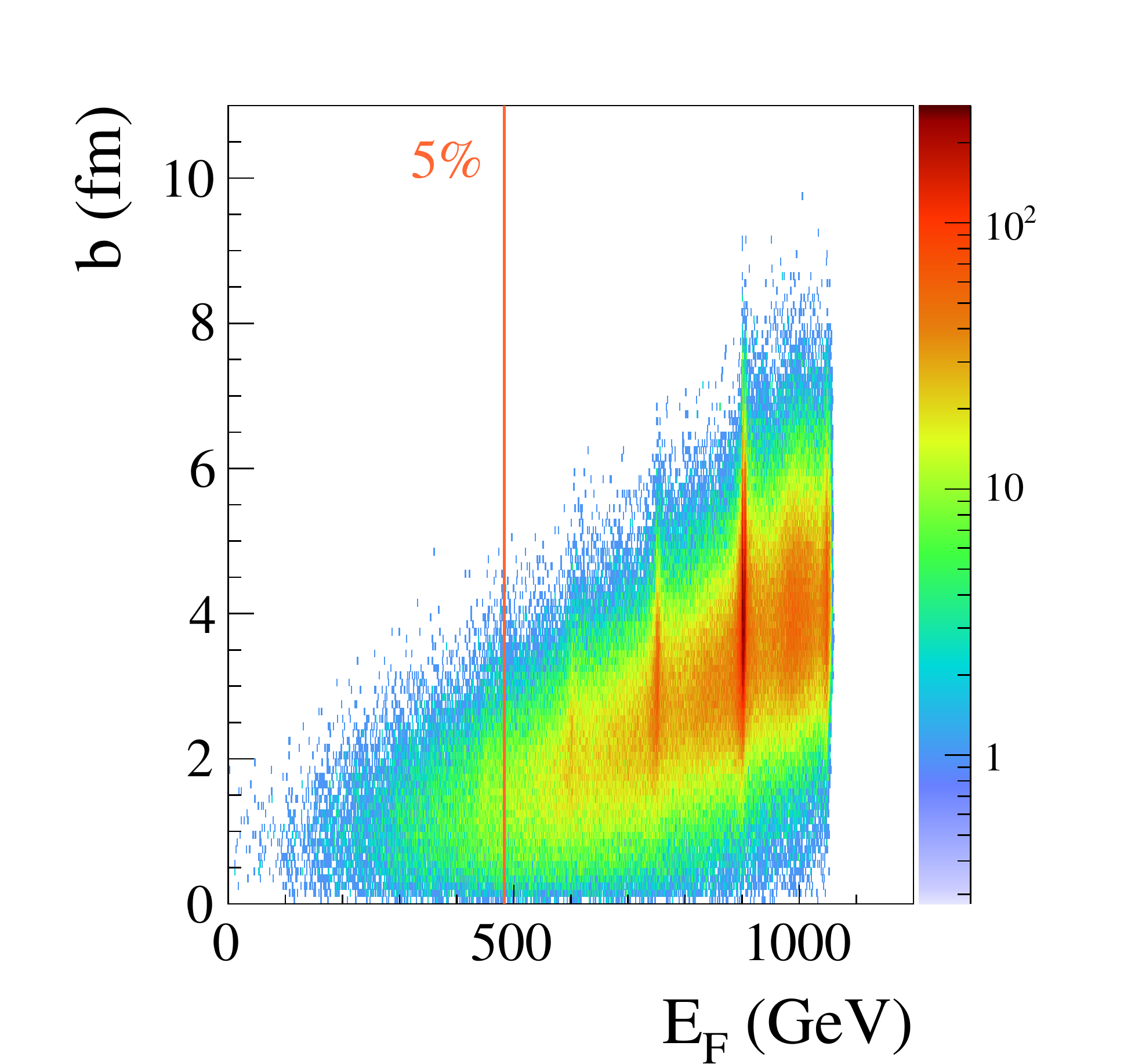}\\
  \caption{Examples of the distribution of the number of wounded nucleons $W$ (\textit{top,left}) 
           and impact parameter $b$ (\textit{top, right}) for events with the 5\%
           smallest forward energies $E_F$ and $E_F$ versus $W$ (\textit{bottom, left})
           and $E_F$ versus $b$ (\textit{bottom, right})
           at beam momentum of 150\AGeVc simulated 
           with \Epos using the acceptance map provided in~Ref.~\cite{PSD_acceptance}.}
\label{fig:woundedDistribution}
\end{figure}

Comparisons of particle yields in collisions of different size nuclei usually employs the
average number of wounded nucleons $\langle W \rangle$ in the respective reactions.
For estimating the average number of wounded nucleons corresponding to the
selected \textit{central} collisions \Epos 1.99 (version \Crmc 1.5.3)~\cite{Werner:2005jf} was employed
which uses the Glauber model and a parton ladder mechanism to generate the interactions.
\Epos was modified~\cite{Pierog:2018} to provide the values of $W$ of its internal Glauber model calculation. The results on $\langle W \rangle$ for the 5\% most \textit{central} collisions 
from the \Epos  model are listed in Table~\ref{tab:w}. 
Fluctuations of the listed values are due to the integer nature of $W$. 
As \Epos simulates all particles of the final state a more realistic estimate of $\langle W \rangle$ is obtained by selecting \textit{central} collisions based on the energy $E_F$.
The resulting mean number of wounded nucleons and the mean impact parameter are also listed in Table~\ref{tab:w}. 
Values of $\langle W \rangle$ for the two selection procedures differ by about two units. 
Examples of the distributions of $W$ and $b$ for the 5\% most \textit{central} collisions are shown in Fig.~\ref{fig:woundedDistribution}.
As the nucleon density is low in the Be nucleus these distributions are quite broad.
This emphasises that for model
comparisons it is important to use equivalent \textit{centrality} selection procedures to obtain a meaningful result.
 
\begin{table*}
 \centering
 \footnotesize
 \begin{tabular}{l|cccccc}
  Momentum (\AGeVc) & & 19 & 30 & 40 & 75 & 150\\
  \\
  \hline
  \hline
  \\
  \Epos WNM & $\langle W \rangle$ & $11.8$ & $11.8$ & $11.8$ & $11.8$ & $11.8$\\
  & $\sigma$ & $1.0$ & $1.0$ & $1.0$ & $1.0$ & $1.0$ \\
  \Epos $E_F$ & $\langle W \rangle$ & $9.54$ & $9.44$ & $9.67$ & $9.61$ & $9.51$\\
  & $\sigma$ & $2.4$ & $2.4$ & $2.3$ & $2.4$ & $2.4$ \\
  & $\langle b \rangle$ & $1.44$ & $1.54$ & $1.32$ & $1.26$ & $1.32$\\
  & $\sigma$ & $0.7$ & $0.8$ & $0.7$ & $0.6$ & $0.7$ \\
   \hline
  & $c_{cent}$ & $1.019$ & $1.029$ & $1.001$ & $1.005$ & $1.009$ \\
 \end{tabular}
 \caption{Average number of wounded nucleons $\langle W \rangle$  and average impact parameter $\langle b\rangle$
          in the 5\% most \textit{central} Be+Be collisions estimated from simulations using the \Epos~\cite{Werner:2005jf} model.
          The values of $\sigma$ denote the widths of the distributions of $W$ and $b$.
          Results \Epos~WNM are for \textit{centrality} selection using the smallest number of spectators, \Epos~$E_F$
          using the forward energy $E_F$ within the acceptance map in Ref.~\cite{PSD_acceptance}}
 \label{tab:w}
\end{table*}

\subsection{Event and track selection}\label{sec:cuts}

\subsubsection{Event selection}

\begin{figure}[!ht]
\centering
\includegraphics[width=1.0\textwidth]{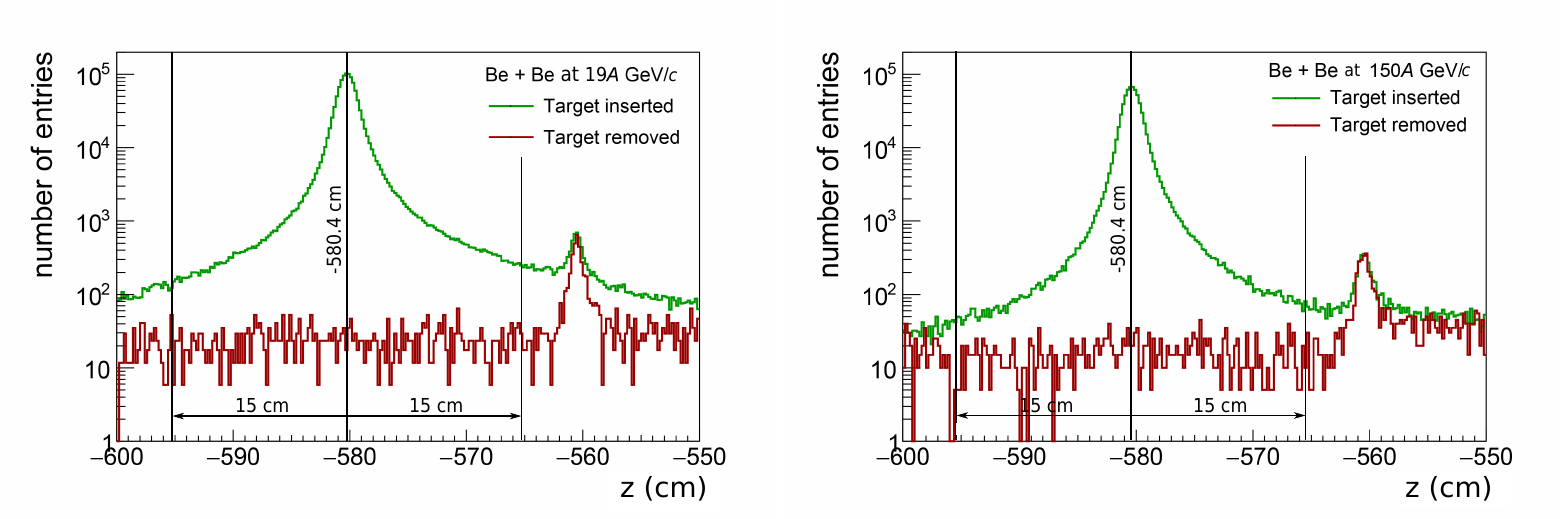}
\caption{Distribution of fitted vertex \coordinate{z} coordinate for the 20\% most \textit{central} $^7$Be+$^9$Be interactions
with target inserted (green histogram) and target removed (red histogram).
 \emph{(Left)}: 19\AGeVc. \emph{(Right)}: 150\AGeVc. Target position and cut values are marked. Target is installed in the box filled with He gas to minimise background interactions. Smaller peak on the right hand side of the plots corresponds to interactions with a target holder window.}
\label{fig:vertexZzoomout}
\end{figure}

For the analysis Be+Be events were selected using the following criteria:
\begin{enumerate}[(i)]

    \item four units of charge measured in S1, S2, and Z counters as well as BPD3
          (this requirement also rejects most interactions upstream of the Be target),
    \item no off-time beam particle detected within a time window of $\pm$4.5$~\mu$s
          around the trigger particle,
    \item no other event trigger detected within  a time window of $\pm$25$~\mu$s
          around the trigger particle,
    \item beam particle trajectory measured in at least three planes out of four
          of BPD-1 and BPD-2 and in both planes of BPD-3,
    \item charge measured in the GTPC smaller than that of Be (applied at 
          40$A$, 75$A$ and 150\AGeVc),
    \item a well reconstructed interaction vertex with \coordinate{z} position (fitted using
          the beam trajectory and TPC tracks) not farther away than 15~cm
          from the center of the Be target (see Fig~\ref{fig:vertexZzoomout};
          the cut removes less than 0.4\% of \textit{central} interactions),
    \item the energy $E_{PSD}$ measured in the subset of the PSD modules smaller than an upper limit 
          (55, 73, 104, 165, 442~\GeV for collisions at 19$A$, 20$A$, 30$A$, 40$A$, 75$A$ and 150\AGeVc, respectively) 
          in order to select the 5\% most \textit{central} collisions 
          (see discussion in Sec.~\ref{sec:centrality}).
\end{enumerate}
The event statistics after applying the selection criteria
are summarized in Table~\ref{tab:eventStat}.

\subsubsection{Track selection}

In order to select tracks of primary charged hadrons and to reduce the contamination
of tracks from secondary interactions, weak decays and off-time interactions,
the following track selection criteria were applied:

\begin{enumerate}[(i)]
    \item track momentum fit at the interaction vertex should have converged,
    \item fitted \coordinate{x} component of particle rigidity at the vertex $\left(p_{\text{lab},\text{\coordinate{x}}}/q\right)$ is positive.
          This selection minimizes the angle between the track trajectory and the TPC
          pad direction for the given magnetic field direction, reducing uncertainties
          of the reconstructed cluster position, energy deposition and track parameters,
    \item total number of reconstructed points on the track should be greater than 15,
    \item sum of the number of reconstructed points in VTPC-1 and VTPC-2 should be greater than 15
          or greater than 4 in the GTPC,
    \item the distance between the track extrapolated to the interaction plane and the
          interaction point (track impact parameter) should be smaller than 4~cm in the horizontal
          (bending) plane and 2~cm in the vertical (drift) plane,
    \item electron tracks were excluded by a cut on the measured particle energy loss \dEdx in the TPCs.
\end {enumerate}
 
\subsection{Corrections}
\label{sec:corrections}

In order to determine the mean multiplicity of \textit{primary} $\pi^-$ mesons produced in \textit{central} Be+Be collisions 
a set of corrections was applied to the extracted raw results. The main effects are detector acceptance,
loss of events due to the cut on reconstructed vertex position,
reconstruction efficiency, contributions of particles from weak decays (feed-down), 
and contribution of \textit{primary} hadrons other than negatively charged pions (mostly $K^-$ mesons).
The contamination of events occurring outside the target was negligible.

A simulation of the \NASixtyOne detector is used to correct the data for acceptance,
reconstruction efficiency and contamination. Only Be+Be interactions in the target material
were simulated and reconstructed. The \Epos model~\cite{Werner:2005jf,Pierog:2018} was selected to generate
the primary interactions as it best describes the \NASixtyOne measurements. A \GeantThree based program chain
was used to track particles through the spectrometer, generate decays and secondary interactions and
simulate the detector response (for more detail see Ref.~\cite{Abgrall:2013pp_pim}). Simulated events were 
then reconstructed using the \NASixtyOne reconstruction chain and reconstructed tracks were
matched to the simulated particles based on the cluster positions. The same event selection
procedure was used as for data (cut on the summed energy in the subset of PSD modules used to select the 5\% most \textit{central} collisions).
Particles which were not produced in the primary interaction can amount to a significant fraction of the selected
track sample. Thus a careful effort was undertaken to evaluate and subtract this contribution.

The correction factor $c_{yp_T}$ for primary $\pi^-$, based on the event and detector simulation was calculated for each
\y and \pt bin as:

\begin{equation}
c_{yp_T} =  n[\pi^-]^\mathrm{MC}_\mathrm{gen}~/~n[h^-]^\mathrm{MC}_\mathrm{sel} 
\label{eq:correctiongeo}
\end{equation}

where $ n[h^-]^\mathrm{MC}_\mathrm{sel} $ is the mean multiplicity of reconstructed  negatively charged 
particles after the event and track selection criteria and 
$ n[\pi^-]^\mathrm{MC}_\mathrm{gen} $ 
is the mean multiplicity of \textit{primary} negatively charged pions from the $E_{PSD}$-selected Be+Be collisions generated by the \Epos model.


The corrected multiplicities were then calculated as:
\begin{equation}
n[\pi^-]^{corr} = c_{yp_T} \cdot n[\pi^-]^{raw}
\end{equation}

The final results in bins of \y and \pt are shown in Fig.~\ref{fig:2dSpectra}.

\subsection{Statistical uncertainties}

Statistical uncertainties of the yields receive contributions from the finite statistics of both
the data and the simulated events used to obtain the correction factors. The dominating contribution
is the uncertainty of the data which were calculated assuming a Poisson probability distribution for the
number of entries in a bin. Compared to the statistics of the data the statistics of the simulation were
much higher and the statistical uncertainties of the latter were neglected.  

\subsection{Systematic uncertainties}

Systematic uncertainties presented in this paper were calculated taking into
account contributions from the following effects.

                
\begin{enumerate}[(i)]
 \item \label{item:syst_cuts1} Possible biases due to track cuts which are not corrected for. These are:
 \begin{enumerate}[(a)]
  \item a possible bias due to the \dEdx cut applied to remove electrons,
  \item a possible bias related to the removal of events with off-time beam
        particles close in time to the trigger particle.
 \end{enumerate}
 Their magnitude was estimated by varying the values of the corresponding cut for data selection. The possible bias due to 
 the \dEdx cut  was changed by $\pm 0.01$~\dEdx units (where 1
 corresponds to a minimum ionising particle, and 0.04 is a typical width of the \dEdx distribution for $\pi^-$), 
 and the off-time interactions cut was varied from a $\pm 3.5~\mu$s to a $\pm 5.5~\mu$s time window.
 The assigned systematic uncertainty was calculated as the maximum of the
 absolute differences between the results obtained for lower and upper values.
The estimated bias is on the level of 1-3\%.

This uncertainty is listed in the tables including
numerical values and it is visualised by a shaded band around
the data points in plots presenting the results.
Systematic biases in different bins are correlated,
whereas statistical fluctuations are almost independent.

  \item \label{item:syst_cuts2} Uncertainty  related to the track cuts which were corrected for. 
It was estimated by varying the track selection cuts used for data and Monte Carlo events: removing the
impact parameter cut and decreasing the minimum number of required points to 12
(total) and 10 (in VTPCs).
The potential bias is below 2\%.

\item
 \label{item:syst_hminus} Uncertainty of the correction
      for contamination of the primary $\pi^-$ mesons.
 There was no data available to adjust
 the simulated spectra. To estimate a possible bias the simulated spectra were instead adjusted to preliminary \NASixtyOne data on the $K^{-}/\pi^{-}$ ratio~\cite{Pulawski:2017veg}, and the difference between the results with adjusted and standard correction of order 2\% was assigned as relative potential systematic uncertainty. Since $K^{-}$ are the main contribution to the $h^{-}$ correction and the absolute correction is small, this contribution was finally neglected in the systematic uncertainty estimate.
\end{enumerate}

The total systematic uncertainty was calculated by adding in quadrature
the contributions \mbox{(i) - (iii)}:
\begin{equation}
\sigma_\text{sys} = 
\sqrt{\sigma_\text{i}^2+\sigma_\text{ii}^2+\sigma_\text{iii}^2}.
\end{equation}
This uncertainty is listed in the tables including numerical values and it is visualised by a shaded
band around the data points in plots presenting the results. Systematic biases in different bins 
are correlated, whereas statistical fluctuations fluctuations are almost independent.
\section{Experimental results}\label{sec:results}

This section presents results on negatively charged pion spectra at 19$A$, 30$A$, 40$A$, 75$A$ and 150\AGeVc
beam momentum in the 5\% most \textit{central} $^7$Be+$^9$Be collisions
(see Sec.~\ref{sec:centrality} for definition of \textit{central} collisions). 
The results refer to primary pions produced by strong interaction and decay processes and in electromagnetic
decays of neutral hadrons.


\subsection{Double-differential (\y, \pt) and (\y, $m_T-m_{\pi}$) yields}

\begin{figure}[ht!]
\centering
\includegraphics[width=0.3\linewidth]{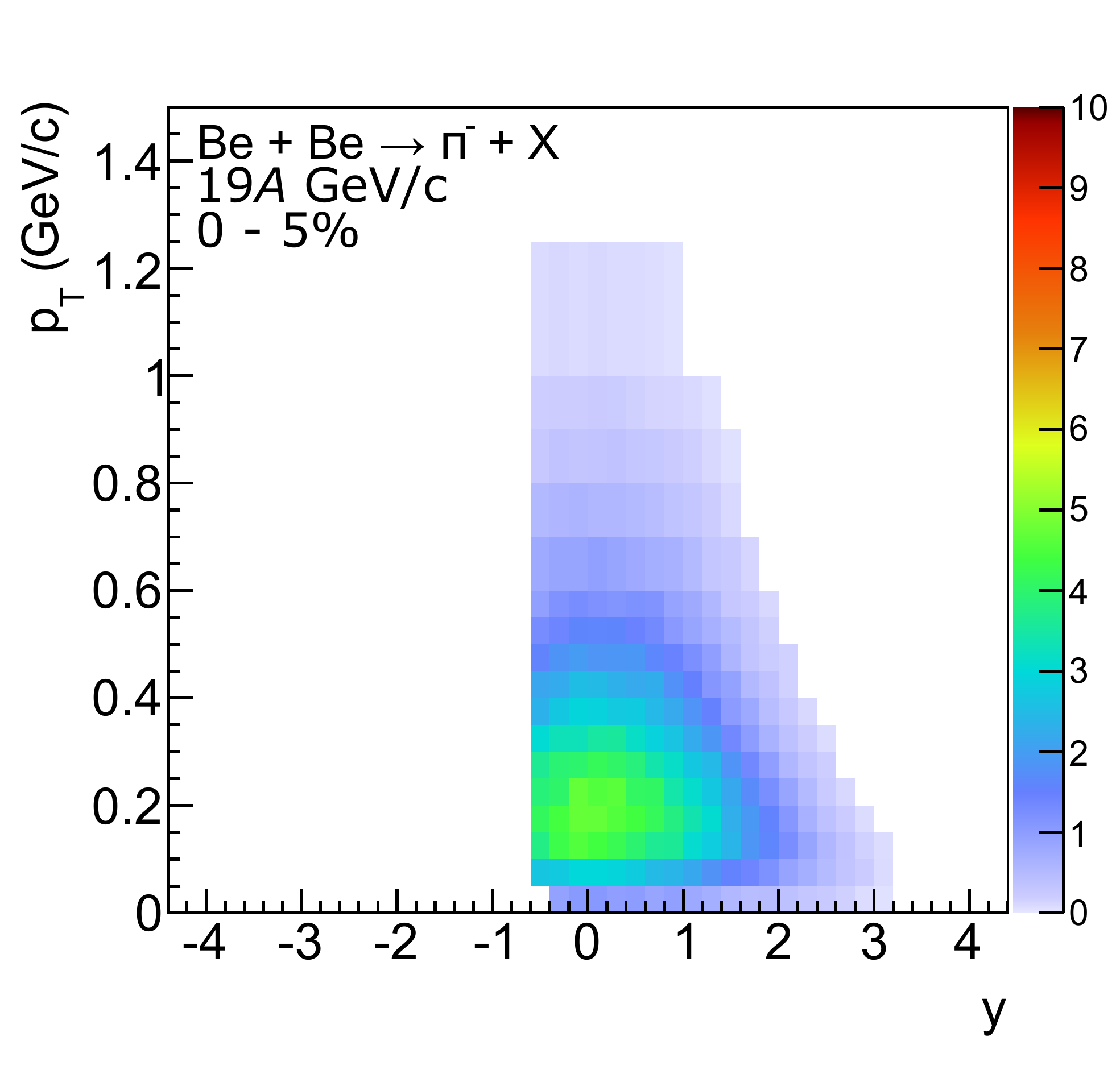}
\includegraphics[width=0.3\linewidth]{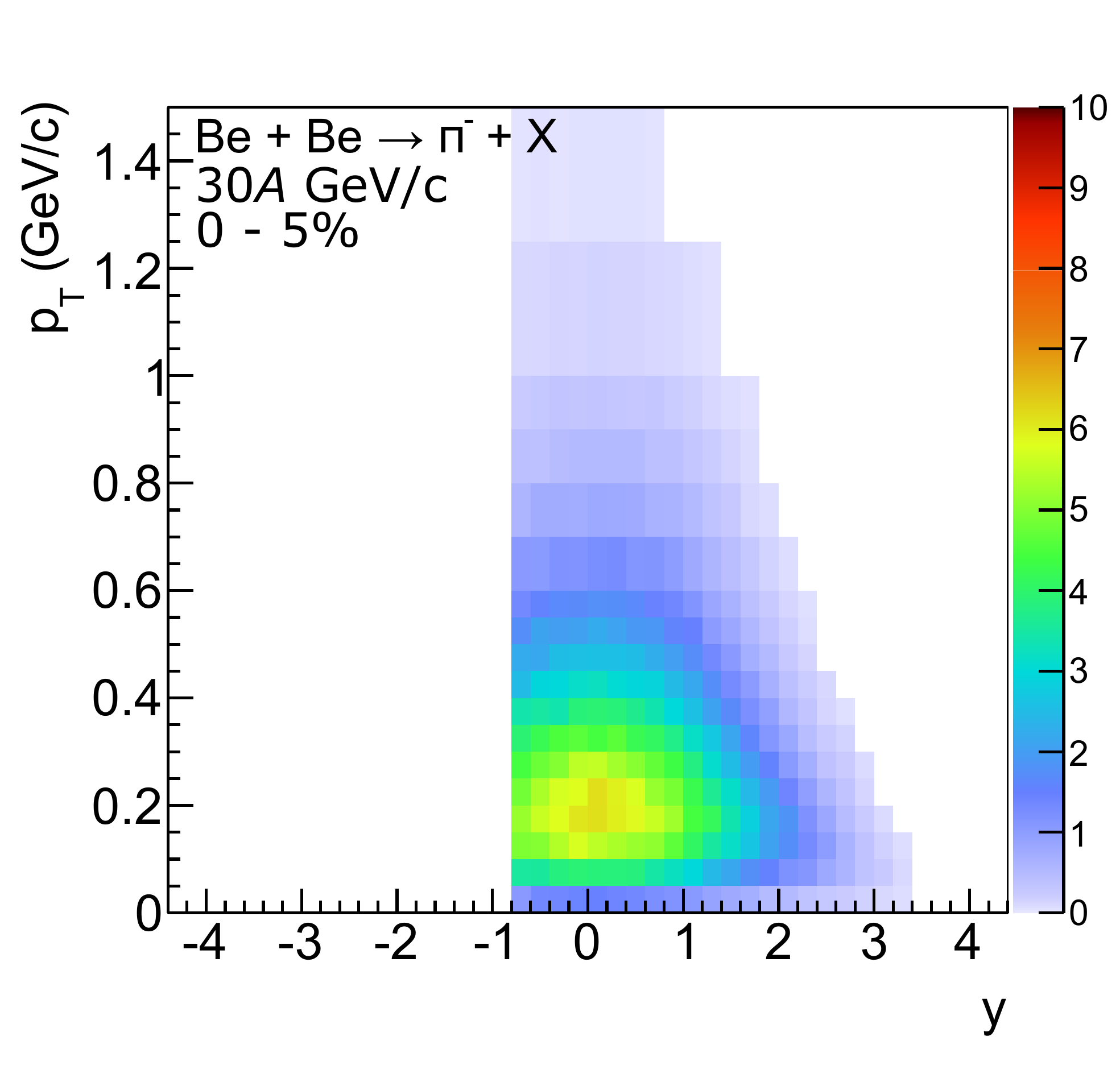}
\includegraphics[width=0.3\linewidth]{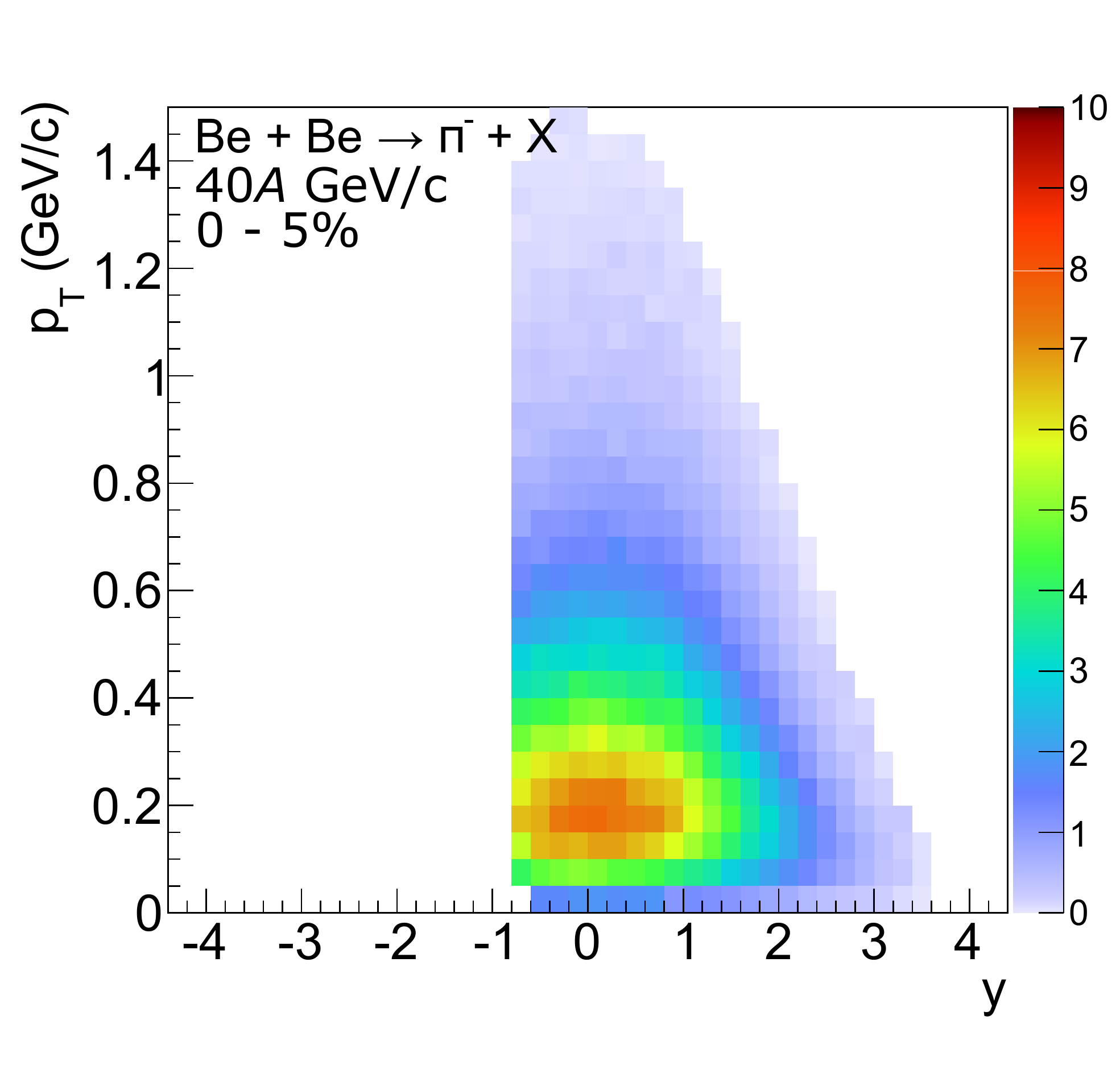}\\
\includegraphics[width=0.3\linewidth]{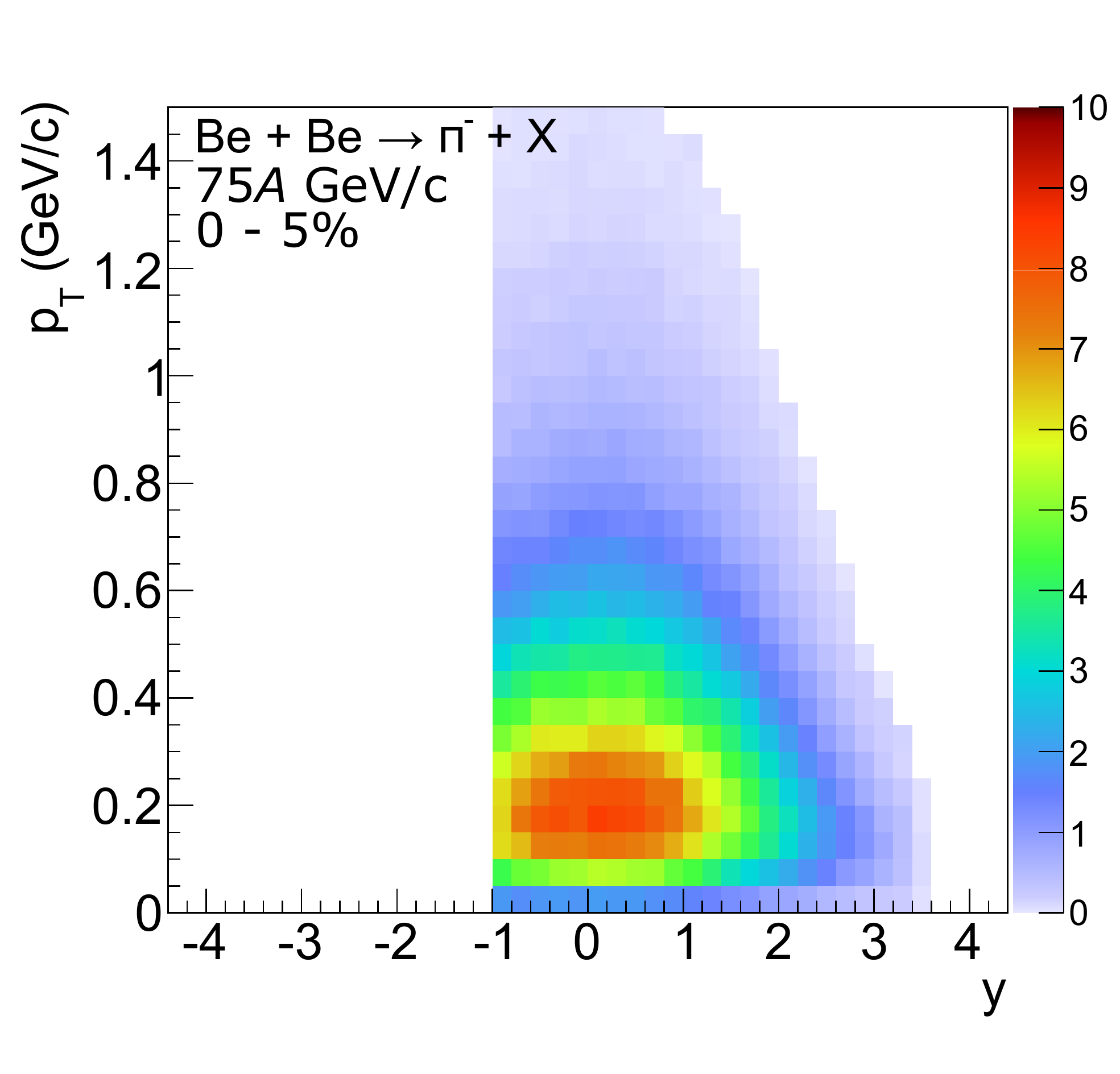}
\includegraphics[width=0.3\linewidth]{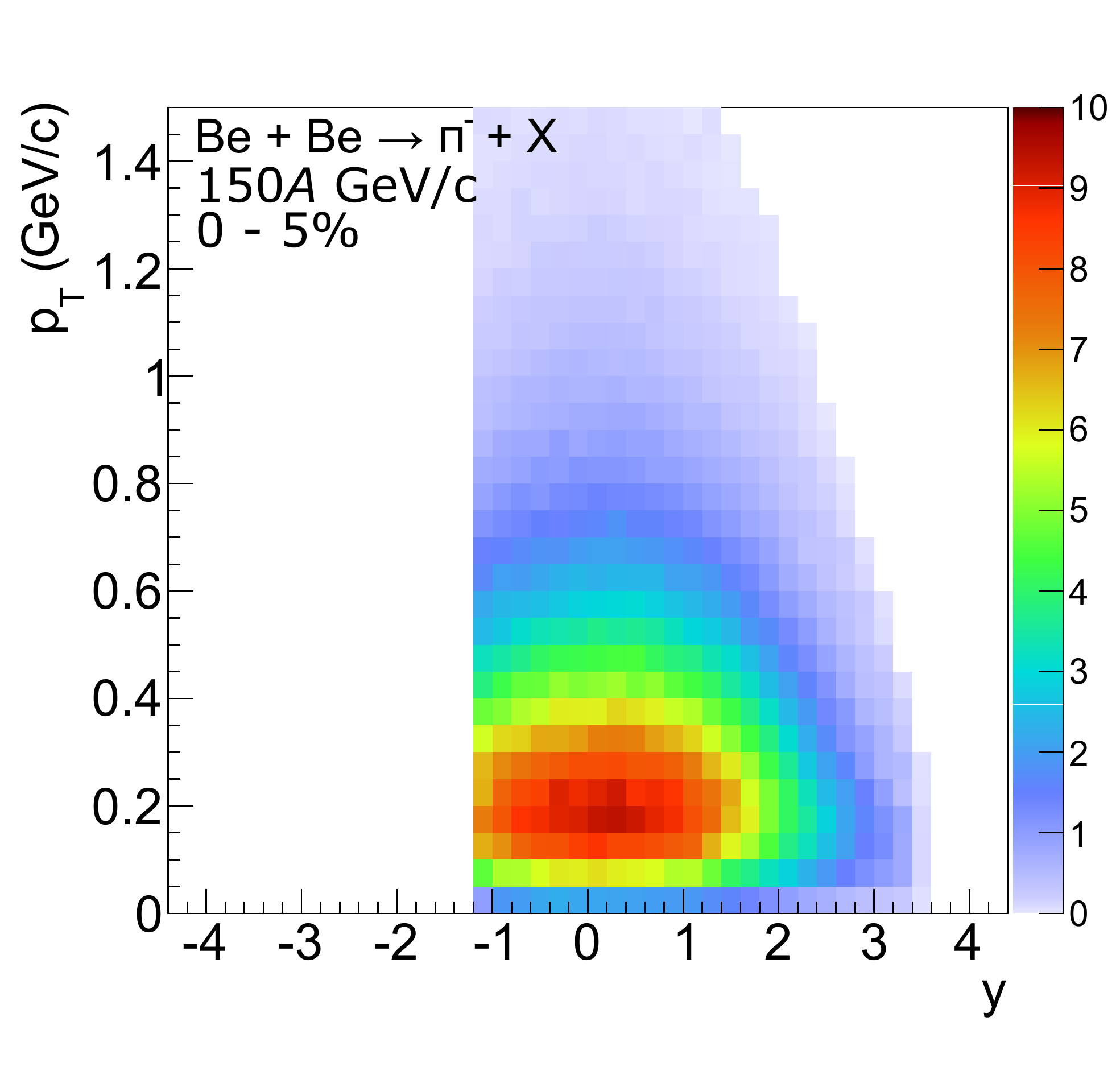}
\caption{Double-differential spectra $\frac{d^2n}{d\y d\pt}$ of negatively 
       charged pions produced in the 5\% most \textit{central} Be+Be collisions at beam momenta
       of 19$A$, 30$A$, 40$A$, 75$A$ and 150\AGeVc.
       }
\label{fig:2dSpectra}
\end{figure}

Figure~\ref{fig:2dSpectra} shows fully corrected  double-differential
(\y, \pt) distributions $\frac{d^2n}{d\y d\pt}$ of $\pi^{-}$ measured in Be+Be interactions and illustrates the wide
phase space acceptance of the detector. Rapidity bins with limited acceptance ($y < -0.6$ for 19\AGeV, $y < -0.8$ for 30\AGeV, $y < -0.8$ for 40\AGeV, $y < -1.2$ for 75\AGeV and $y < -1.4$ for 150\AGeVc) are not used in the subsequent analysis.

\subsection{Transverse momentum and transverse mass spectra}

\begin{figure}[ht]
                \begin{center}
                \includegraphics[width=0.3\textwidth]{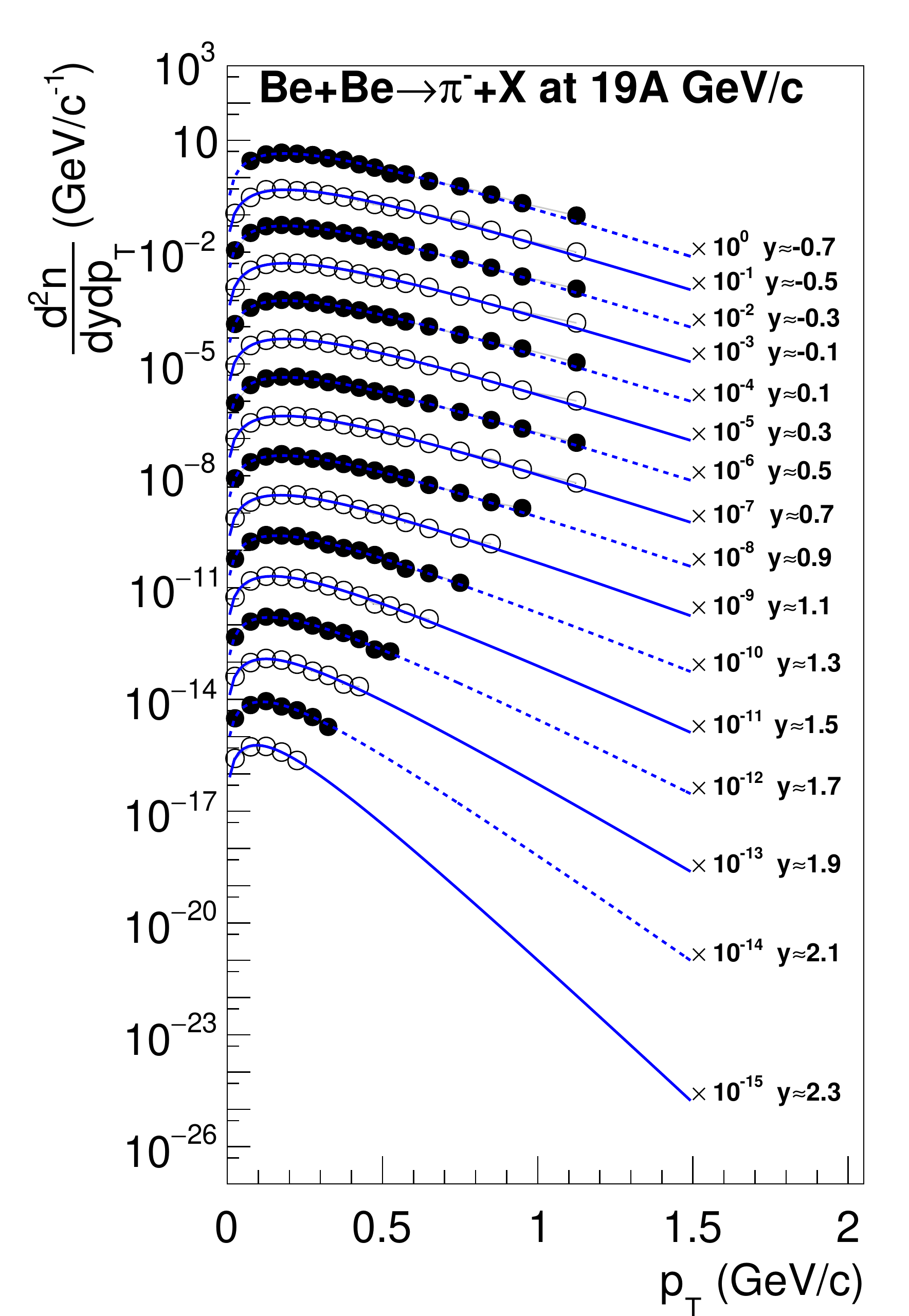}
                \includegraphics[width=0.3\textwidth]{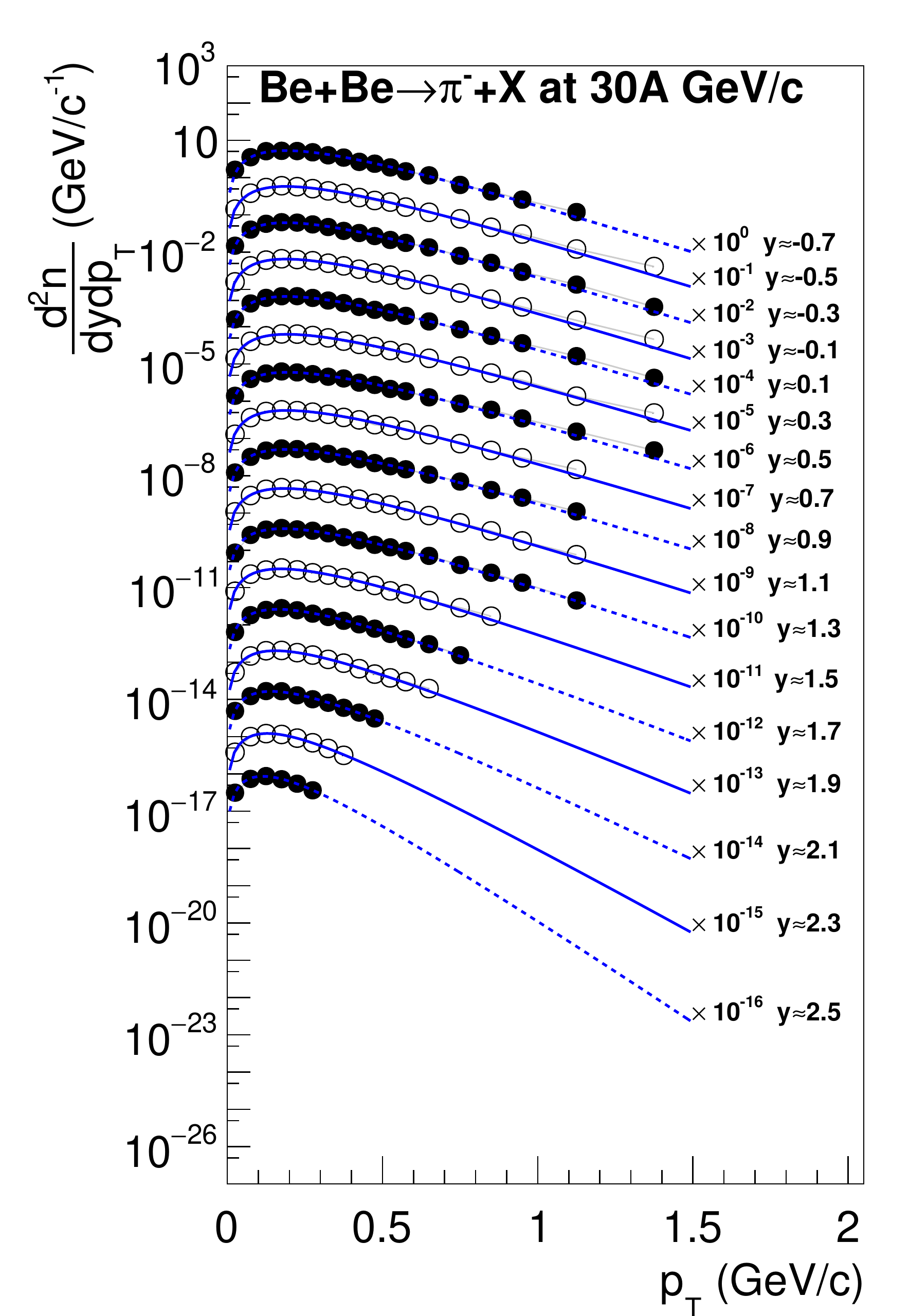}\\
                \includegraphics[width=0.3\textwidth]{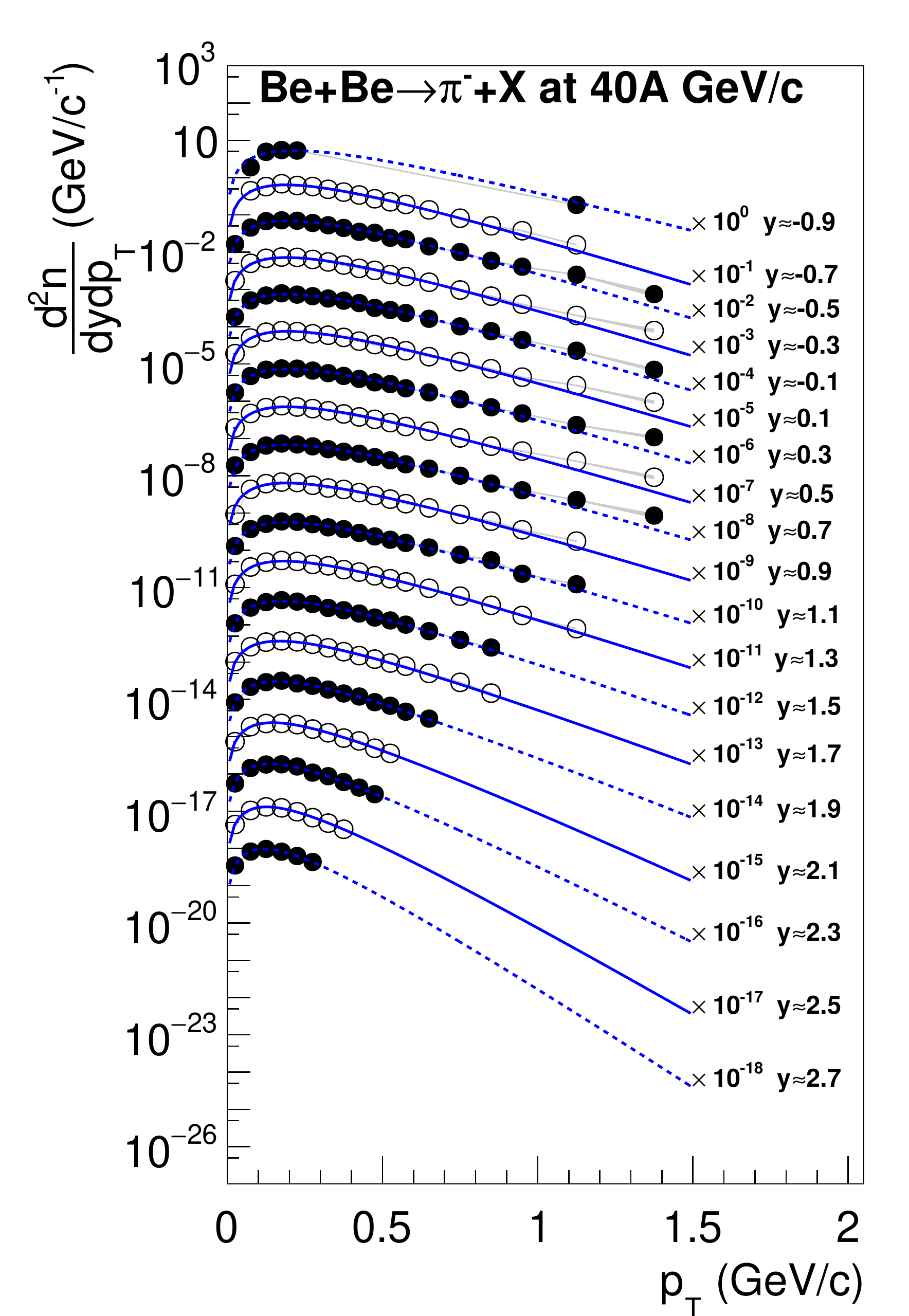}
                \includegraphics[width=0.3\textwidth]{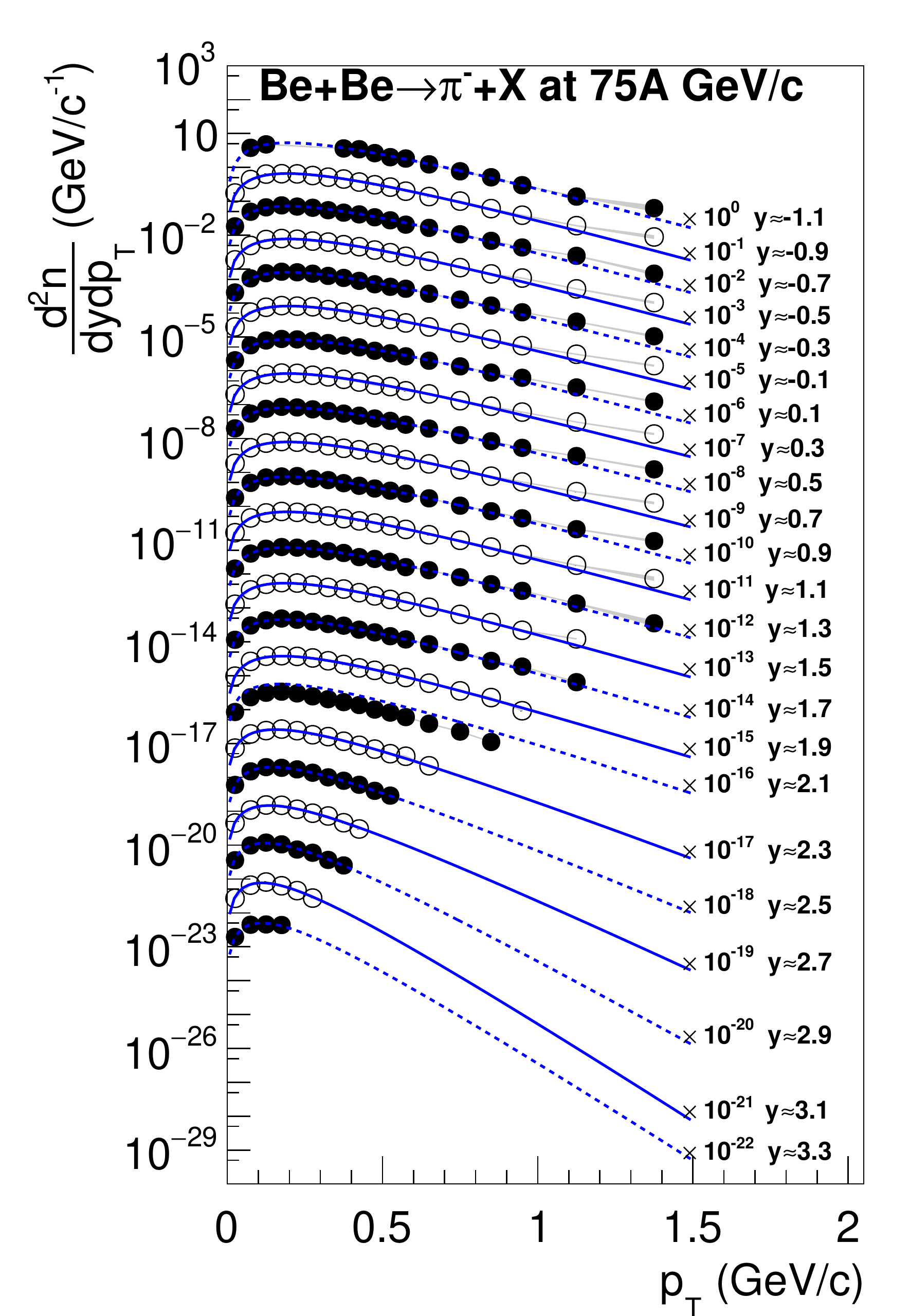}
                \includegraphics[width=0.3\textwidth]{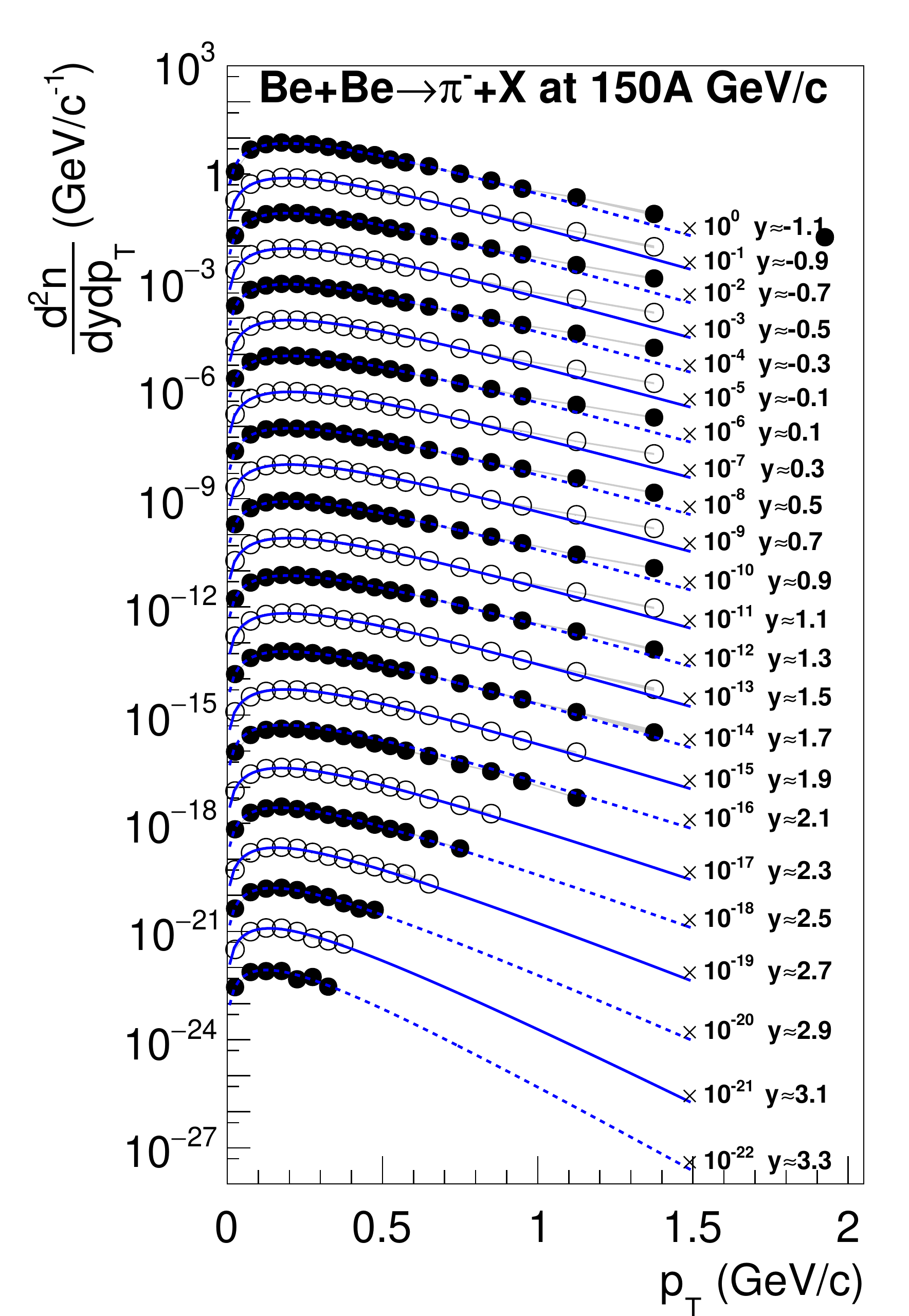}
                \end{center}
                \caption{Transverse momentum spectra of $\pi^{-}$ in rapidity slices produced
                         in the 5\% most \textit{central} Be+Be collisions. Rapidity values given in the
                         legends correspond to the middle of the corresponding interval. Data
                         points for consecutive rapidity slices are scaled down by factors of 10.
                         Shaded bands show systematic uncertainties. Curves depict thermal model
                         motivated fits with Eq.~\ref{eq:ptFit}.
                         }
                \label{fig:ptSpectra}
\end{figure}

Spectra of transverse momentum $p_T$~in slices of rapidity \y are plotted in Fig.~\ref{fig:ptSpectra}.
Superimposed curves show the results obtained from fitting the function

\begin{equation}          
 f(p_T)=C \cdot p_T \cdot \exp\left(\frac{-\sqrt{{(c p_T})^2+m^2}}{T}\right)   , 
\label{eq:ptFit}
\end{equation} 

motivated by the thermal model, where the inverse slope parameter $T$ and the normalisation 
constant $C$ are the fit parameters.

\begin{figure}[h]
  \centering
    \includegraphics[width=0.7\textwidth]{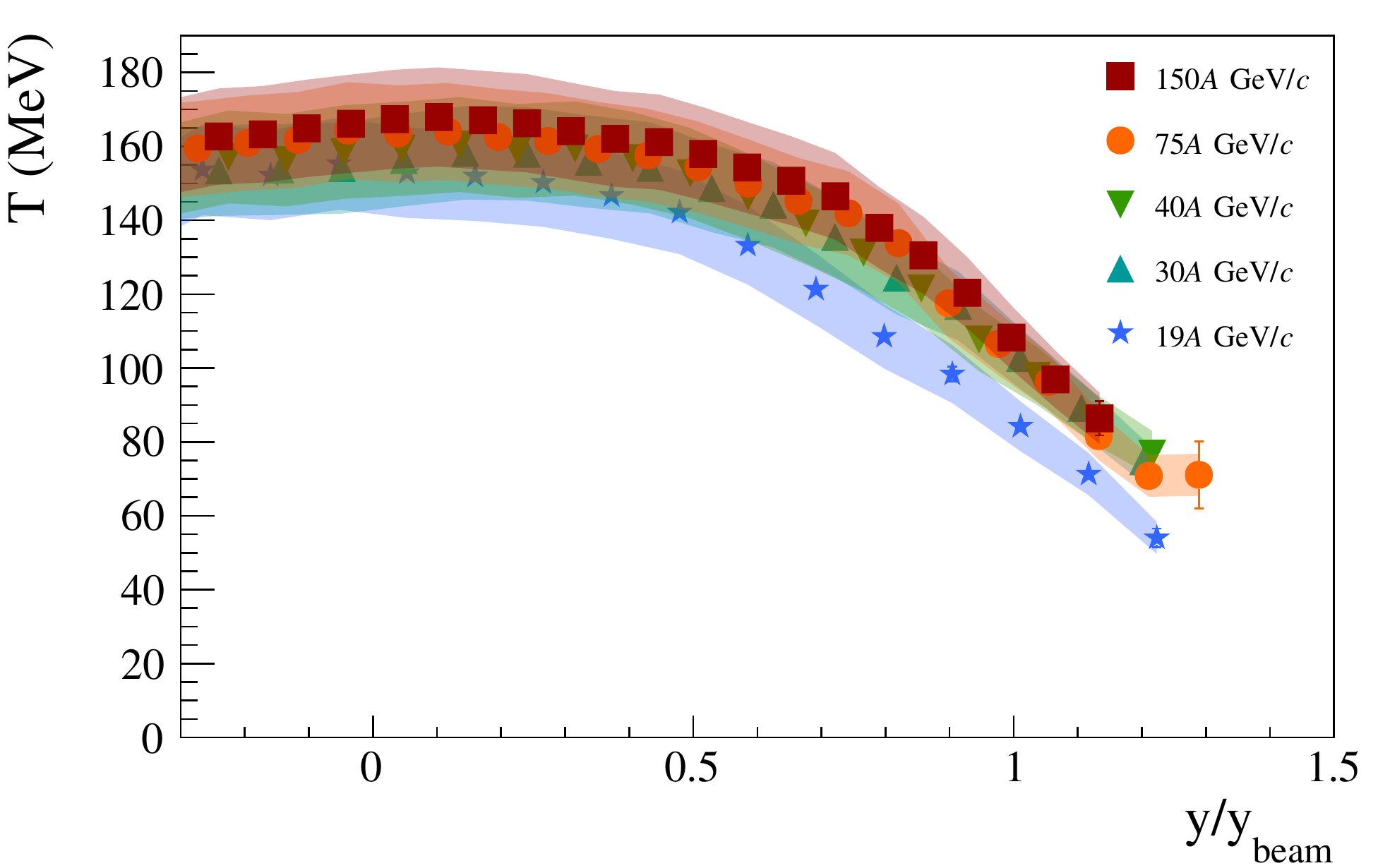}
  \caption{The inverse slope parameter $T$ of the transverse mass spectra of negatively charged pions 
           in \textit{central} Be+Be collisions at the SPS energies as a function of 
           rapidity divided by the beam rapidity. The fit range is $0.2<m_T-m_{\pi}<0.7$~\GeV. 
           Statistical uncertainties are mostly smaller than marker size, systematic uncertainties 
           are indicated by shaded bands.
           }
  \label{fig:invSlope}
\end{figure}

A summary of the fitted values of the inverse slope parameter $T$ are shown in Fig.~\ref{fig:invSlope}
plotted versus rapidity divided by beam rapidity. The decrease of $T$ towards larger rapidities and 
the obtained values are close to those found for inelastic \pp interactions~\cite{Abgrall:2013pp_pim}.
Numerical values of $T$ at \y~$\approx$~0 are given in Table~\ref{tab:T_dndy_y0}.

\begin{table}
   \caption{
   Inverse slope parameter $T$ near mid-rapidity fitted in the interval $0.2<m_T-m_{\pi}<0.7$~\GeV
   and mid-rapidity yield $dn/dy$ of $\pi^-$ mesons in the 5\% most \textit{central} Be+Be collisions. For comparison inverse slope parameter $T_{\pp}$ near mid-rapidity fitted in the same interval of $\pi^-$ mesons in the \pp interactions at close beam momenta~\cite{Abgrall:2013pp_pim}.
   }
\vspace{0.3cm}
\centering
  \begin{tabular}{ c | c | c || c }
    $p_\text{beam}$ (\AGeVc) &  $T$ (MeV) & $dn/dy_{(y=0)}$ &  $T_{p+p}$ (MeV) \\
    \hline 
    19  & 150 $\pm$ 1 $\pm$ 6  & 2.00 $\pm$ 0.06 $\pm$ 0.12 & 149.1 $\pm$ 5.0 $\pm$ 4.8\\
    30  & 158 $\pm$ 1 $\pm$ 7  & 2.57 $\pm$ 0.08 $\pm$ 0.16 & 153.3 $\pm$ 2.2 $\pm$ 1.2\\
    40  & 160 $\pm$ 1 $\pm$ 6  & 3.02 $\pm$ 0.09 $\pm$ 0.19 & 157.7 $\pm$ 1.7 $\pm$ 2.1\\
    75  & 163 $\pm$ 1 $\pm$ 8  & 3.44 $\pm$ 0.10 $\pm$ 0.21 & 159.9 $\pm$ 1.5 $\pm$ 4.1\\
    150 & 167 $\pm$ 1 $\pm$ 8  & 3.80 $\pm$ 0.10 $\pm$ 0.23& 159.3 $\pm$ 1.3 $\pm$ 2.6
  \end{tabular}
  \label{tab:T_dndy_y0}
\end{table}  

\begin{figure}[ht!]
        \centering
        \includegraphics[width=0.32\linewidth]{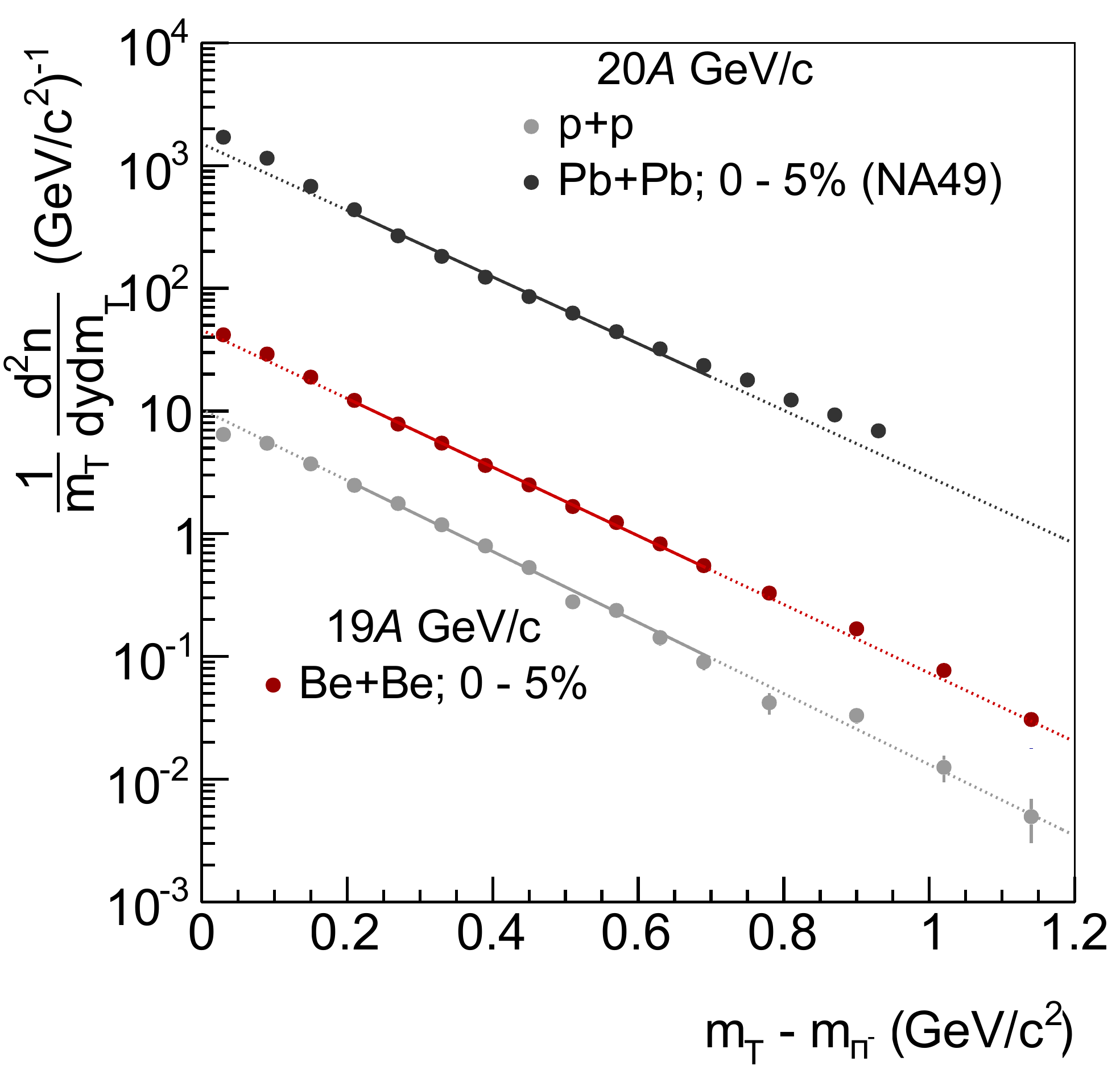}
        \includegraphics[width=0.32\linewidth]{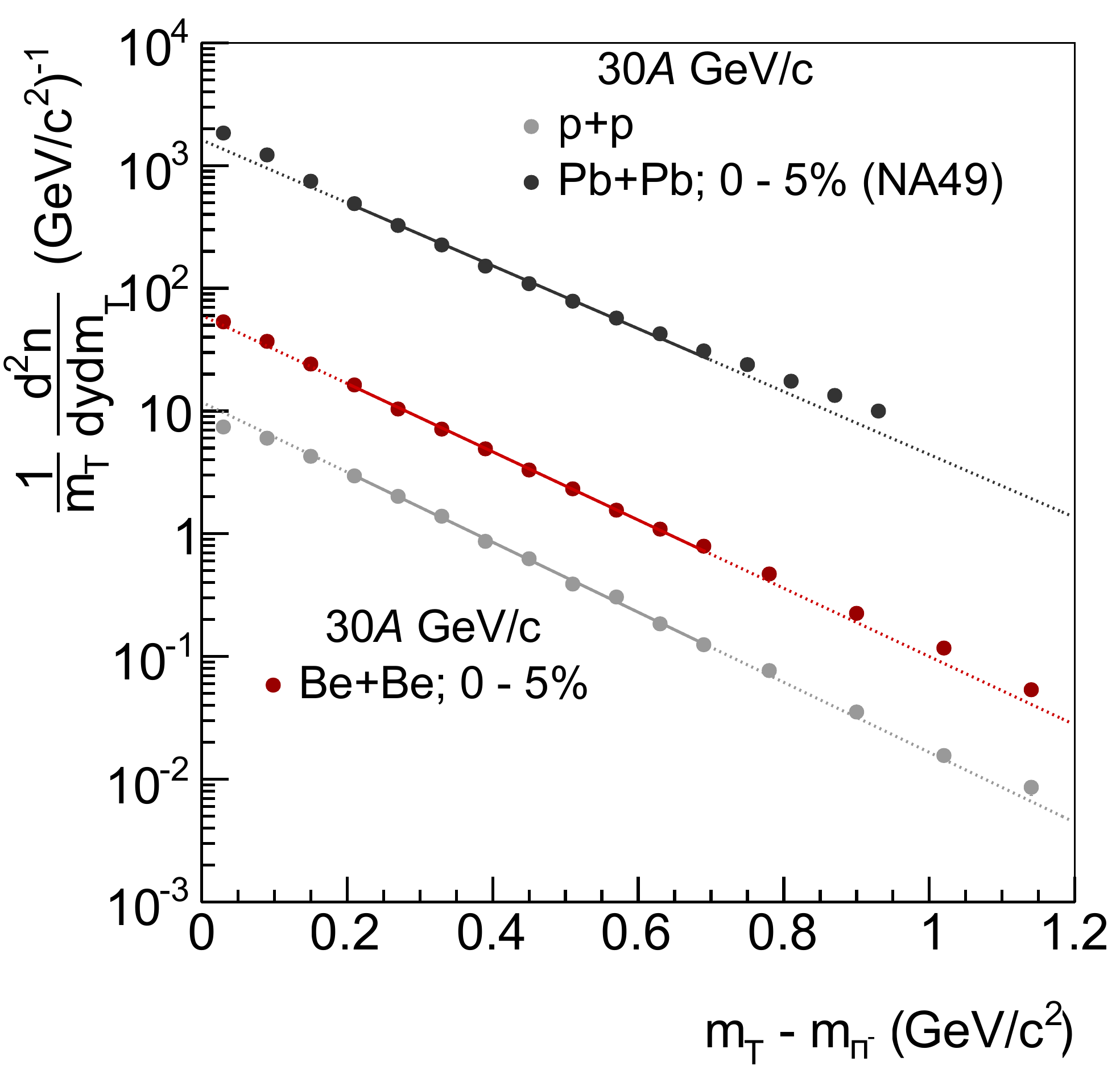}
        \includegraphics[width=0.32\linewidth]{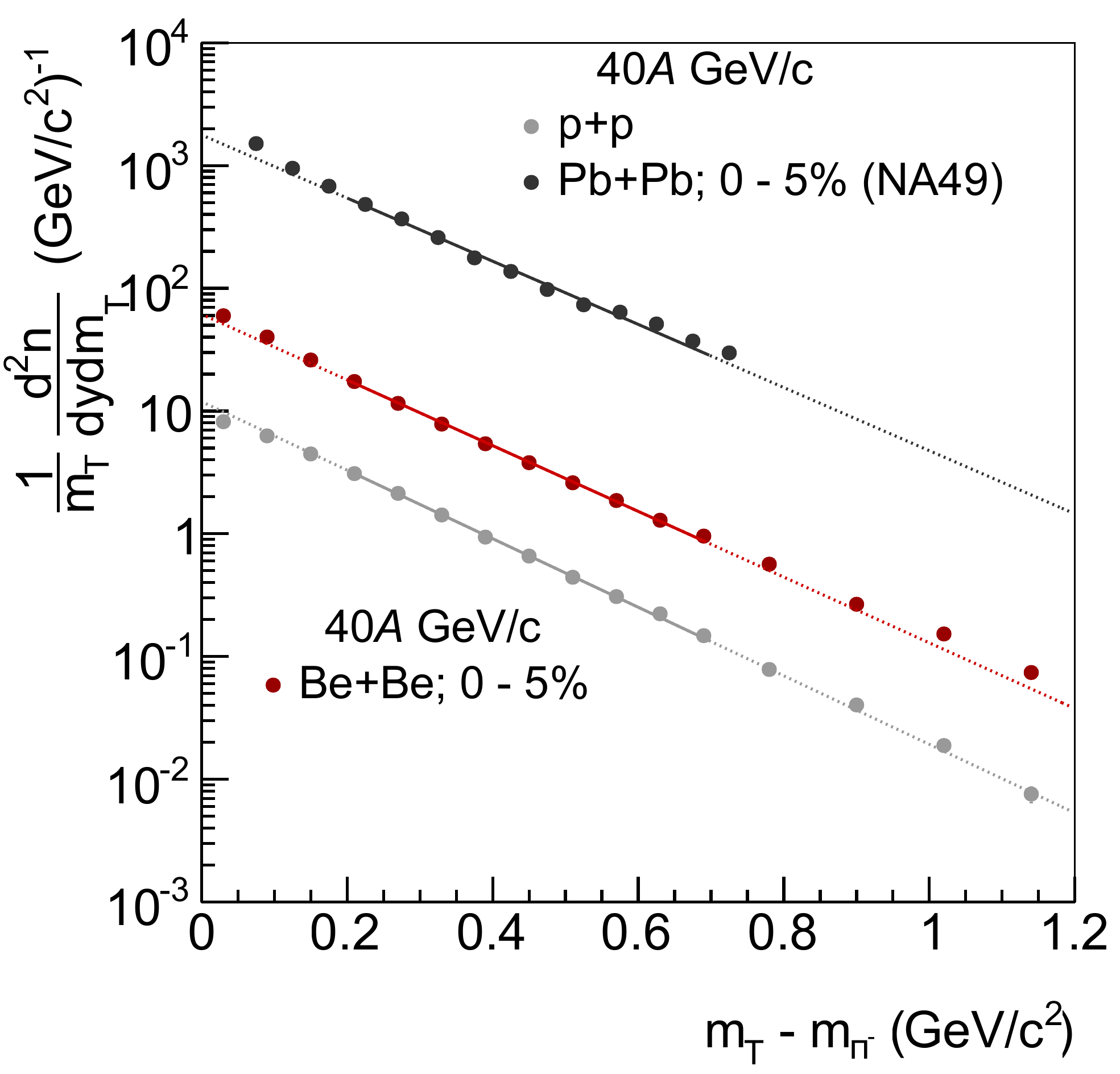}\\
        \includegraphics[width=0.32\linewidth]{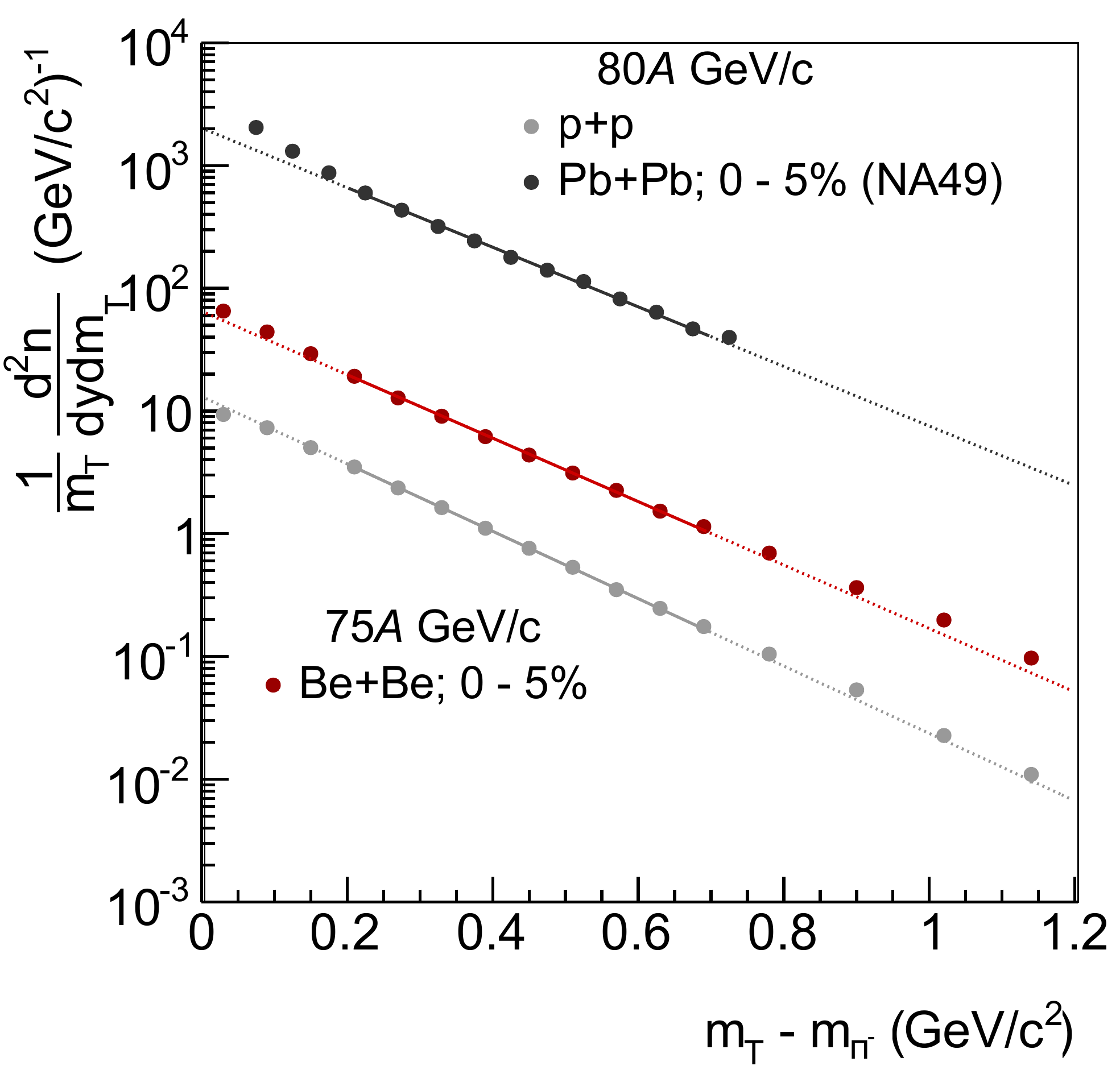}
        \includegraphics[width=0.32\linewidth]{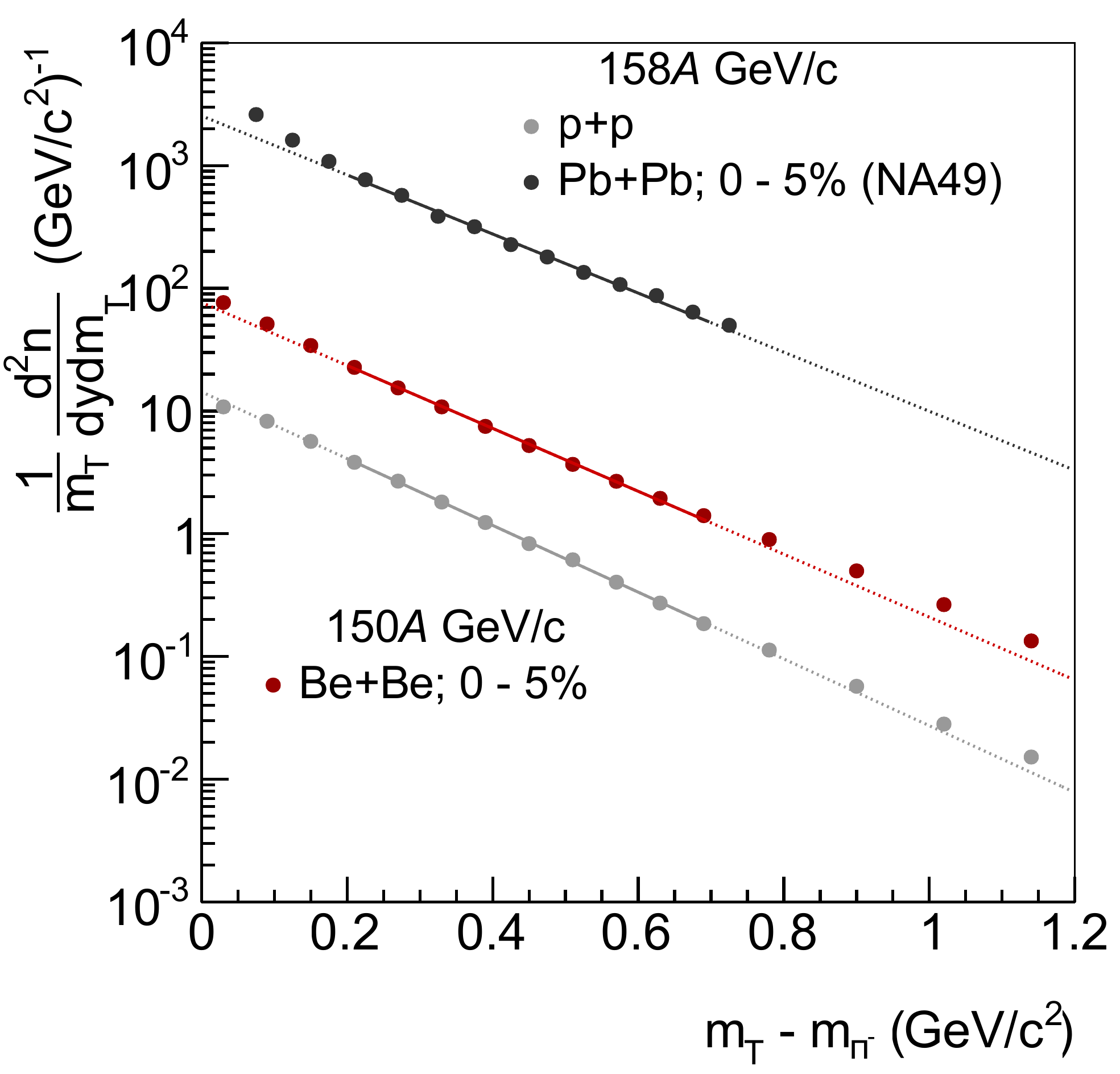}

        \caption{Transverse mass spectra $\frac{1}{m_T} \frac{d^2n}{d\y dm_T}$
           of negatively charged pions produced in \textit{central} Be+Be collisions at the SPS energies. 
           Statistical errors are smaller than the size of the points. 
           Results from inelastic \pp interactions~\cite{Abgrall:2013pp_pim} and central
           Pb+Pb interactions~\cite{Afanasiev:2002mx,Alt:2007aa} are shown for comparison.
           Lines show exponential fits with Eq.~\ref{eq:ptFit}.
           }
        \label{fig:mtSpectra}
\end{figure}

\begin{figure}[ht!]
        \centering
                \begin{minipage}[c]{0.8\textwidth}
        \includegraphics[width=0.49\linewidth]{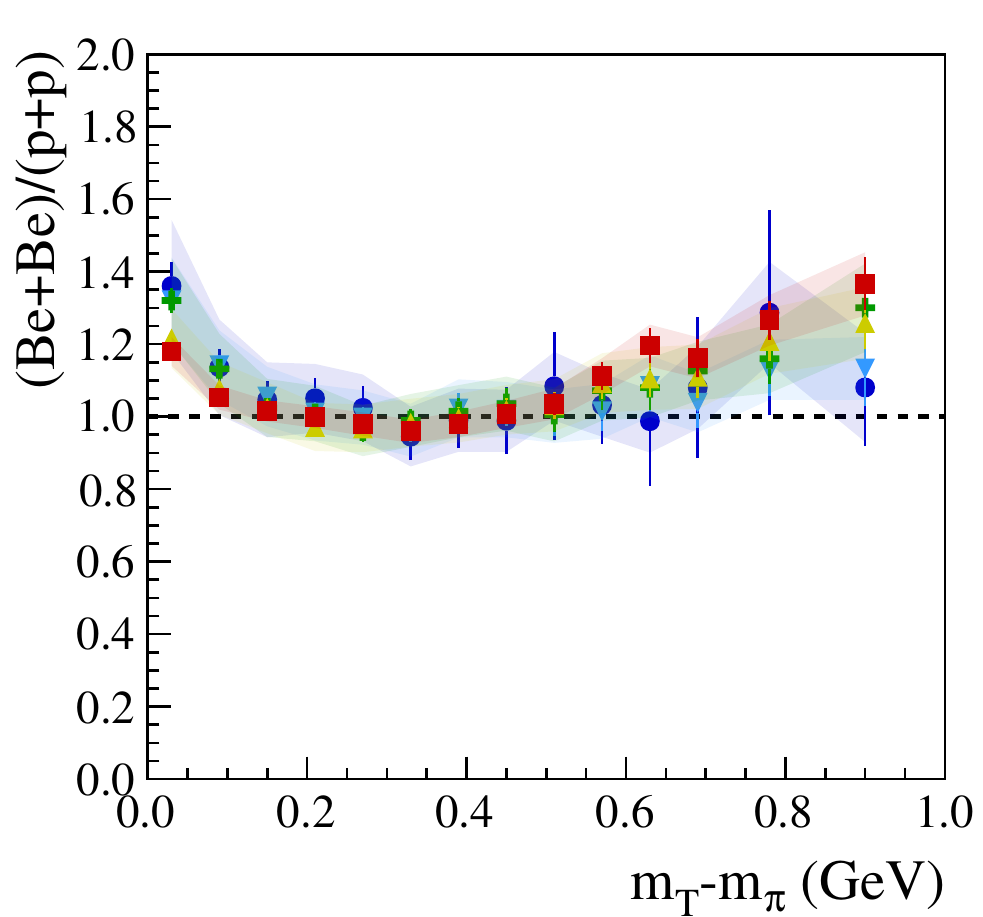}
        \includegraphics[width=0.49\linewidth]{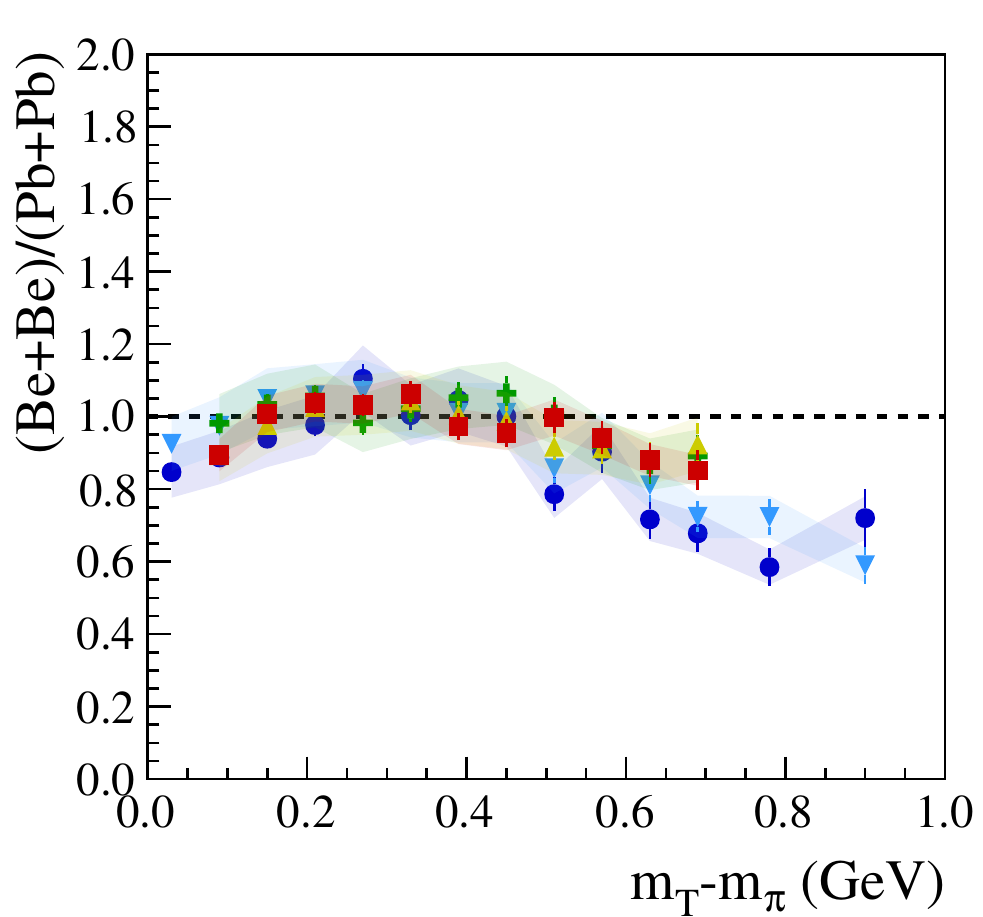}
        \end{minipage}
        \begin{minipage}[c]{0.18\textwidth}
            \begin{itemize}
                \item[\tiny\textcolor{color150Comparison}{\markerOneFiftyComparison}] 150\AGeVc
	    		\item[\tiny\textcolor{color75Comparison}{\markerSeventyFifeComparison}] 75\AGeVc
			    \item[\tiny\textcolor{color40Comparison}{\markerFourtyComparison}] 40\AGeVc
			    \item[\tiny\textcolor{color30Comparison}{\markerThirtyComparison}] 30\AGeVc
			    \item[\tiny\textcolor{color19Comparison}{\markerNineteenComparison}] 19\AGeVc
            \end{itemize}
        \end{minipage}
        \caption{Ratio of normalized transverse mass spectra: Be+Be/\pp \emph{(left)} and Be+Be/Pb+Pb \emph{(right)} at the SPS energies.
        }
        \label{fig:mtSpectraDiv}
\end{figure}

Spectra of transverse mass $m_T-m_{\pi}$ at mid-rapidity ($0<\y<0.2$) are shown in Fig.~\ref{fig:mtSpectra}
for the 5\% most \textit{central} Be+Be collisions and for inelastic \pp interactions~\cite{Abgrall:2013pp_pim}
as well as central Pb+Pb collisions~\cite{Afanasiev:2002mx,Alt:2007aa}.
The \pp data follow exponential distributions as shown by the lines fitted in the range 
$0.24 < m_T - m_{\pi^-} < 0.72$ using Eq.~\ref{eq:ptFit} expressed in $m_T$. Interestingly, 
relative to the exponential fits the spectra from nucleus-nucleus interactions develop 
enhancements at low and high transverse mass which increase with the size of the collision system. 
To compare in more detail 
the transverse mass spectra between systems, each spectrum was normalized to the integral of the spectrum 
in the range of $0.24 < m_T - m_{\pi^-} < 0.72$. The normalized Be+Be spectra were then divided by the
corresponding \pp and Pb+Pb spectra used as a reference. The resulting ratios of the normalised spectra 
are presented in Fig.~\ref{fig:mtSpectraDiv}.


The shape of $m_T$ spectra in \textit{central} Be+Be collisions is significantly different from the one observed in inelastic \pp interactions (Fig.~\ref{fig:mtSpectraDiv} \textit{left}). However, it is important to note that the 
Be+Be system is isospin symmetric whereas \pp has $I_3 =1$. Comparing Be+Be to Pb+Pb (Fig.~\ref{fig:mtSpectraDiv} \textit{right}) reveals that both shapes are similar. Note that Pb+Pb is to a large extent isospin symmetric.

\subsection{Rapidity spectra and mean multiplicities}

To extract one-dimensional rapidity spectra from the two-dimensional \y-$p_T$ spectra 
the contribution from the missing high $p_T$ acceptance has to be accounted for. 
The transverse momentum spectrum for each rapidity bin was parametrized with Eq.~\ref{eq:ptFit}.
An additional constraint was added to the fit 
to ensure that the integral of the fitted function agrees with the integral (sum) of the
measurements over the interval where data are available. The $p_T$ extrapolation 
increases the value of the summed measurements by $\approx 0.1\%$. Only for $\y > 3$ the 
extrapolation effect rises to around 1\%.

\begin{figure}[ht!]
        \centering
        \includegraphics[width=0.4\linewidth]{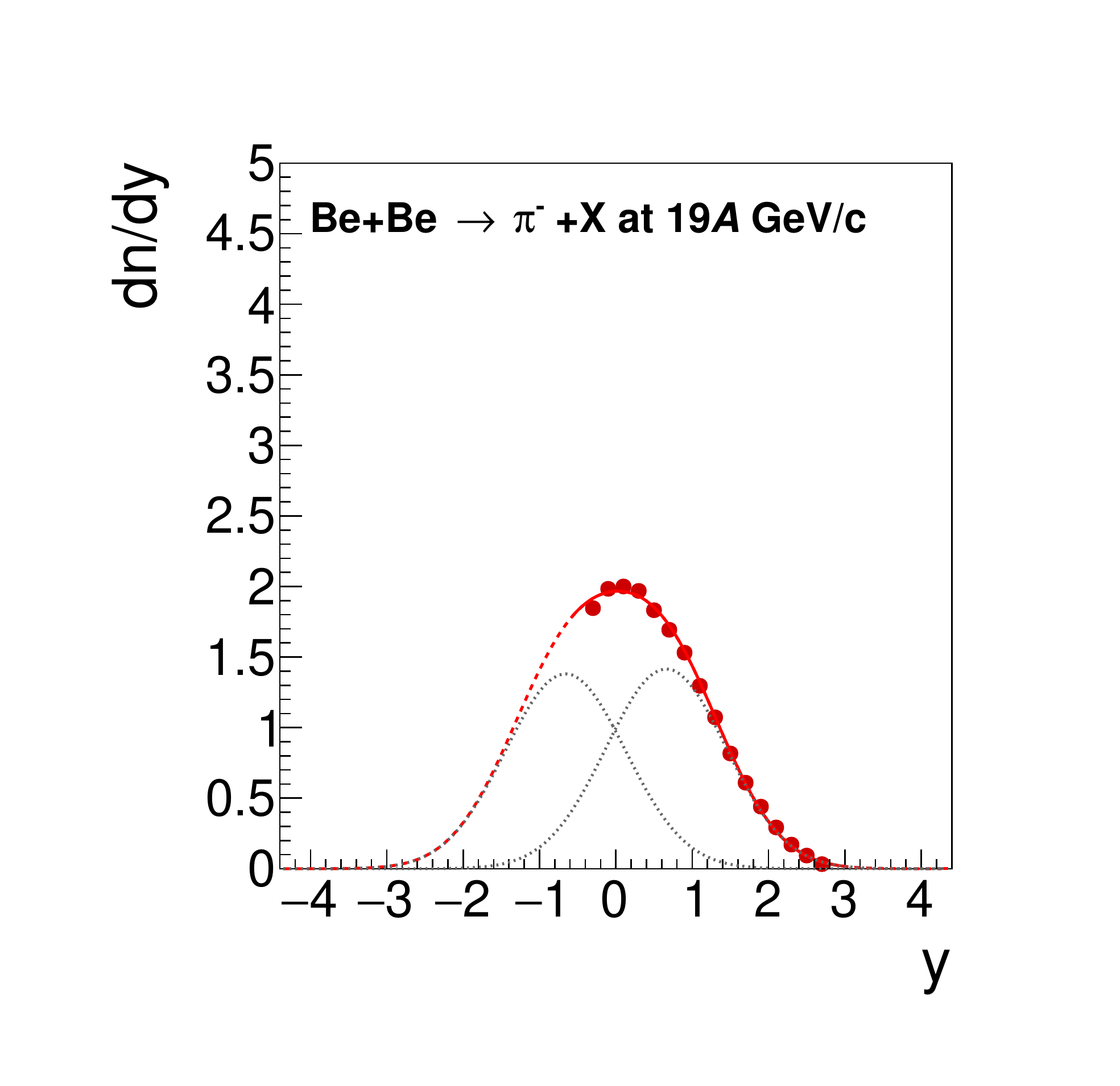}
        \includegraphics[width=0.4\linewidth]{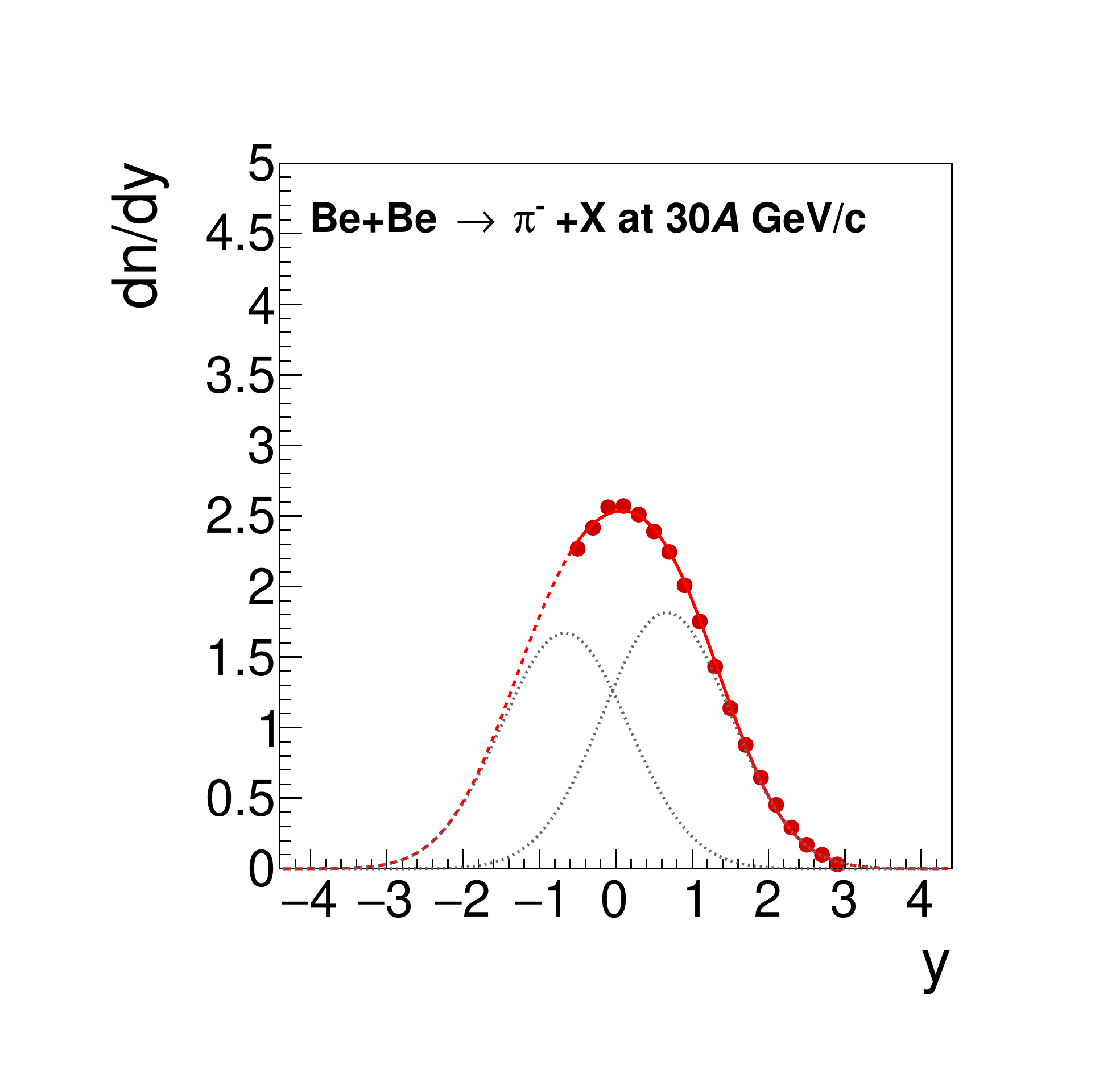} 
        \includegraphics[width=0.4\linewidth]{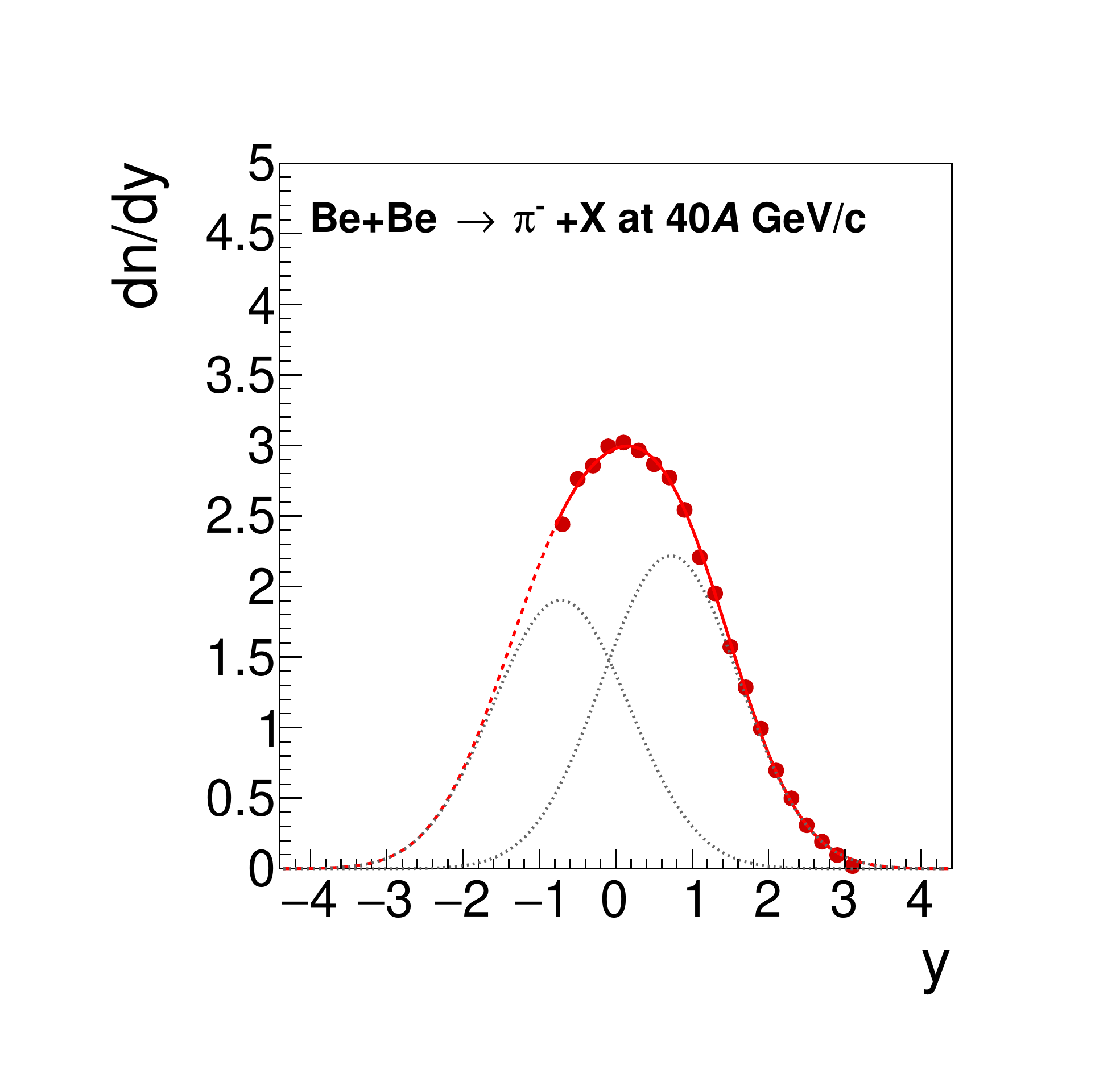}
        \includegraphics[width=0.4\linewidth]{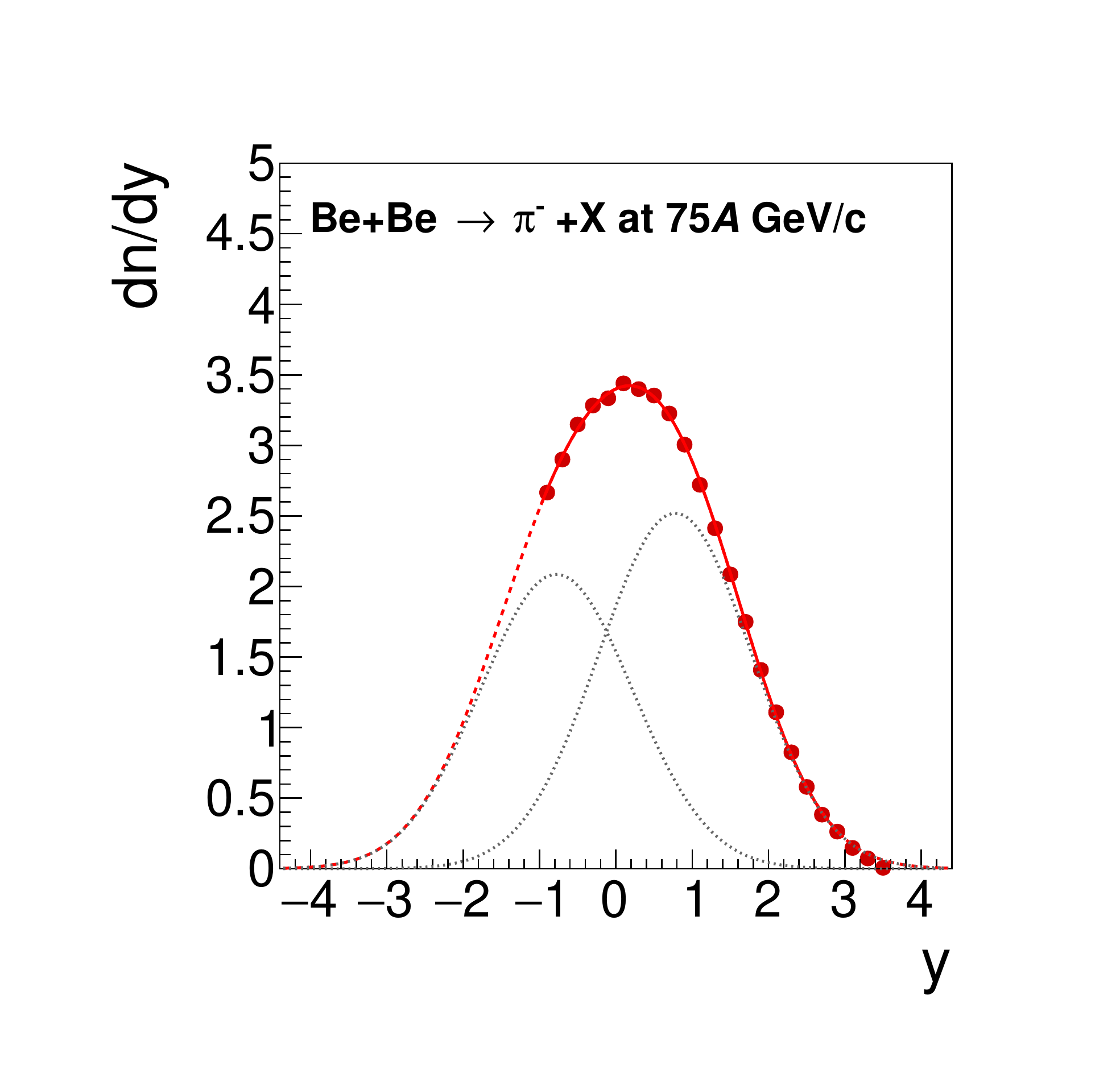} 
        \includegraphics[width=0.4\linewidth]{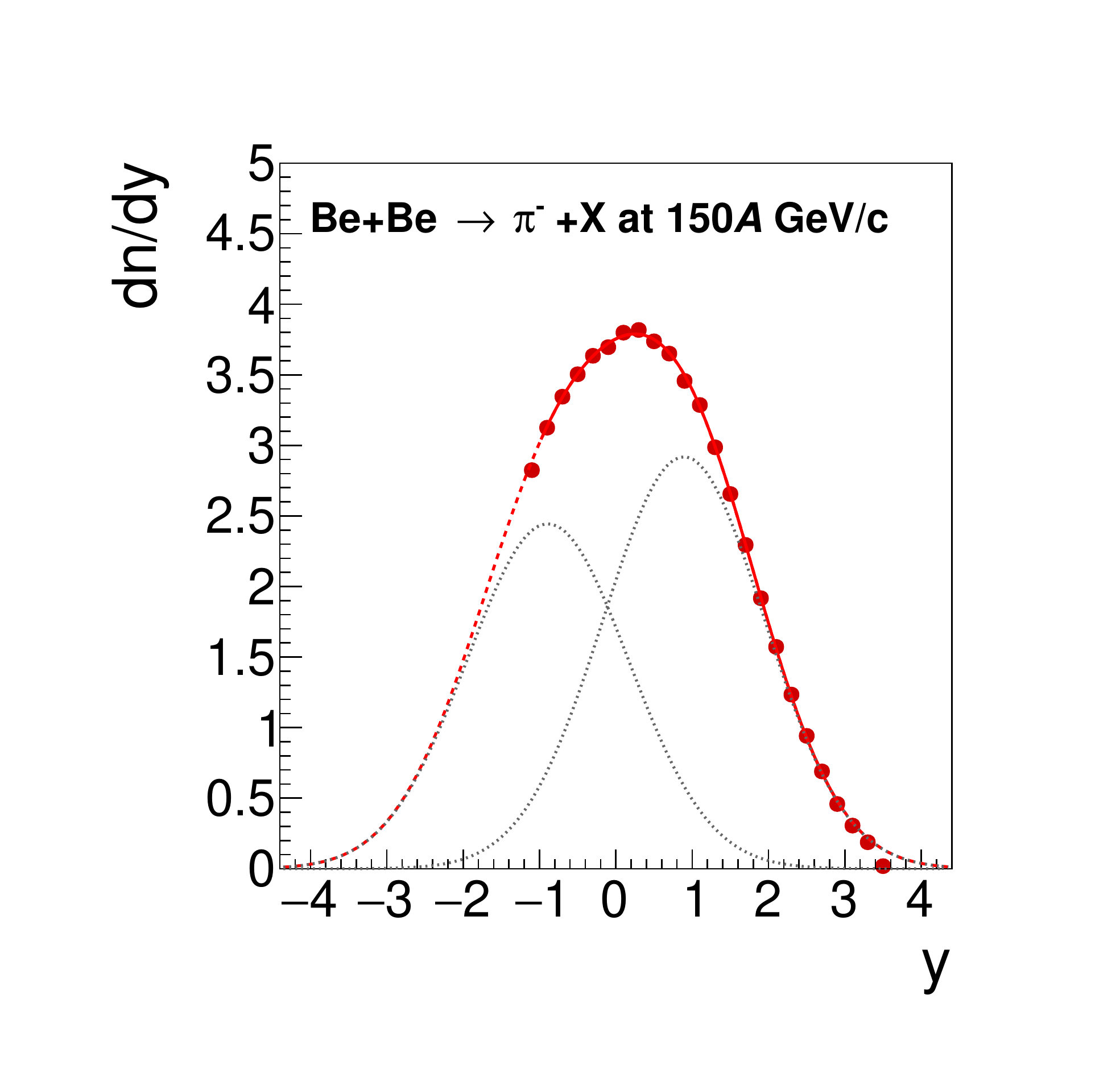}
        \caption{
        Rapidity distributions of negatively charged pions in \textit{central} Be+Be collisions at the SPS energies. The parametrization of the spectra by Eq.~\ref{eq:rapidity} is shown. The solid line shows the fitted function in the range of the fit, and the red dashed line 
        depicts the extrapolation of the fitted function. The two Gaussian functions 
        constituting the fitted function are represented by the black dashed lines.
                 }
        \label{fig:fittedRapidity}
\end{figure}

The rapidity spectra are plotted in Fig.~\ref{fig:fittedRapidity}. A closer look reveals an asymmetry
of the spectra with respect to mid-rapidity. To quantify the amount of asymmetry the spectra
were parametrized with the sum of two Gaussian functions with the same width and mean value displaced from mid-rapidity by the same amount:
\begin{equation}
 g(\y)=\frac{A \cdot A_\text{rel}}{\sigma\sqrt{2\pi}}\exp\left(-\frac{(\y-\y_0)^2}{2\sigma^2}\right)
      +\frac{A_0}{\sigma\sqrt{2\pi}}\exp\left(-\frac{(\y+\y_0)^2}{2\sigma^2}\right)   ,
\label{eq:rapidity}
\end{equation}
where $A$ is the normalization parameter, $A_\text{rel}$ is the relative amplitude of the Gaussian distributions, $\sigma$ is the common width and $\y_0$ is the displacement from mid-rapidity. Results for 
the fitted functions are presented in Fig.~\ref{fig:fittedRapidity}. Numerical values of the
fitted parameters $A_\text{rel}$, $\y_0$ as well as the RMS width $\sigma_\y$ of the rapidity distribution calculated from

\begin{equation}
\label{eq:y_width}
\sigma_\y = \sqrt{\sigma^2 + \y_0^2}
\end{equation}

are listed in Table~\ref{tab:gauss_params}.

\begin{table*}
 \centering
 \footnotesize
 \begin{tabular}{l|cccccc}
  Momentum (\AGeVc) & 19 & 30 & 40 & 75 & 150 \\
  \\
  \hline
  \\
        $A_\text{rel}$ & $0.975$ & $0.919$ & $0.858$ & $0.828$ & $0.837$\\
        $\delta(A_\text{rel})$ & $0.0319$ & $0.0172$ & $0.0172$ & $0.0100$ & $0.0076$\\
        $y_0$& $0.659$ & $0.667$ & $0.720$ & $0.778$ & $0.891$\\
        $\delta(y_0)$ & $0.0032$ & $0.0065$ & $0.0078$ & $0.0062$ & $0.0047$\\
        $\sigma_y$ & $1.025$ & $1.067$ & $1.148$ & $1.265$ & $1.385$\\
        $\delta(\sigma_y)$ & $0.048$ & $0.052$ & $0.061$ & $0.059$ & $0.057$\\
 \end{tabular}
 \caption{Fitted parameters $A_\text{rel}$, $\y_0$ of the double Gaussian fit and RMS width $\sigma_\y$ 
          of the rapidity distribution calculated from Eq.~\ref{eq:y_width}.}
 \label{tab:gauss_params}
\end{table*}
 

The relative amplitude of the Gaussian distributions decreases slowly with increasing beam momentum, i.e the asymmetry
increases. This deviation of $A_\text{rel}$ from unity signals a forward-backward asymmetry of 
the rapidity distribution which may be explained by the asymmetry of the collision system and the event
selection procedure:
\begin{enumerate}[(i)]
        \item Asymmetric collisions of a $^7$Be beam with a larger mass 
              $^9$Be target may lead to enhanced particle production at backward rapidity,
        \item a larger number of neutrons in the $^9$Be target nuclei might result in a difference in
              the ratio of \pim to \pip in the backward and forward rapidity regions,
        \item selection of \textit{central} collisions by requiring the forward energy $E_F$ below a cut value.
\end{enumerate}

The asymmetry was studied using the Wounded Nucleon Model (WNM)~\cite{Bialas:1976ed}, where production 
of particles in the backward hemisphere is proportional to the number of wounded nucleons in the target 
and production of particles in the forward hemisphere is proportional to the number of wounded nucleons 
in the projectile. In the WNM the effect of the asymmetric system leads to a small enhancement of 
of the particle yield below mid-rapidity, which is opposite to what is seen in the data. On the other hand, 
the effect of the \textit{centrality} selection based only on the forward energy is  
enhancing particle production at forward rapidity. The data show that the latter effect dominates.

\begin{figure}[ht!]
        \centering
    \includegraphics[width=0.4\linewidth]{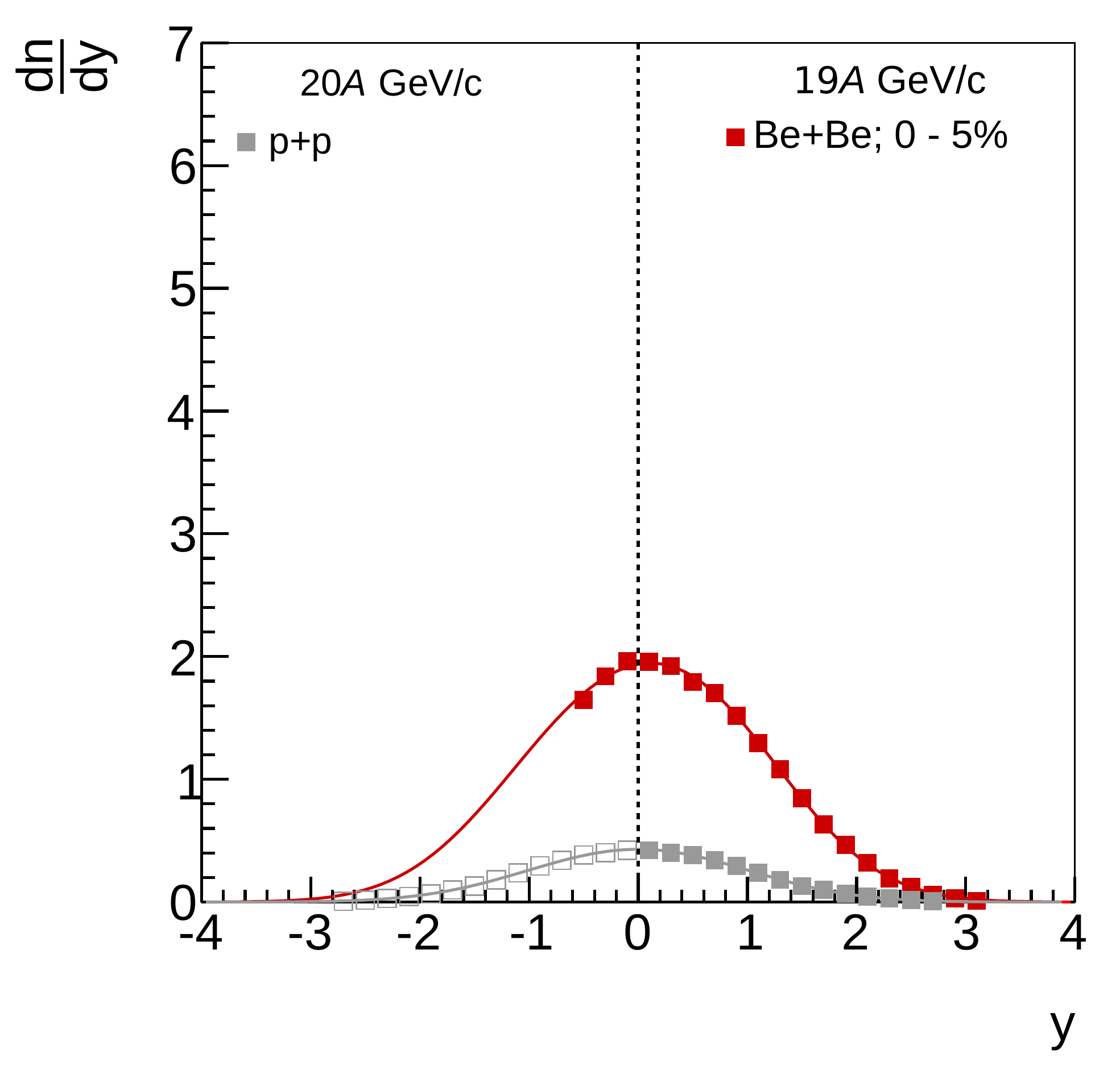}
    \includegraphics[width=0.4\linewidth]{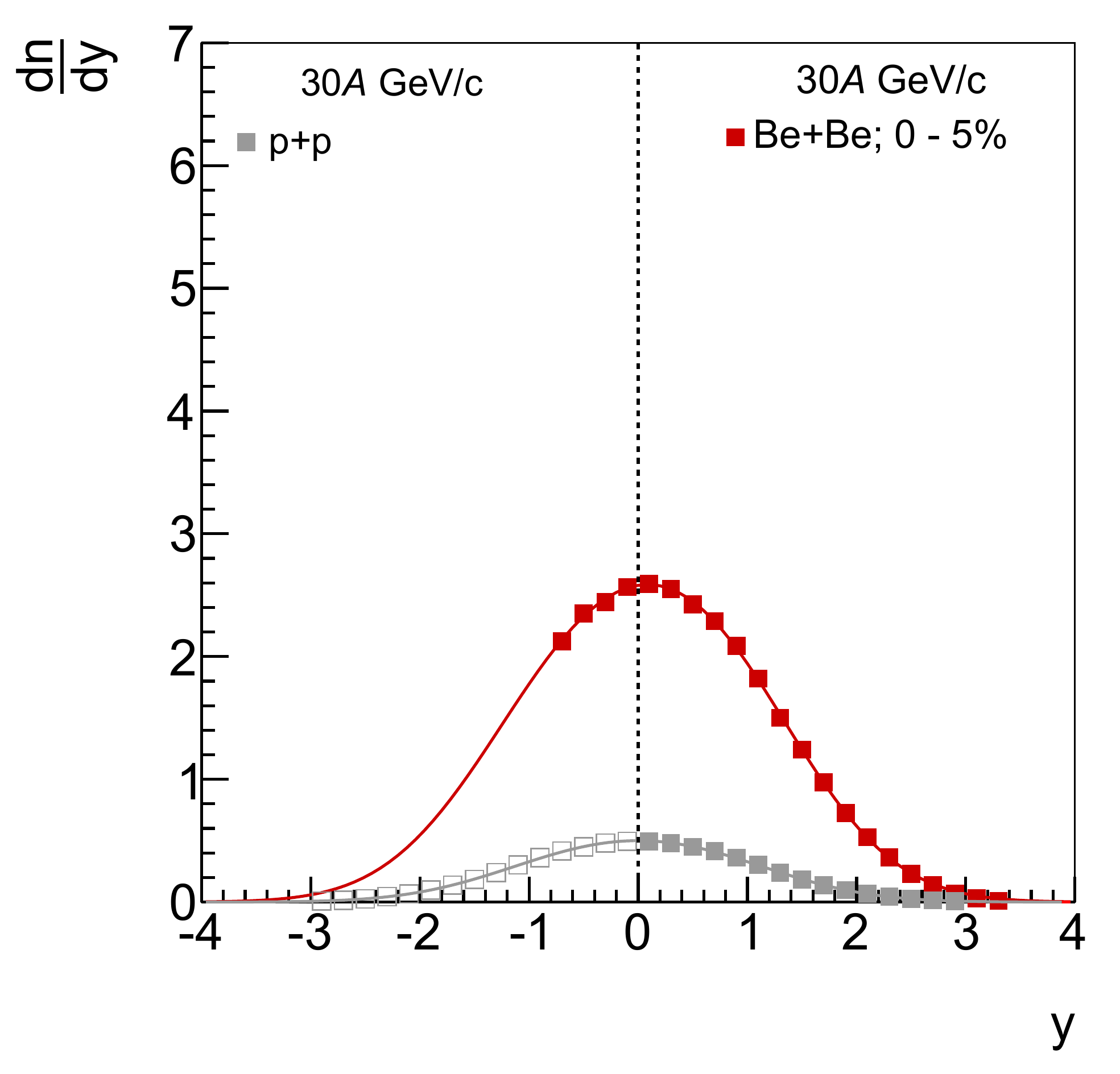} 
    \includegraphics[width=0.4\linewidth]{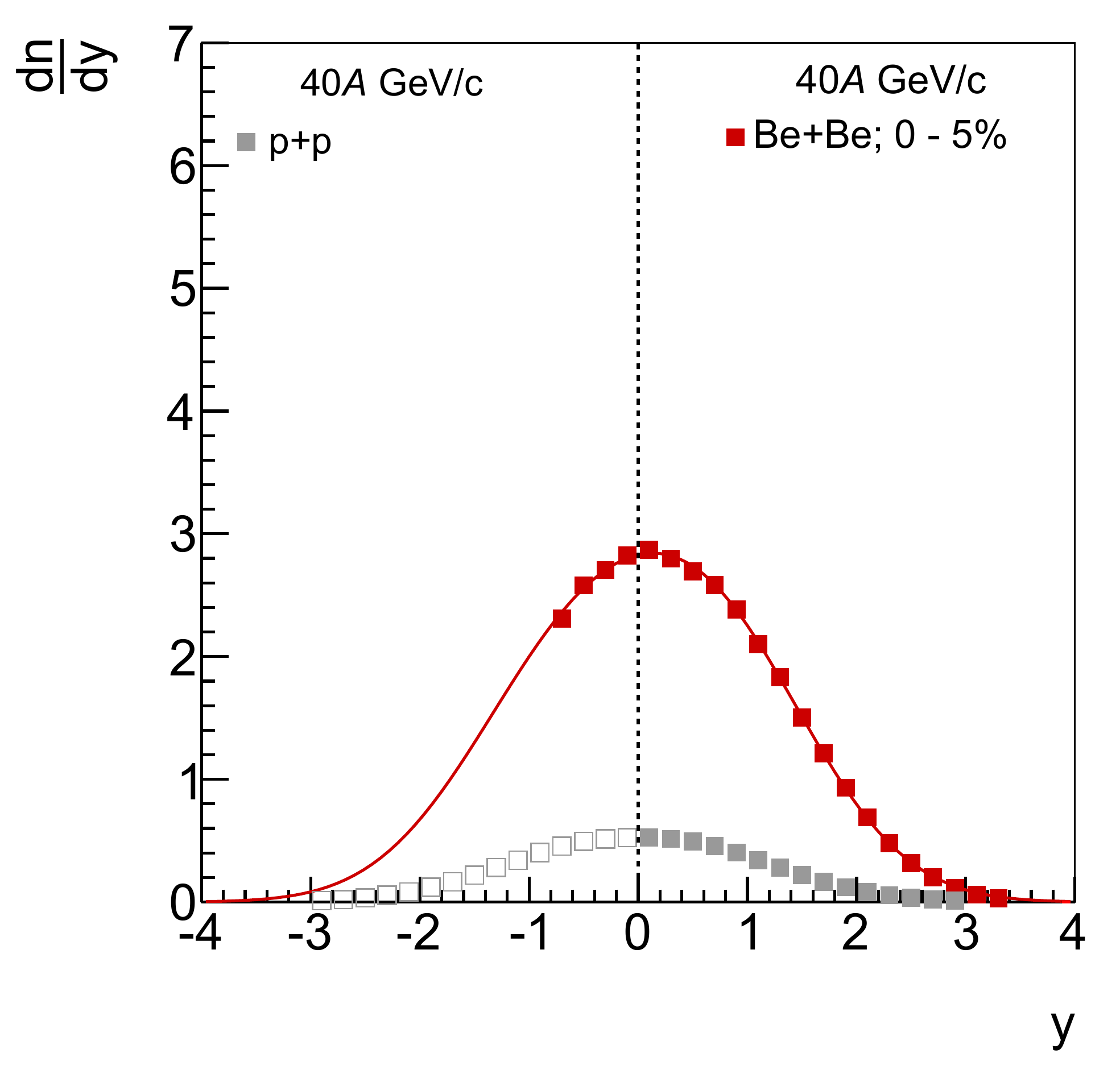}
    \includegraphics[width=0.4\linewidth]{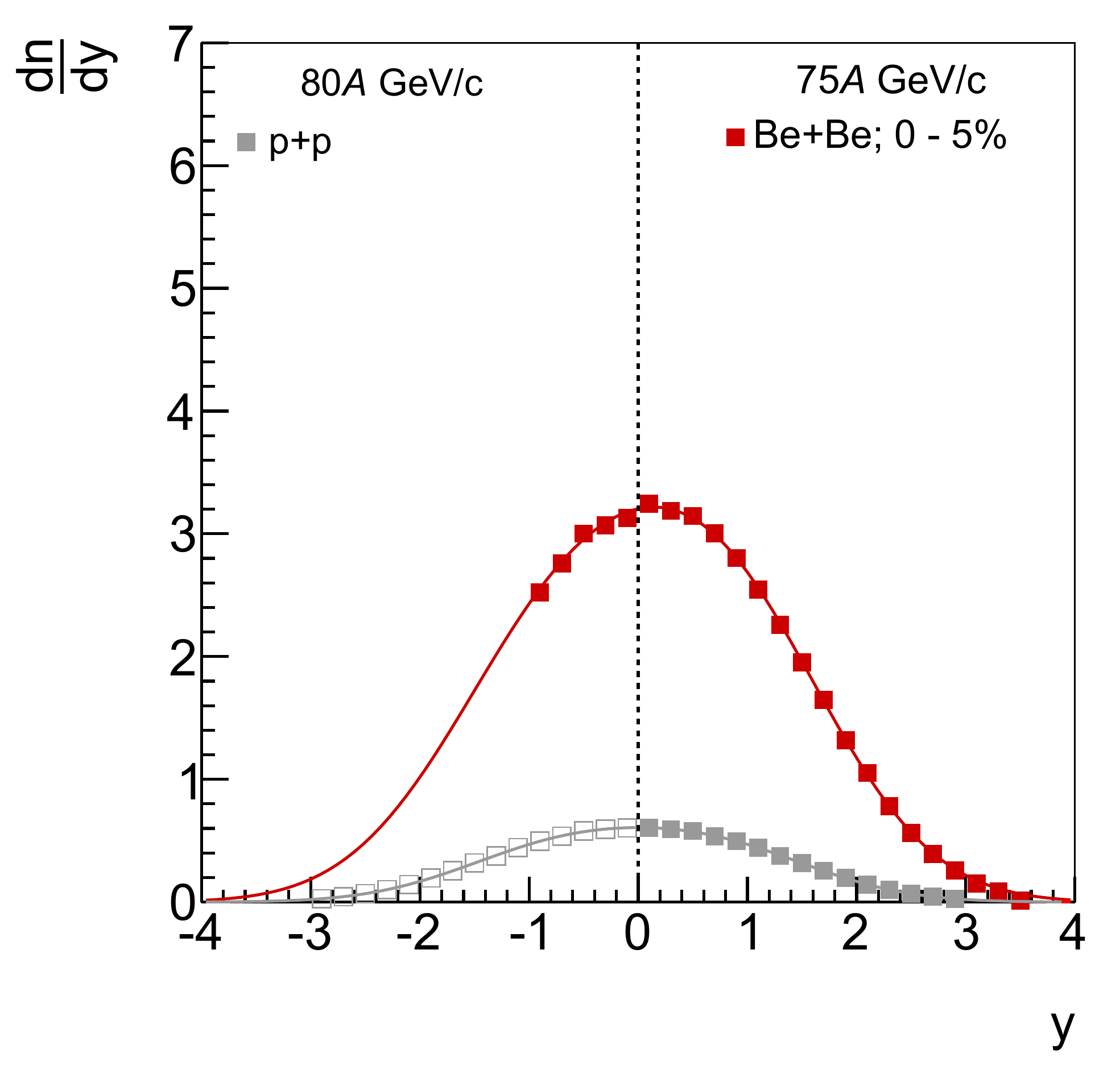}
    \includegraphics[width=0.4\linewidth]{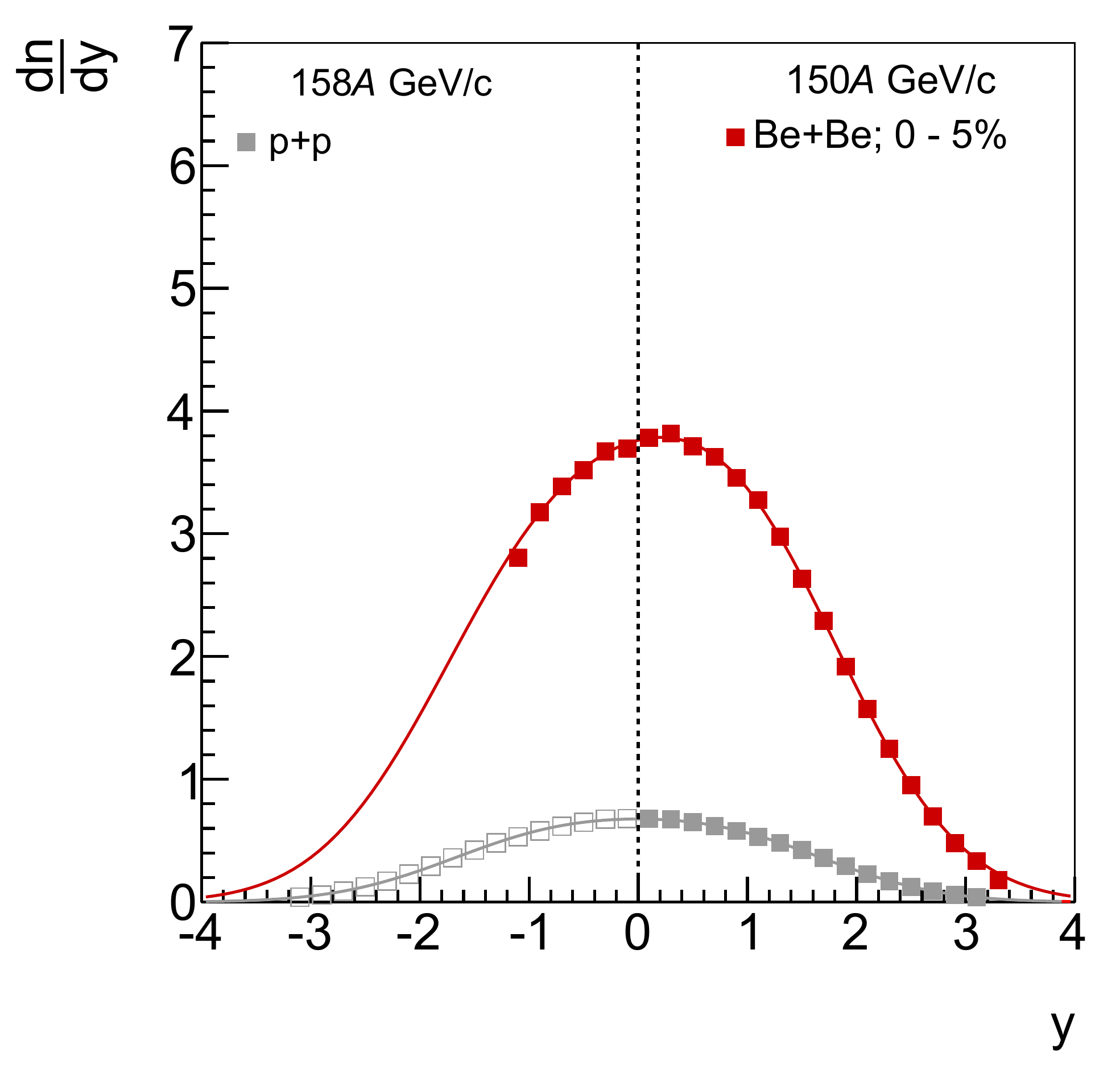}

        \caption{Rapidity spectra of \pim produced in the 5\% most \textit{central} Be+Be collisions at the SPS energies. 
           Results from inelastic \pp interactions~\cite{Abgrall:2013pp_pim} are shown for comparison. Statistical errors 
           are smaller than the size of the markers.
           }
        \label{fig:rapidity}
\end{figure}

Rapidity spectra in \textit{central} Be+Be collisions are compared to results from
inelastic \pp interactions~\cite{Abgrall:2013pp_pim} in Fig.~\ref{fig:rapidity}.
Mean negative pion multiplicities $\langle\pi^-\rangle$ were obtained by summing the measured
data points and adding a contribution from the fitted function Eq.~\ref{eq:rapidity} for the
unmeasured region. Half of the contribution added based on the fit is added to systematic uncertainty.  The results are listed in Table~\ref{tab:piMultiplicity}. 

\begin{table*}
 \centering
 \footnotesize
 \begin{tabular}{l|cccccc}
  \\
  Momentum (\AGeVc) & 19 & 30 & 40 & 75 & 150 \\
  \\
  \hline
        $\langle \pim \rangle$ & $5.33$ & $7.61$ & $8.75$ & $11.98$ & $14.32$\\
        $\delta_{\text{stat}}(\langle \pim \rangle)$ & $\pm 0.11$ & $\pm 0.08$ & $\pm 0.09$ & $\pm 0.07$ & $\pm 0.09$\\
        $\delta_{\text{sys}}(\langle \pim \rangle)$& $\pm 0.63$ & $\pm 0.90$ & $\pm 0.95$ & $\pm 1.20$ & $\pm 1.37$\\
        $\langle \pi \rangle$ / $\langle W \rangle$ & 1.66 & 2.39 & 2.69 & 3.70 & 4.48 \\
 \end{tabular}
 \caption{Mean \pim multiplicities in the 5\% most \textit{central} Be+Be collisions with statistical and 
 systematic uncertainties as well as ratios of mean $\pi$ multiplicities to average number of wounded nucleons.
 }
 \label{tab:piMultiplicity}
\end{table*}

\begin{figure}[htbp]
        \centering
        \includegraphics[width=0.45\linewidth]{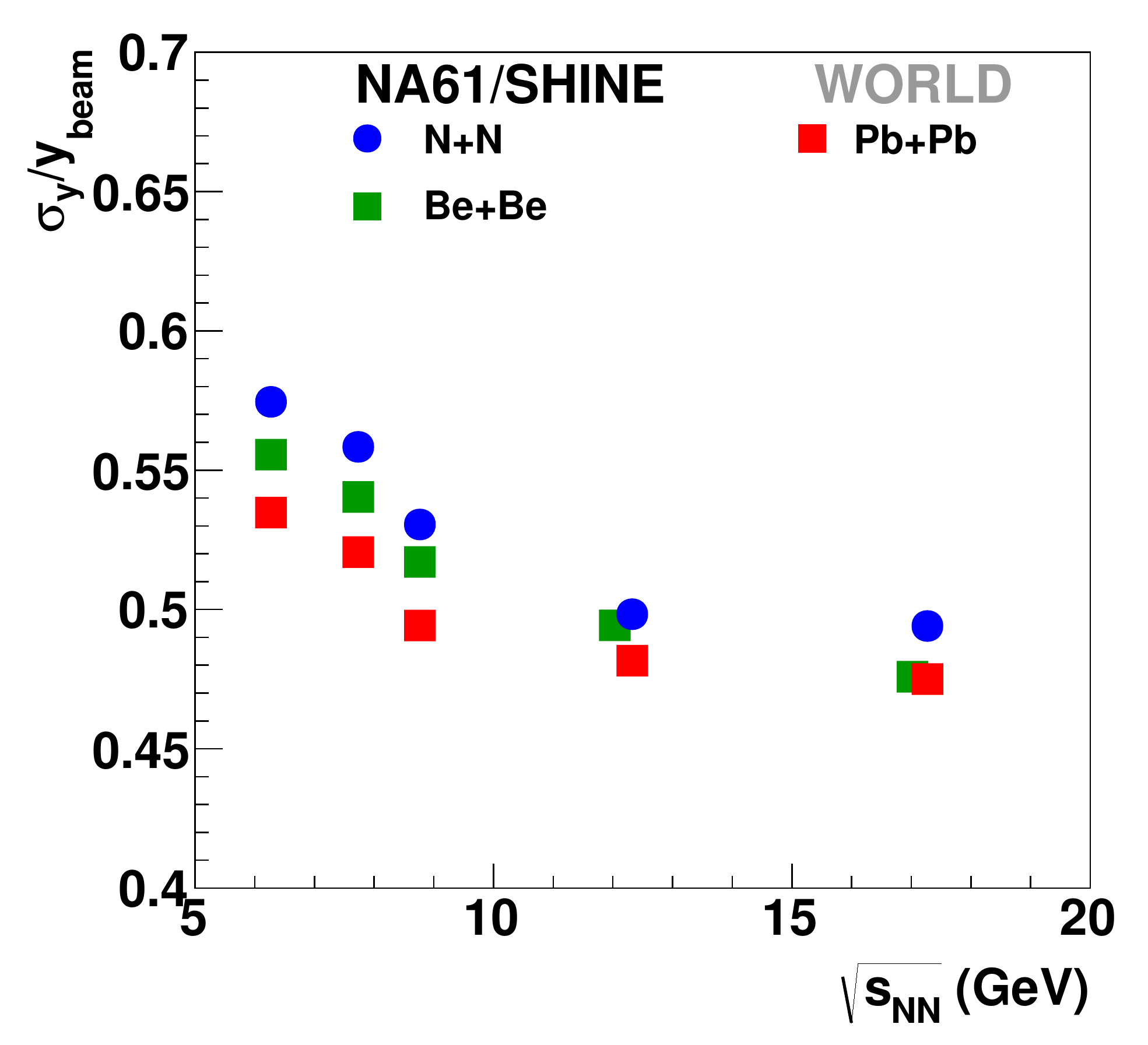}
        \caption{Collision energy dependence of the ratio of the width $\sigma_y$ of the rapidity distribution to the 
                 beam rapidity $y_{beam}$ for \textit{central} Be+Be and Pb+Pb collisions and inelastic \NN interactions.
                 }
        \label{fig:isoSigmaY}
\end{figure}

The widths of the rapidity distributions were calculated 
from~Eq.\ref{eq:y_width} and are listed in Table~\ref{tab:gauss_params}. The beam energy dependence of 
the width of the rapidity distribution divided by the beam rapidity $\sigma_\y/\y_\text{beam}$ is presented 
in Fig.~\ref{fig:isoSigmaY}. For Be+Be and Pb+Pb interactions the ratio was calculated for \pim mesons. 
Since the \pp collision system is not isospin symmetric the isospin average (\pim + \pip)/2
was plotted for comparison. These results are referred to as results for nucleon-nucleon (\NN) collisions~\cite{Gazdzicki:1991ih}. 
For all system sizes the relative width decreases monotonically with beam energy and system size.

\section{Discussion}
\label{sec:discussion}

Several features of $\pi$ meson production were predicted to be sensitive to the onset 
of deconfinement, namely the energy dependence of 
the transverse mass distribution, the width of the rapidity distribution - both due to the softening of the equation of state~\cite{Hung:1994eq,Rischke:1995pe,Brachmann:1999mp} - and the mean multiplicity due to the increasing entropy during the transition from the hadronic to the partonic phase~\cite{Gazdzicki:1995ze}. The data presented in this paper are discussed in the context of these predictions in the following.

In the collision energy range in which the mixed hadron and parton matter is created
a stalling of the expansion of the system is expected~\cite{Gorenstein:2003cu}. This results in a slowing of the increase of
radial flow of the produced particles and in a step-like structure in the energy dependence
of the inverse slope parameter of the transverse mass spectra $T$~\cite{Gazdzicki:1998vd}. This feature was clearly observed 
for $K$ mesons in central Pb+Pb collisions and was interpreted as one of the indications of the onset of
deconfinement~\cite{Afanasiev:2002mx,Alt:2007aa}.

\begin{figure}[ht!]
        \centering
        \includegraphics[width=0.6\linewidth]{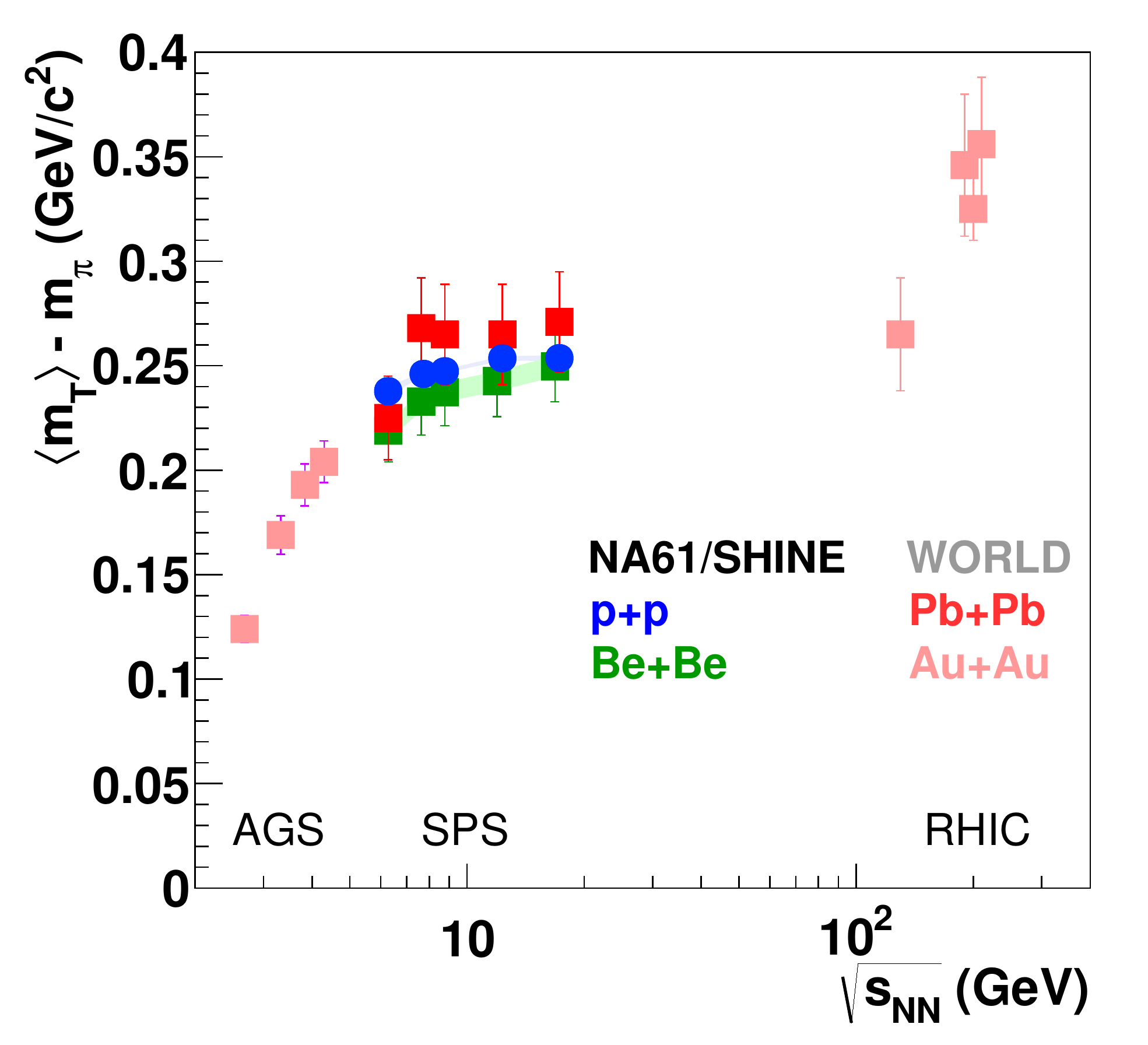}
        \caption{
        Energy dependence of the mean transverse mass of \pim measured at
        mid-rapidity in \textit{central} Be+Be, Pb+Pb~\cite{Afanasiev:2002mx, Alt:2007aa}  and Au+Au~\cite{Klay:2003zf, Bearden:2004yx} collisions and inelastic \pp interactions~\cite{Abgrall:2013qoa}.
        }
        \label{fig:pionStep}
\end{figure}

Pion transverse mass spectra deviate significantly from the exponential function Eq.~\ref{eq:ptFit} used to fit the inverse slope parameter. This is attributed to a large contribution of pions from resonance decays and possible effect of 
transverse collective flow~\cite{Gorenstein:2003cu}.
Thus in order to avoid model-dependence of results $\pi$ transverse mass spectra
are characterize by the mean transverse mass $\langle m_T \rangle - m$.
Figure~\ref{fig:pionStep} shows the "step" plot for \pim in \textit{central} Be+Be collisions compared to central Pb+Pb interactions
and inelastic \pp reactions. The values and energy dependence measured in Be+Be collisions are surprisingly similar 
to those in inelastic \pp interactions and there is only a small increase towards \textit{central} Pb+Pb collisions. 
This suggests that the average transverse mass of \pim mesons is only weakly sensitive to the transverse flow and 
thus it is not a discriminating observable for the onset of deconfinement. 

The Landau hydrodynamical model of high energy collisions~\cite{Landau:1953,Belenkij:1956cd} predicts rapidity distributions 
of Gaussian shapes. In fact this prediction is approximately confirmed by the experimental data,
see Ref.~\cite{Blume:2005ru} and references therein. Moreover, the collision energy dependence
of the width was derived by Shuryak~\cite{Shuryak:1972zq} from the same model under simplifying assumptions and reads:
\begin{equation}
\label{eq:cs}
\sigma_\y^2(\pi^-) = \frac{8}{3}\frac{c^2_\text{s}}{1-c^4_\text{s}}\ln{\left(\frac{\sqrt{s_{NN}}}{2m_p}\right)}~,
\end{equation}
where $c_\text{s}$ denotes the speed of sound, and $c^2_\text{s} = 1/3$ for an ideal gas of massless particles.

\begin{figure}[ht!]
        \centering
        \includegraphics[width=0.48\linewidth]{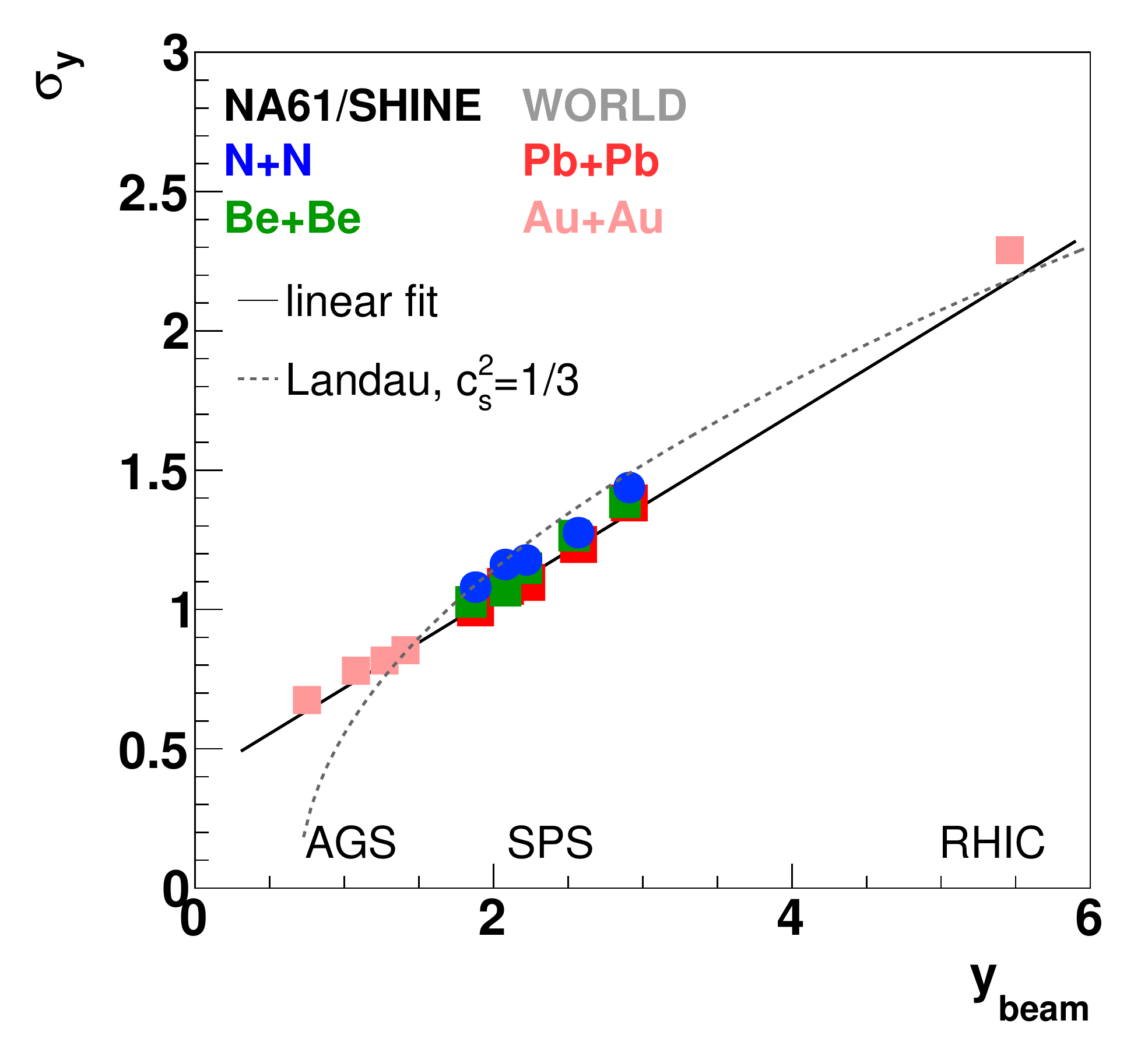}
        \includegraphics[width=0.48\linewidth]{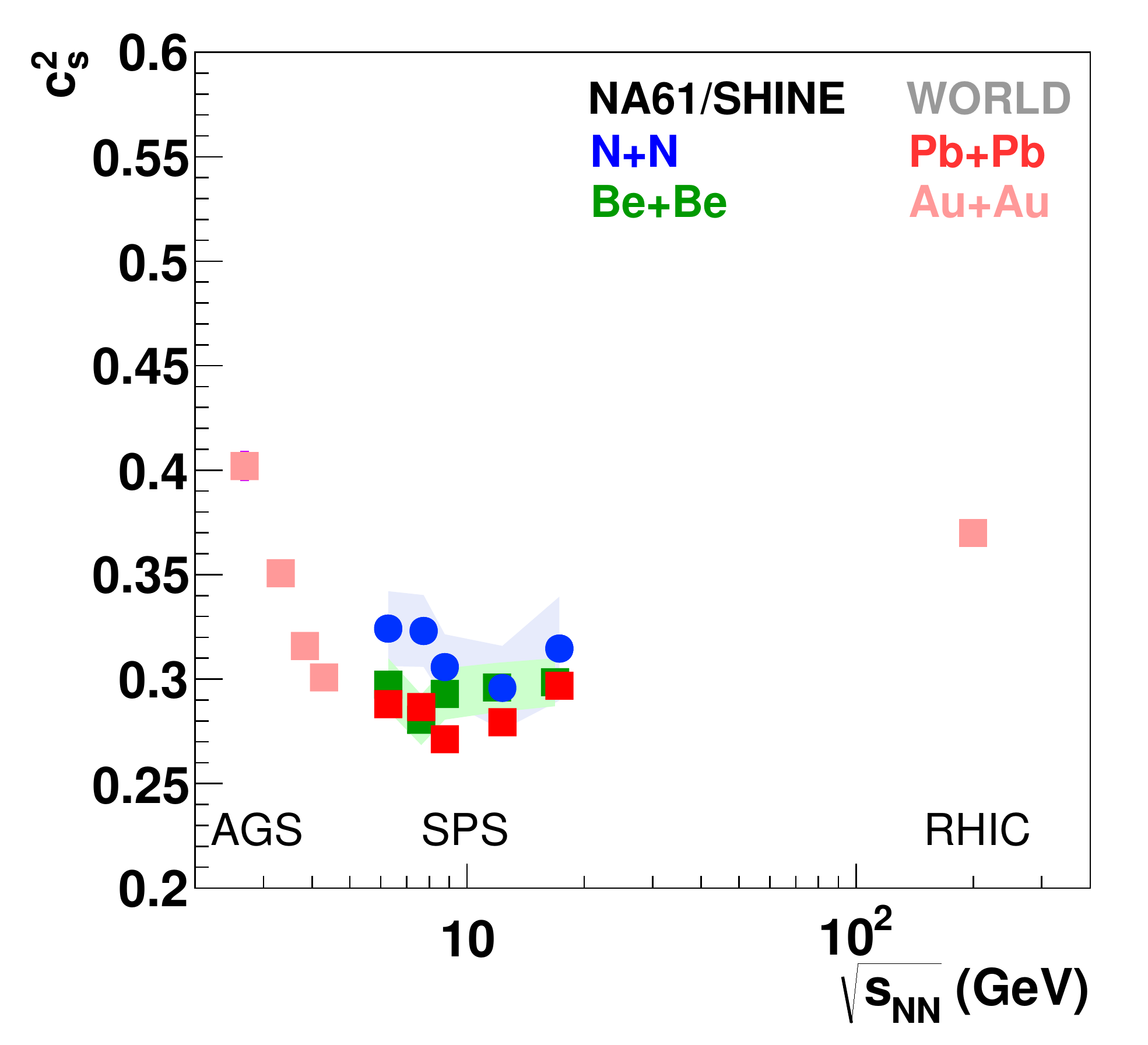}
        \caption{
        Comparison of the Landau hydrodynamical model with rapidity distributions
        of charged pions produced in central nucleus-nucleus collisions and inelastic nucleon-nucleon interactions. \textit{Left:} The width $\sigma_y$ of the rapidity
        distributions of negatively charged pions in \textit{central} Be+Be, Pb+Pb~\cite{Afanasiev:2002mx, Alt:2007aa}  (Au+Au~\cite{Klay:2003zf, Bearden:2004yx}) reactions and the width $\sigma_y$ of the average of rapidity
        distributions of positively and negatively charged pions in \pp~\cite{Aduszkiewicz:2017sei, Abgrall:2013qoa} (denoted as \NN) as a function of the beam rapidity $y_{beam}$.  The dotted line indicates the Landau model
        prediction with $c^2_\text{s} = 1/3$, while the full line shows a linear fit through the data
        points. \textit{Right:} The speed of sound as a function of beam energy 
        as extracted from the data using Eq.~\ref{eq:cs}. 
        }
        \label{fig:Landau}
\end{figure}

The above prediction is compared with the experimental data on the width $\sigma_y$ of the rapidity
distributions of $\pi^-$ mesons produced in central nucleus-nucleus collisions and
inelastic nucleon-nucleon interactions in Fig.~\ref{fig:Landau}~(\textit{left})\footnote{For \pp interactions
the figure shows isospin symmetrised values denoted as \textit{N+N}~\cite{Abgrall:2013pp_pim}}.
The model calculations are close to the measured dependence on the beam rapidity $y_{beam}$.
However, linear increase with $y_{beam}$ provides a better fit to the measurements as shown by the
straight line fit. The deviations were attributed to the changes in the equation of 
state~\cite{Bleicher:2005tb,Petersen:2006mp}, which can be effectively parametrised by allowing the speed 
of sound to be dependent on collision energy. Clearly the measured values of $\sigma_y$ 
differ very little between the studied reactions in the SPS energy range.

By inverting Eq.~\ref{eq:cs} one can express $c^2_\text{s}$
in the medium as a function of the measured width of the rapidity distribution.
The energy dependence of the sound velocities extracted from the data
are presented in Fig.~\ref{fig:Landau}~(\textit{right}). The energy range for results from Be+Be collisions
and inelastic \textit{p+p} reactions is too limited and the fluctuations in the data too large to allow 
a significant conclusion about a possible minimum. Data on \textit{central} Pb+Pb collisions, in combination with results from AGS and RHIC on central Au+Au collisions,  cover a much wider energy range. Here the sound velocity exhibits a clear minimum (usually called the softest point) 
at $\sqrt{s_{NN}} \approx 10$~GeV consistent with the reported onset of deconfinement~\cite{Afanasiev:2002mx,Alt:2007aa}. 

Pions are the most copiously produced hadrons ($\approx90\%$) in collisions of nucleons and nuclei at
SPS energies. Their multiplicity is closely related to the entropy produced in
such interactions~\cite{Fermi:1950jd}. 
Within the Fermi and Landau models the entropy is expected to increase as
\begin{equation}
    S \sim F~,
\end{equation}
where the Fermi collision energy measure is defined as 
\begin{equation}
F=\left[(\sqrt{s_\textit{NN}}-2m_{\text{N}})^3/\sqrt{s_{\textit{NN}}}\right]^{1/4}~.
\end{equation}

Since the number of degrees of freedom $g$ is higher
for the quark-gluon plasma than for confined matter, it is also expected that the entropy
density of the produced final state at given energy density should also be higher 
in the first case. The following simple relation describes the expected dependence~\cite{Gazdzicki:1995ze}:
\begin{equation}
    S/V \sim g^{1/4} ~ F~.
    \label{eq:entropy}
\end{equation}

Therefore, the entropy and information regarding the state of matter formed in the early stage
of a collision should be reflected in the number of produced pions normalized to the volume
of the system. This intuitive argument was quantified in the
Statistical Model of the Early Stage (SMES)~\cite{Gazdzicki:1998vd}. The increase with
collision energy of the mean number of produced pions $\langle \pi \rangle$, normalized
to the number of wounded nucleons $\langle W \rangle$~\cite{Bialas:1976ed} is expected
to be linear when plotted against $F$.
The rate of increase is related to the number of degrees of freedom as given by Eq.~\ref{eq:entropy}.
This simple prediction is modified at low collision energies when
absorption of pions in the hadronic matter is expected to significantly decrease the final pion yield~\cite{Gazdzicki:1998vd}.

\begin{figure}[h]
  \centering
	\begin{minipage}{0.75\textwidth}
    \centering
    \includegraphics[width=0.98\textwidth]{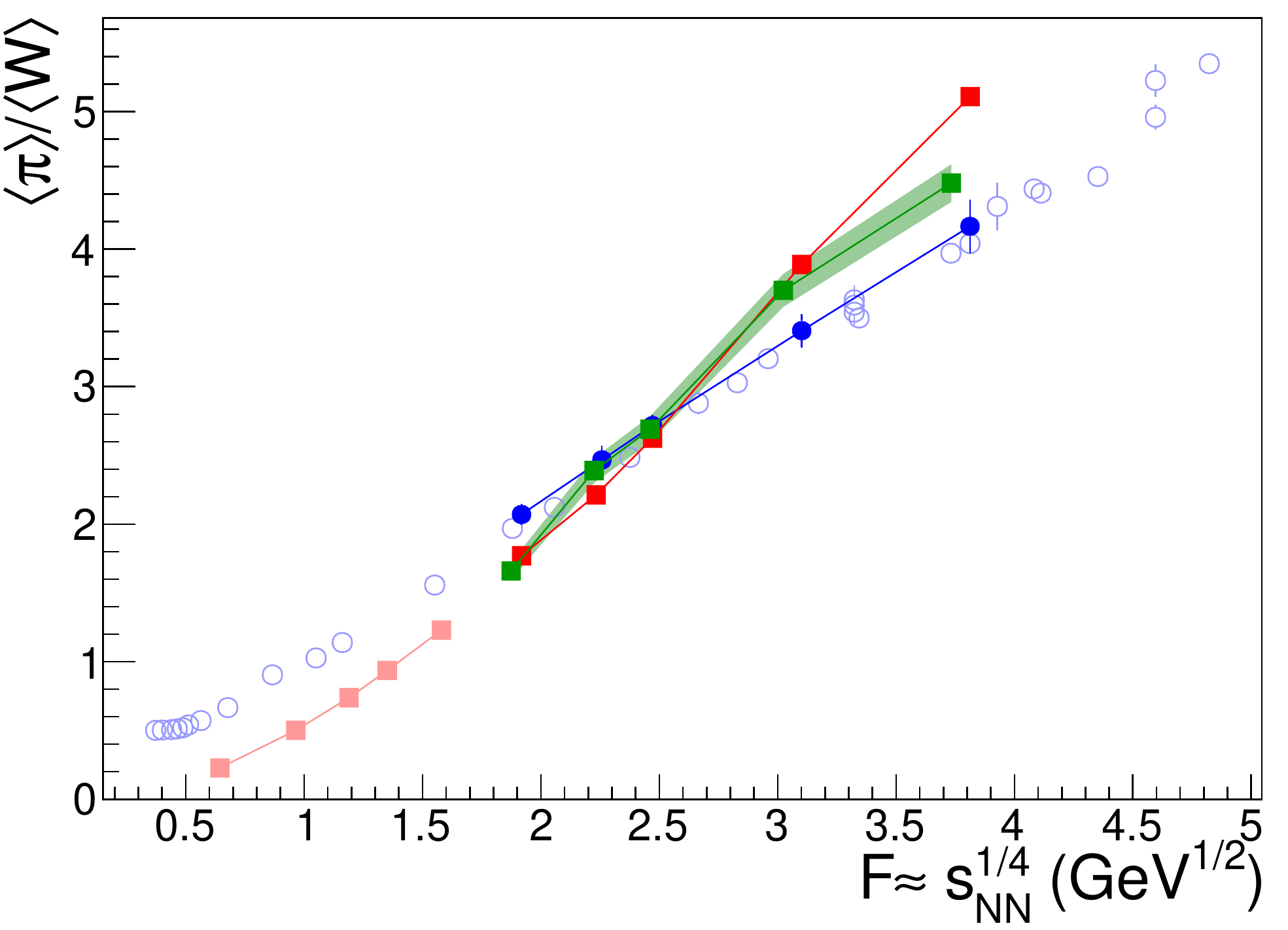}
	\end{minipage}
	\begin{minipage}{0.22\textwidth}
		\normalsize
		NA61/SHINE
	  \begin{itemize}
	    \item[\tiny\textcolor{kGreen}{\SquareSolid}] Be+Be
	    \item[\tiny\textcolor{kBlue}{\CircleSolid}] \textit{N+N}~\cite{Aduszkiewicz:2017sei}
		\end{itemize}
		WORLD
	  \begin{itemize}
	    \item[\tiny\textcolor{kBlue}{\CircleShadow}] \textit{N+N}~\cite{Afanasiev:2002mx}
	    \item[\tiny\textcolor{kRed}{\SquareSolid}] Pb+Pb~\cite{Afanasiev:2002mx,Alt:2007aa}
	    \item[\tiny\textcolor{kAuAu}{\SquareSolid}] Au+Au~\cite{Gazdzicki:1996ak,Gorenstein:2003cu}
		\end{itemize}

	\end{minipage}

  \caption{The "kink" plot showing the ratio of pion multiplicity $\langle \pi \rangle$ to number of wounded nucleons $\langle W \rangle$
           versus the Fermi energy variable $F \approx s_{\textit{NN}}^{1/4}$ . Results from \textit{central} Be+Be collisions are
           compared to measurements in for inelastic nucleon-nucleon reactions and \textit{central} collisions of heavier nuclei. Results of Be+Be collisions are shown with statistical (vertical lines) and systematic (shaded band) uncertainties. All other results are presented with total uncertainty.}
  \label{fig:kinkNew}
\end{figure}

Figure~\ref{fig:kinkNew} displays the ratio of mean pion multiplicitys\footnote{The mean number 
of produced pions $\langle \pi \rangle$ is calculated as
$\langle \pi \rangle = 3 \cdot \langle \pim \rangle$ for nucleus-nucleus collisions and
$\langle \pi \rangle = 1.5 \cdot (\langle \pip \rangle + \langle \pim \rangle)$ for \NN interactions} to the number of wounded nucleon
as a function of $F$.

The new measurements of the $\avg{\pi}/\avg{W}$ ratio in \textit{central} Be+Be collisions are compared to a compilation of results from central 
Pb+Pb (Au+Au) collisions and inelastic nucleon-nucleon interactions.
Above $F \approx 2$~\GeV$^{1/2}$ the slope of the Pb+Pb dependence is about a factor 1.3 higher than for
nucleon-nucleon interactions. The ratio  in \textit{central} Be+Be collisions  follows the ratio 
for central Pb+Pb collisions up to $F \approx 3$~\GeV$^{1/2}$. Thus in this energy range the Be+Be 
slope is also by a factor 1.3 higher than in \NN collisions.  
However this behaviour seems to change at the top SPS energy where the slope for  
the Be+Be ratio decreases to the one observed in \NN interactions. Note, that $\avg{W}$ is not 
a measured quantity, but has to be derived from models. Here the \Epos model was used (see Sec.~\ref{sec:wounded}).
Therefore the ratio $\avg{\pi}/\avg{W}$ is directly model dependent and this dependence increases with decreasing
nuclear mass number of colliding nuclei. For Be+Be collisions using different models leads to variation of the ratio
of up to 10\%, which is comparable to the difference between results on Be+Be and Pb+Pb collisions.

New results on Ar+Sc and Xe+La collisions 
from \NASixtyOne will be available soon and are expected to clarify the energy and system size dependence 
of $\avg{\pi}/\avg{W}$ which is a measure of the entropy of the produced fireball.
\section{Summary and conclusions}\label{sec:summary}

The \NASixtyOne experiment at the CERN SPS measured spectra and multiplicities 
of $\pi^{-}$ mesons produced in the 5\% most \textit{central} $^7$Be+$^9$Be collisions 
at beam energies of 19$A$, 30$A$, 40$A$, 75$A$ and 150\AGeVc using the 
so-called $h^-$ method. This is the first step in the systematic study of the phase diagram
of hadronic matter and the first such measurement in Be+Be collisions.

The normalized width of the rapidity distribution $\sigma_y/y_{beam}$ decreases with increasing collision energy and the values lie between the results for inelastic nucleon-nucleon 
and central Pb+Pb collisions. The average transverse mass $\langle m_T \rangle - m$ versus
collision energy shows a plateau in the SPS energy range at a similar level like in
inelastic nucleon-nucleon and central Pb+Pb collisions. The mean multiplicity of pions per wounded nucleon 
in \textit{central} $^7$Be+$^9$Be collisions rises linearly with the Fermi energy variable $F$ 
and is close to the one in \textit{central} Pb+Pb collisions expect for the top SPS energy, where it is closer to the \NN ratio. 

The results are discussed in the context of predictions for the onset of deconfinement at the CERN SPS
collision energies.

\clearpage
\section*{Acknowledgments}
We would like to thank the CERN EP, BE, HSE and EN Departments for the
strong support of NA61/SHINE. We thank A. Kubala-Kukuś and D. Banaś from the Institute of Physics,
Jan Kochanowski University for the target purity measurements with the
WDXRF technique.

This work was supported by
the Hungarian Scientific Research Fund (grant NKFIH 123842\slash123959),
the Polish Ministry of Science
and Higher Education (grants 667\slash N-CERN\slash2010\slash0,
NN\,202\,48\,4339, NN\,202\,23\,1837 and DIR\slash WK\slash 2016\slash 2017\slash 10-1), the National Science Centre Poland (grants~2014\slash14\slash E\slash ST2\slash00018, 2014\slash15\slash B\slash ST2 \slash\- 02537 and
2015\slash18\slash M\slash ST2\slash00125, 2015\slash 19\slash N\slash ST2 \slash01689, 2016\slash23\slash B\slash ST2\slash00692,
2017\slash\- 25\slash N\slash\- ST2\slash\- 02575,
2018\slash 30\slash A\slash ST2\slash 00226,
2018\slash 31\slash G\slash ST2\slash 03910),
the Russian Science Foundation, grant 16-12-10176 and 17-72-20045,
the Russian Academy of Science and the
Russian Foundation for Basic Research (grants 08-02-00018, 09-02-00664
and 12-02-91503-CERN),
the Russian Foundation for Basic Research (RFBR) funding within the research project no. 18-02-00086,
the National Research Nuclear University MEPhI in the framework of the Russian Academic Excellence Project (contract No.\ 02.a03.21.0005, 27.08.2013),
the Ministry of Science and Higher Education of the Russian Federation, Project "Fundamental properties of elementary particles and cosmology" No 0723-2020-0041,
the European Union's Horizon 2020 research and innovation programme under grant agreement No. 871072,
the Ministry of Education, Culture, Sports,
Science and Tech\-no\-lo\-gy, Japan, Grant-in-Aid for Sci\-en\-ti\-fic
Research (grants 18071005, 19034011, 19740162, 20740160 and 20039012),
the German Research Foundation (grant GA\,1480/8-1), the
Bulgarian Nuclear Regulatory Agency and the Joint Institute for
Nuclear Research, Dubna (bilateral contract No. 4799-1-18\slash 20),
Bulgarian National Science Fund (grant DN08/11), Ministry of Education
and Science of the Republic of Serbia (grant OI171002), Swiss
Nationalfonds Foundation (grant 200020\-117913/1), ETH Research Grant
TH-01\,07-3 and the Fermi National Accelerator Laboratory (Fermilab), a U.S. Department of Energy, Office of Science, HEP User Facility managed by Fermi Research Alliance, LLC (FRA), acting under Contract No. DE-AC02-07CH11359 and the IN2P3-CNRS (France).

\clearpage

\bibliographystyle{include/na61Utphys}
{\footnotesize\raggedright
\bibliography{include/na61References}
}


\end{document}